\RequirePackage{snapshot}
\documentclass[%
 reprint,
amsmath,amssymb,
aps,
prx,
nobalancelastpage,
onecolumn,
nofootinbib
]{revtex4-2}

\usepackage{graphicx}
\usepackage{dcolumn}
\usepackage{bm}
\usepackage[caption=false]{subfig}
\usepackage{url}
\usepackage{overpic}
\usepackage[caption=false]{subfig}
\usepackage{xcolor}
\usepackage{enumitem}
\usepackage{placeins}

\DeclareMathAlphabet{\mymathbb}{U}{bbold}{m}{n} 

\makeatletter
\def\bibsection{%
   \par
   \begingroup
    \baselineskip26\p@
    \bib@device{\hsize}{72\p@}%
   \endgroup
   \nobreak\@nobreaktrue
   \addvspace{19\p@}%
  }%
\makeatother

\begin{document}

\newcommand{\ket}[1]{| #1 \rangle}
\newcommand{\bra}[1]{\langle #1 |}
\newcommand{\1}{\mymathbb{1}} 

\title{Quantum data centres: a simulation-based comparative noise analysis} 

\author{K. Campbell}
\email{elkmc@leeds.ac.uk}
\author{A. Lawey}
\author{M. Razavi}
\email{m.razavi@leeds.ac.uk}
\affiliation{School of Electronic and Electrical Engineering, University of Leeds, Leeds, LS29JT, United Kingdom}

\begin{abstract}
Quantum data centres (QDCs) could overcome the scalability challenges of modern quantum computers. Single-processor monolithic quantum computers are affected by increased cross talk and difficulty of implementing gates when the number of qubits is increased. In a QDC, multiple quantum processing units (QPUs) are linked together over short distances, allowing the total number of computational qubits to be increased without increasing the number of qubits on any one processor. In doing so, the error incurred by operations at each QPU can be kept small, however additional noise will be added to the system due to the latency cost and errors incurred during inter-QPU entanglement distribution. We investigate the relative impact of these different types of noise using a classically simulated QDC with two QPUs and compare the robustness to noise of the two main ways of implementing remote gates, cat-comm and TP-comm. We find that considering the quantity of gates or inter-QPU entangled links is often inadequate to predict the output fidelity from a
quantum circuit and infer that an improved understanding of error propagation during distributed quantum circuits may represent a significant optimisation opportunity for compilation.
\end{abstract}

\maketitle

\section{Introduction} \label{sec:introduction}

Quantum computing offers the chance to  revolutionise many fields, from chemistry and economics to cybersecurity and manufacturing \cite{commercialApplicationsQC},  through its ability to simulate otherwise inimitable quantum systems \cite{quantumSimulation} and efficiently solve certain problems significantly faster than conventional, classical, computers \cite{ShorAlogrithm,GroverAlgorithm}. However, tapping into the potential of quantum computing requires more qubits than are available in current devices. 

A key limitation on the number of qubits to date has been the difficulty in avoiding unwanted interactions between qubits as the number of qubits or parallel gates within the same control system is increased. The correlated errors arising from such interactions are known as cross talk errors \cite{Sarovar2020detectingcrosstalk} and present a significant barrier to quantum error correction, which typically assumes errors to be localised and small \cite{QuantumErrorCorrectionIntroductoryGuide}. Without quantum error correction, errors propagate rapidly, making it challenging to obtain a meaningful output from circuits acting on data sets large enough to have useful applications \cite{LandauerArgumentAgainstQC, QCFactoringAndDecoherence, 9qubitShorErrorCorrection}. Therefore, simply adding more qubits to a quantum computer is not useful if the errors associated with those qubits are not adequately small.

A promising way of increasing the number of high-quality computational qubits is distributed quantum computing. A distributed quantum computer links together multiple distinct quantum computers, which are often referred to as quantum processing units (QPUs) when used in this way. The linking of QPUs creates a larger effective device with more qubits than any of its constituent components. This avoids the need to have many qubits share the same control system and so reduces the number of unwanted interactions. Thus, distributed quantum computing offers a scalable way of increasing the number of qubits while retaining their quality and the quality of local (intra-QPU) gates between qubits on the same QPU.

The price of these improvements relative to the single-processor, or monolithic, quantum computing paradigm is that the need for inter-QPU communication imposes additional sources of error. Whenever a multi-qubit gate acts on qubits that belong to different QPUs, some communication must occur between those QPUs. In most existing proposals for distributed quantum computers, this inter-QPU communication will consist of classical communication and distributing entanglement between qubits on each QPU. 

\begin{figure}
    \centering
    \begin{overpic}[scale=1.0]{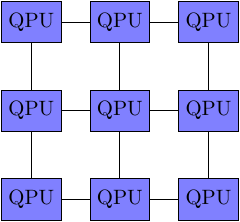}%
        \put(-15, 80){(a)}    
    \end{overpic}\hspace{4em}%
    \begin{overpic}[scale=1.0]{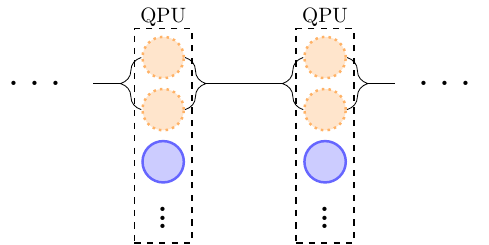}
        \put(18, 40){(b)}
    \end{overpic}
    \caption{(a) A quantum data centre. (b) The linking of two QPUs in a QDC. Communication qubits are shown in orange with a dotted border. The communication qubits on each QPU are linked together via a quantum-classical connection through which entanglement is distributed. Also present on each QPU are processing qubits, shown in blue with a solid border. Processing qubits are  manipulated with local gates, and are used in much the same way as qubits on a monolithic device. The number of communication qubits is typically more limited than the number of processing qubits.}
    \label{fig:QDC}
\end{figure}

Distributing entanglement between spatially separated entities is technologically challenging. One such challenge is that loss will occur when qubits are transported between QPUs. However, if increasing computational power is the main concern, there is little need to have a large inter-QPU separation, and so channel loss can be kept small. We shall refer to a distributed quantum computer where the distance between its constituent QPUs is sufficiently small that all of the QPUs can be kept in a single warehouse as a quantum data centre (QDC) \footnote{Alternative definitions of quantum data centres exist \cite{altQDCdef}.}.  An example of a QDC is shown in Fig. \ref{fig:QDC}(a). %
In the future, QDCs may be used to provide cloud computing services, but here, we focus on scenarios where the entire QDC is used to implement a single algorithm. 

The linking of QPUs is shown in more detail in Fig. \ref{fig:QDC}(b). As illustrated in Fig. \ref{fig:QDC}(b), it is often helpful to segregate qubits into three different types: processing, communication, and flying qubits. Processing and communication qubits are static and exist within the confines of a QPU. Processing qubits \footnote{Processing qubits are often referred to as data qubits in the literature \cite{TimeSlicedPartitioning, AutoComm, Ferrari, QuComm, ModularDQCcompilingFramework}. However, we feel that the term data qubits is somewhat of a misnomer, implying that the qubits in question are used only for storage. Therefore, we use the term processing qubits throughout this work.} are exclusively used to carry out the same function that qubits in a monolithic quantum computer would do. They store quantum information and are manipulated using quantum gates. Communication qubits are reserved for being entangled with communication qubits on other QPUs. They are also manipulated with quantum gates as part of inter-QPU gate implementations and may participate in multi-qubit gates with processing qubits. Flying qubits are able to move between QPUs and are used to facilitate entanglement distribution between pairs of communication qubits. Most existing schemes use optical photons as flying qubits, and photons are likely to be uniquely well suited for this role, given their unparalleled speed in low-noise media and extensive use in classical telecommunications. Therefore, we assume photons are the flying qubits in the rest of this work. 

Another challenge of distributing entanglement between QPUs is that, due to noise during the entanglement distribution process, the entangled state distributed will not be maximally entangled. Consequently, there is an inherent trade-off in any QDC between increasing the intra-QPU and inter-QPU errors. Intra-QPU errors can be reduced by keeping the individual QPUs small. However, more QPUs would then be required to obtain a given number of computational qubits. Splitting the computation between more QPUs typically means that there will be more inter-QPU gates. As such there will be more inter-QPU communication and the inter-QPU errors will increase. Conversely, when trying to obtain a given number of computational qubits, inter-QPU errors can be reduced by using larger QPUs, leading to higher intra-QPU errors.

To explore this trade off, it is important to understand the relative impact of different errors in various contexts. To this end, we make the following contributions: 

\begin{enumerate}
    \item We demonstrate the limitations of first-order error analysis relative to full classical simulation in the QDC context. 
    \item We determine the relative impact of noise in inter-QPU entanglement, intra-QPU CNOT gates, and decohering qubits (as time evolves) on QDC performance.
    \item We compare different approaches to inter-QPU (remote) gate operation by calculating the output fidelity for each approach.
    \item We provide a pessimistic estimate for the number of inter-QPU gates that could realistically be implemented on a QDC in the upcoming years while retaining a useful output.
\end{enumerate}

The rest of the paper is organised as follows. Firstly, we discuss the relevant literature in Sec. \ref{sec:related_work}. Following that,  the problem at hand is elucidated in Sec. \ref{sec:problem_set_up}. Then, the results of our analysis are presented and discussed in Secs. \ref{sec:single_remote_gates} and \ref{sec:larger_quantum_circuits} for individual inter-QPU gates and larger quantum circuits, respectively. Finally, we summarise our findings and discuss their implications in Sec. \ref{sec:conclusion}.

\section{Related literature and problem motivation}\label{sec:related_work}

In a QDC, careful consideration must be given to the implementation of multi-qubit gates. In particular, however a quantum circuit is partitioned between QPUs, it is likely that there will be multi-qubit gates that act on qubits assigned to different QPUs. We will henceforth refer to such inter-QPU gates as remote gates and refer to intra-QPU gates as local gates.

Most methodologies for carrying out remote gates boil down to one of two schemes. In the first scheme, which we call TP-comm (as in Ref. \cite{AutoComm}), quantum teleportation \cite{teleportationProposal, firstExperimentalTeleportation} is used to transfer quantum states from one QPU to another, and then the gate is conducted locally. Alternatively, as advocated for in Ref. \cite{heightIncreaseNonLocalGates}, an entangled state can be maintained between the two QPUs while the local gate is applied. This latter idea is extended in Ref. \cite{ generalizedGHZandDQC}, where it is shown that the remote gate methodology proposed in Ref. \cite{heightIncreaseNonLocalGates} can be decomposed into subprotocols \cite{generalizedGHZandDQC}, called cat-entanglement and cat-disentanglement, between which one or more local gates are conducted. In doing so, one can implement Shor's algorithm using fewer inter-QPU entangled pairs \cite{catCommDistributedShor}---which are referred to as entangled bits or ebits \footnote{Technically, ebits could refer to intra-QPU entangled pairs too but in this paper, it is the inter-QPU ebits that are relevant. Consequently, ebit is used to refer to inter-QPU entangled pairs specifically in this work.}. The term cat-comm \cite{generalizedGHZandDQC, catCommDistributedShor} is used here to describe implementing a remote gate or gates in this way. 
 
Subsequent work \cite{reordingPartitionDQCcircuitsTPcommOnly,OptimisedQuantumCircuitPartitioning,TimeSlicedPartitioning, AutomatedDistributionCircuitsViaHypergraph, EfficientDistributionQuantumCircuits, ConnectivityMatrixCompiler, OptimisingTeleportaionCost,davis2023dqcPartitioning,Ferrari,AutoComm,QuComm,ModularDQCcompilingFramework}, compiles arbitrary quantum circuits into a form suitable for distributed quantum computing, with a view to minimising the number of ebits used. Many compiler proposals \cite{Ferrari, ModularDQCcompilingFramework, AutoComm} involve a step we call communication assignment \cite{Ferrari, ModularDQCcompilingFramework, AutoComm}, in which cat-comm or TP-comm is assigned to remote gates [27] or groups of remote gates [14, 29]. Current work on communication assignment \cite{Ferrari, ModularDQCcompilingFramework, AutoComm} compares cat-comm or TP-comm with respect to the number of ebits they each use or their latency cost. Both metrics are intimately related due to the disproportionate impact of entanglement distribution on latency. Therefore, using either metric implicitly assumes that latency is the most significant optimisation parameter for distributed quantum computing or that entanglement error is the most impactful source of noise \footnote{Latency considerations are complicated in the quantum computing context by the greater difficulty correcting errors relative to classical computing. In current quantum computers, more noise means more runs of the circuit are required to obtain a given precision in the result, increasing the latency. Moreover, quantum error correction, which is expected to be incorporated into future quantum computers, only works if the errors are small \cite{QuantumErrorCorrectionIntroductoryGuide}. Therefore, simply arguing that a process takes a long time relative to other processes does not automatically ensure that the overall latency will be lower.}. Such assumptions may or may not be true in the QDC setting. Moreover, there are many cases where cat-comm and TP-comm have the same ebit and/or latency cost for implementing a remote gate or group of remote gates \cite{AutoComm}, and so previous work gives no clear way of distinguishing between cat-comm and TP-comm in such circumstances.

In our work, we instead consider the output fidelity, $F_{\rm out}$, after a remote gate or quantum circuit. $F_{\rm out}$ indicates the similarity between the ideal output, $\rho_{\rm ideal}$, and the actual output, $\rho_{\rm noisy}$, obtained in the presence of noise. Specifically \cite{fidelityDef},%
\begin{equation} \label{eq:fidelity_definition}
F_{\rm out} = \left( \rm Tr\sqrt{(\rho_{\rm ideal})^{\frac{1}{2}} \rho_{\rm noisy} (\rho_{\rm ideal})^{\frac{1}{2}}}\right)^2.
\end{equation}%
Crucially, this allows us to differentiate between cat-comm and TP-comm when the number of ebits is the same, and to understand the robustness of each scheme to different types of error. Using $F_{\rm out}$, we can also compare the relative impact of different error types. With this far more detailed understanding of cat-comm, TP-comm, and errors, we hope to facilitate more optimal compilation heuristics in the future.

Our work is not the first to consider the errors in the QDC setting. Some of the earliest work to do so is Ref. \cite{DQCoverNoisyChannels}. However, the results of Ref. \cite{DQCoverNoisyChannels} are specific to the quantum phase estimation algorithm \cite{quantumPhaseEstimation, NielsenChuang} and fewer types of noise are considered. The only local error considered is time-dependent dephasing, for which effective suppression techniques exist \cite{dynamicalDecouplingSurvey}, and comparisons are conducted not between different types of noise, as we do, but between different implementations of the quantum phase algorithm---one in which QPUs can only communicate classically and another in which QPUs can also share quantum entanglement. Consequently, it may be difficult to ascertain from Ref. \cite{DQCoverNoisyChannels} where optimisation efforts for hardware and compilation should focus. Moreover, the figure of merit used by Ref. \cite{DQCoverNoisyChannels} is a cost function which treats the `cost' of running a computation on each QPU and the `cost' of sending measurement outcomes to a central QPU as abstract parameters and relates these `costs' to the number of QPUs. This degree of abstraction makes the a priori linking of results to experiment challenging and implicitly assumes that intra-QPU errors are uncorrelated with the number of QPUs. In other words, they assume that the number of qubits and local gates on each QPU are fixed. This neglects the fact that cat-comm and TP-comm require additional local gates, rather than just ebits.  By contrast, $F_{\rm out}$, which we use, can be experimentally measured and has clear physical significance.

Later work \cite{decoherenceSharedEntanglementDQCphaseEstimation} uses classical simulation to find the output fidelity of two different distributed implementations of the quantum phase estimation algorithms on a QDC. In one implementation, cat-entanglement and cat-disentanglement are conducted for each remote gate, whereas in the other implementation several remote gates are amalgamated by deferring cat-disentanglement until later in the circuit. The authors consider the impact of adding depolarising noise to inter-nodal entanglement and find that $F_{\rm out}$ declines less quickly with increasing depolarisation probability when the remote gates are amalgamated than when they are not. They also quantify the decrease in output fidelity with the number of devices for different depolarisation probabilities. However, no comparison between cat-comm and TP-comm is made and only entanglement error is considered.

A further limitation of Refs. \cite{DQCoverNoisyChannels, decoherenceSharedEntanglementDQCphaseEstimation} is the specialisation to the quantum phase estimation algorithm. To  generalise results, it is necessary to understand how the building blocks of circuits work, as well as how errors propagate through circuits with different structures. For this reason, we consider the impact of errors on individual remote gates, as well as on a wide variety of larger circuits. In this way, we are able to more deeply and generally understand the impacts of different errors and communication schemes than previous work.   

\section{Quantum data centres: main components}
\label{sec:problem_set_up}

To understand where optimisation of future QDC manufacturing and compilation efforts should focus, it is important to understand the impact of different types of noise on a QDC. In this work, we quantify the extent to which the output of distributed quantum circuits is degraded by the different types of noise present in a QDC. The metric we use to do this is the output error, characterised by $1 - F_\mathrm{out}$. We also compare the robustness of cat-comm and TP-comm to error.

For simplicity, we assume that just two QPUs are present in the QDC we consider. We assume that all remote gates in a given quantum circuit are distributed using the same communication scheme and compare the impact of different communication schemes on the output error. Further details on the schemes considered are provided in Sec. \ref{sec:cat_comm_and_tp_comm}. We also consider different types of noise, which we model using the methods described in Sec. \ref{subsec:error_models}. Additional information about how entanglement is distributed between QPUs and how entanglement error could be reduced are discussed in Sec. \ref{subsec:entanglement_distribution}.

\subsection{Cat-comm and TP-comm circuits} \label{sec:cat_comm_and_tp_comm}

Comparing cat-comm and TP-comm is not as simple as it may appear at first glance.  In particular, there is some ambiguity in what it means to implement a remote gate using TP-comm. Therefore, in this work, we consider cat-comm and three different versions of TP-comm, which would be used in different scenarios within larger quantum circuits.

\begin{figure}
    \centering
    \begin{overpic}[scale=0.6]{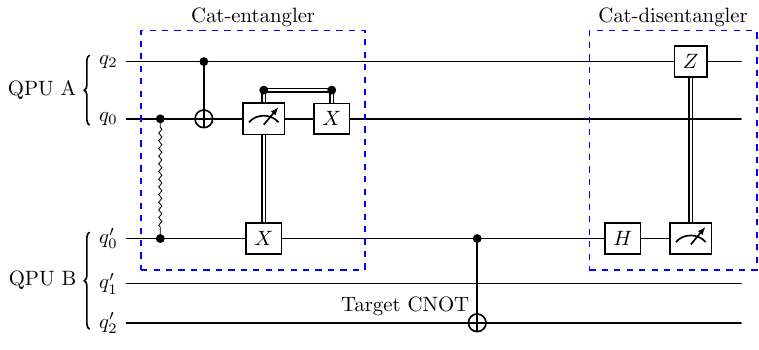}
        \put(-2.5, 43){(a)}
    \end{overpic}\hspace{3em}%
    \begin{overpic}[scale=0.6]{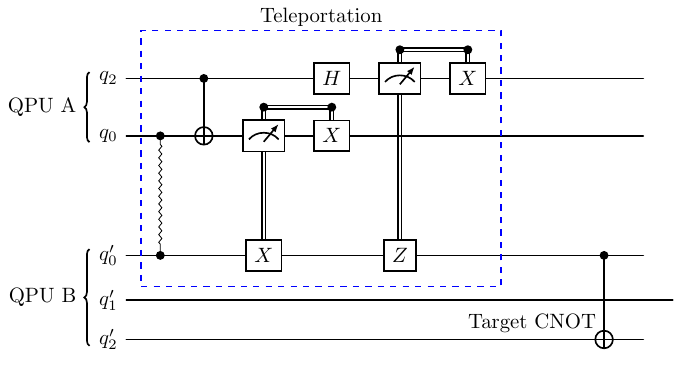}
        \put(-8.8, 47){(b)}
    \end{overpic}\\%
    \begin{overpic}[scale=0.6]{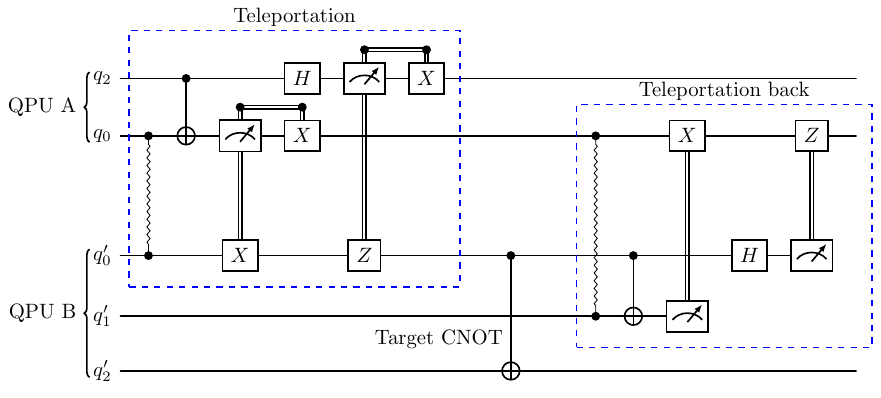}
        \put(5.2, 40){(c)}
    \end{overpic}%
    \begin{overpic}[scale=0.6]{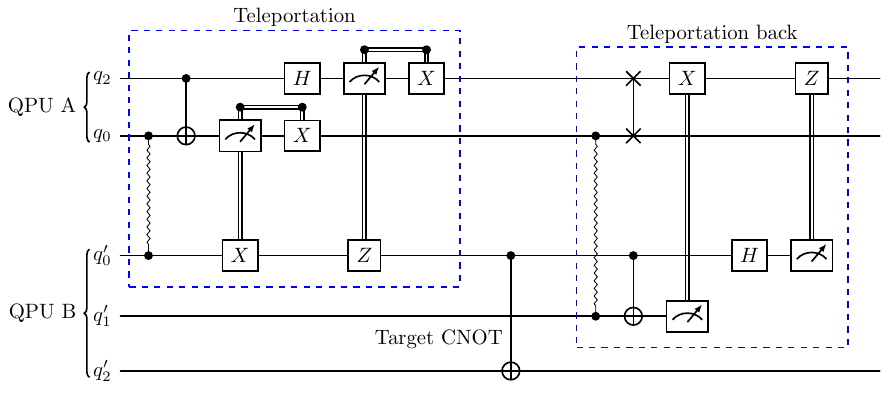}
        \put(5, 40){(d)}
    \end{overpic}%
    \caption{A remote CNOT gate implemented using: (a) cat-comm \cite{generalizedGHZandDQC, catCommDistributedShor}; (b) 1TP; (c) 2TP; and (d) TP-safe.  Zigzags represent ebits, which here, in the ideal case, are Bell pairs in the state $|\Phi^+\rangle = \frac{1}{\sqrt{2}}(|00\rangle + |11\rangle)$. Double lines represent classical communication. Gates classically connected to a measurement device activate on a measurement result of `1' only.}
    \label{fig:remote_cnot_implemented_in_different_ways}
\end{figure}

Consider first a remote CNOT gate implemented using cat-comm. The corresponding circuit diagram is shown in Fig. \ref{fig:remote_cnot_implemented_in_different_ways}(a). %
The objective of this circuit is to implement a CNOT gate between QPU A's processing qubit, $q_2$, and QPU B's processing qubit, $q_2'$. $q_2$ is the control qubit and is in the arbitrary initial state $\alpha \ket{0} + \beta \ket{1}$, where $\alpha$ and $\beta$ are complex numbers such that $|\alpha|^2 + |\beta|^2 = 1$. $q_2'$ is the target qubit and is in the arbitary initial state $\ket{\chi}$. The communication qubits $q_0$, from QPU A, and $q_0'$ from QPU B are also used. 

The cat-comm remote CNOT gate consists of three subroutines: cat-entanglement, implementation of the CNOT locally, and then cat-disentanglement. Cat-entanglement entangles $q_2$ with $q_0'$, so that they share the cat-like state, $\alpha \ket{00} + \beta \ket{11}$.  In this way, the state of the control qubit, $q_2$, on QPU A, is shared with $q_0'$, on QPU B. The CNOT gate can then be conducted locally between $q_0'$ and $q_2'$ before the entanglement between $q_2$ and $q_0'$ is destroyed by the cat-disentanglement process shown in figure \ref{fig:remote_cnot_implemented_in_different_ways}(a). A similar process can be used for any controlled-unitary (CU) gate.

 Fig. \ref{fig:remote_cnot_implemented_in_different_ways}(b) shows the circuit diagram for a CNOT gate implmented using our first variant of TP-comm, which we call 1TP. The same qubits that were used in cat-comm are again considered here. In 1TP, the state of $q_2$ is teleported to $q_0'$ and then the CNOT gate is applied locally on QPU B. More generally, any local gate could be conducted after the teleportation. Unlike cat-comm, it is not necessary for the gate to be a CU gate. 

The next TP-comm scheme, 2TP, is depicted in Fig. \ref{fig:remote_cnot_implemented_in_different_ways}(c), again for the specific case of a remote CNOT. 2TP involves first applying 1TP and then teleporting the state of the control qubit back to QPU A. More specifically, the second teleportation is from $q_0'$, on QPU B, to the communication qubit, $q_0$, on QPU A. An additional communication qubit is also needed on QPU B to facilitate the second teleportation. 

Finally, a remote CNOT gate implemented using TP-safe, the final TP-comm scheme, is shown in Fig. \ref{fig:remote_cnot_implemented_in_different_ways}(d). TP-safe is almost identical to 2TP except that the states of $q_0$ and $q_2$ are swapped during the second teleportation, so that the state of the control qubit is restored to where it first started: the processing qubit $q_2$. This is done using a SWAP gate. 

It is not immediately obvious which of the TP-comm schemes discussed here is most comparable with cat-comm. On the one hand, cat-comm and 1TP both use exactly the same quantities and types of ebits, measurements, classical communications, and local gates. On the other hand, after cat-comm, $q_0'$ is free to  host any further ebits needed for additional remote gates. By contrast, after 1TP, $q_0'$ is needed to store the teleported state of the control qubit. To avoid overwriting this crucial quantum information when conducting future remote gates, a second teleportation can be used to free $q_0'$. As we consider only two QPUs, the second teleportation must be back to QPU A, as is done in 2TP. More generally, the teleportation could be to any other QPU. Teleportation back to the original QPU is the least favourable scenario, but it can often be necessary when large quantum circuits are compiled. After 2TP, $q_0'$ on QPU B is left free, but $q_0$ on QPU A is occupied. To circumvent this issue, we can carry out a SWAP gate between $q_0$ and $q_2$ to restore the teleported state back to where it started prior to the remote gate. This is safe to do because we know that $q_2$ was measured and then re-initialised during the first teleportation and so carries no non-trivial quantum information.

Cat-comm and TP-safe can always be used, regardless of what subsequent remote gates are scheduled. 1TP and 2TP cannot always be used but are less resource-intensive than TP-safe. In circumstances where no further remote gates are scheduled on the QPU that holds a teleported state, there is no need to free the communication qubits, and it will be better to use 1TP or 2TP than TP-safe. Otherwise, TP-safe is the only TP-comm scheme that could be used. TP-safe represents the very worst-case scenario that may be necessary during compilation of a distributed quantum circuit using TP-comm, and 1TP is the very best. By considering 1TP and TP-safe, we aim to provide an upper and lower bound on the performance of TP-comm, while 2TP sits somewhere between those two extremes.

\subsection{Error models}
\label{subsec:error_models}

Three principle sources of error are considered in this work: imperfections in the distributed ebits, imperfect implementation of local two-qubit gates, and time-dependent memory decoherence.

In cat-comm and TP-comm, we assume that, in the absence of noise, the ebits used have the pure state $\ket{\Phi^+} = \frac{1}{\sqrt{2}}(\ket{00} + \ket{11})$.  When modelling imperfections in the ebits, we assume that their state is a Werner state of the form \begin{equation}
\label{eq:werner_state}
\begin{split}
    \rho_{\rm w} = & F_{\rm w} \ket{\Phi^+} \bra{\Phi^+} 
    + \frac{1 - F_{\rm w}}{3}\left( \ket{\Phi^-} \bra{\Phi^-} \right. \\%
     & \qquad \left. + \ket{\Psi^+} \bra{\Psi^+}
      + \ket{\Psi^-}\bra{\Psi^-} \right),
\end{split}
\end{equation}
where $0 \leq F_{\rm w} \leq 1$, and  $\ket{\Phi^{\pm}} =  \frac{1}{\sqrt{2}}(\ket{00} \pm \ket{11})$, $\ket{\Psi^{\pm}} = \frac{1}{\sqrt{2}}( \ket{01} \pm \ket{10})$ are the Bell states, which form a basis for all two-qubit entangled states. A pair of qubits which are collectively in a Bell state is referred to as a Bell pair. We refer to $\epsilon_{\mathrm{ebit}} = 1 - F_{\rm w}$ as the entanglement error. 

In reality, the form of the non-ideal entangled states is unlikely to be exactly known and may not be so symmetrical in nature. Nonetheless, any mixture of bipartite entangled states can be transformed into the Werner state via a random bilateral rotation of each of the entangled qubits \cite{distillation}.  Bilateral rotations require only single-qubit gates with relatively low errors and are sometimes done as part of entanglement distillation \cite{distillation, QMultiplexingToNetwork}, which is likely to be necessary in the QDC context. Moreover, less symmetrical imperfections are often advantageous for error correction or mitigation \cite{BenjaminHierarchicalSurfaceCode}. As such, the Werner state is likely to represent a worst-case scenario, which is useful for benchmarking purposes.

Non-ideal application of local gates is modelled by applying a depolarisation channel of the form \cite{imperfectRepeaters}
\begin{equation} \label{eq:depol_2_qubit_gate_error}
    \rho_{\rm in} \rightarrow (1 - \epsilon_{\mathrm{cnot}}) U_{i, j} \rho_{\rm in} U_{i, j}^{\dagger} + 
    \frac{\epsilon_{\mathrm{cnot}}}{4} \mathrm{Tr}_{i,j}(\rho_{\mathrm{in}}) \otimes \1_{i, j},
\end{equation}
to all local two-qubit gates, which in this work are all CNOT gates. Here, $\epsilon_{\mathrm{cnot}}$ is the error probability; $U_{i, j}$ is an ideal two-qubit gate operation on the $i$th and $j$th qubits; $\rho_{\rm in}$ is the input density matrix; $\mathrm{Tr}_{i, j}$ is the partial trace \cite{NielsenChuang} with respect to the subspaces of qubits $i$ and $j$; and $\1_{i, j}$ is the identity operation acting on the these subspaces.  

Memory decoherence on qubit $k$ is modelled using the depolarisation channel 
\begin{equation} \label{eq:mem_depol_channel}
    \rho_{\mathrm{in}} \rightarrow e^{- \Delta t \: r} \rho_{\mathrm{in}} + \frac{(1-e^{- \Delta t \: r})}{2} \mathrm{Tr}_{k} (\rho_{\mathrm{in}}) \otimes \1_k,
\end{equation}
 where $\Delta t$ is the time since the initialisation of qubit $k$, $r$ is the hardware dependent depolarisation rate, and $\rho_{\mathrm{in}}$ is the input density matrix.

\subsection{Entanglement distribution and distillation} \label{subsec:entanglement_distribution}

Entanglement distribution can be carried out in several ways. For instance, photons can be entangled with communication qubits and then measured out, leaving the communication qubits entangled with one another. The measurement process used is typically a Bell state measurement (BSM) \cite{MohsenBookChapter}. A BSM interacts and measures a pair of qubits to ascertain whether those qubits are in one of the Bell states. In principle, a BSM can be conducted using a CNOT gate followed by a hadamard gate on the control qubit \cite{MohsenBookChapter}. This is known as a deterministic BSM. In practice, deterministic BSMs may be difficult to achieve and so partial BSMs using linear optics and photodetectors are frequently used. A partial BSM relies on post-selection of measurement results to probabilistically determine whether a pair of qubits are in one of the Bell states.

Regardless of whether a partial or deterministic BSM is used, most entanglement distribution schemes require the entire entanglement distribution process to be repeated several times. For partial BSMs, this is an intrinsic feature of their probabilistic nature. For deterministic BSMs, repetition is necessitated by the presence of loss in the communication channel. If photons do not reach the detectors in a BSM module, new photons must again be  entangled with communication qubits, and sent until they reach the detectors. In this way, loss is heralded by the lack of proper detection events.

The fidelity of the ebits can also be improved by using a more advanced heralding process and imposing more stringent fidelity requirements on the entangled states distributed \cite{MohsenBookChapter}. This will lower the success rate of entanglement, necessitating a greater number of entanglement distribution attempts. Alternatively, but similarly, entanglement distillation \cite{originalDistillationProposal, PrivacyAmpAndSecurityOverNoisyChannels, EntanglementPurificationForQuantumComms, QMultiplexingToNetwork} could be used. In entanglement distillation, multiple entangled pairs are distributed between QPUs and local gates are conducted between communication qubits, entangling them. Some of the qubits are then measured out and inferences are made about the fidelity of the remaining qubits. This process is repeated iteratively until the fidelity reaches the desired threshold. 

Other options for improving the fidelity of ebits include applying quantum error correction \cite{QuantumErrorCorrectionIntroductoryGuide} locally on the communication qubits for each QPU and encoding the ebit into a larger many-qubit entangled (cluster) state \cite{MohsenBookChapter}.

Regardless of how loss and noise are mitigated, there is a cost to be paid. All of the aforementioned schemes for improving the fidelity of ebits involve a latency cost due to repetitions of the entanglement process, decoding time, or the addition of local gates. Any additional local gates will also be imperfect, adding further noise into the system. Moreover, in many cases, redundant communication qubits and/or photons are also required. Therefore, there is always a trade-off to be made between improving ebit fidelity and paying the price of doing so. 

In this work, we abstract from the details of the entanglement distribution scheme used by assuming that communication qubits are entangled at a fixed rate $R$. The time taken for a given ebit to be distributed is assumed to always be $\frac{1}{R}$ from when the ebit is requested during the compilation process. Ebit requests are generated when the ebit is first needed, which is unfavourable for latency. That said, it does minimise the time that ebits have to decohere after they have been distributed. With these assumptions, all loss is incorporated into the value of $R$ rather than being explicitly modelled. Noise is accounted for by assuming that the ebit is in the Werner state as discussed in Sec. \ref{subsec:error_models}. 

Despite the abstractions we make, it is important to keep in mind the trade-offs inherent to altering $R$ or reducing the entanglement error. Even with current technology, entanglement error could be reduced from the current state-of-the-art values but, as discussed previously, this would come at the expense of a lower entanglement distribution rate and often requires additional qubits and local gates. These trade-offs are important to keep in mind when contemplating our results.

\section{Single remote gates: noise analysis} \label{sec:single_remote_gates}

To understand the behaviour of something as complex as a distributed quantum circuit, it is helpful to first investigate the constituent building blocks. With this in mind, we consider here a single remote CNOT gate, implemented using the schemes in Fig. \ref{fig:remote_cnot_implemented_in_different_ways}. The U3 + CNOT basis \cite{OpenQASM2.0_paper} is used, meaning that the SWAP gate from TP-safe (see Fig. \ref{fig:remote_cnot_implemented_in_different_ways}(d)) is decomposed into three CNOT gates \cite{OpenQASM2.0_paper}. All other remote gate schemes are conducted exactly as shown in Fig. \ref{fig:remote_cnot_implemented_in_different_ways}. Here, we first discuss the error analysis methods that we use in Sec. \ref{subsec:error_analysis_methods}.  Then, we present and analyse the results of our error analysis in Sec. \ref{subsec:single_remote_gate_results}.

\subsection{Error analysis}
\label{subsec:error_analysis_methods}

\subsubsection{Analytical approach}
\label{subsubsec:analytical_results}

A natural starting point for any error analysis is to derive analytical results where possible. Even analytical results related to a small subset of the broader problem can be used to verify the results obtained by other methods. Here, we present closed form analytical results for arbitrary remote controlled-unitary gates implemented using 1TP and cat-comm in the presence of entanglement error only. All other errors are assumed to be zero. We defer the derivation to Appendix \ref{app:analytical_derivations} for brevity. 

For 1TP, we obtain %
\begin{equation}
    \label{eq:1TP_analytical_F_out}
    F_{\rm out} = \frac{1 + 2F_{\rm w}}{3}.
\end{equation} %
Remarkably, this expression is independent of which controlled-unitary operation is applied or the input to that controlled-unitary operation.

For cat-comm, we make the additional limiting assumption that the processing qubits are initially separable, in order to make the calculation tractable. As such, the input state to the remote controlled-unitary gate can be represented as %
\begin{equation}
        (\alpha \ket{0} + \beta \ket{1})_{q_2} \otimes \ket{\chi}_{q_2'},
        \label{eq:general_separable_input_state}
\end{equation} %
where the subscripts $q_2$ and $q_2'$ refer to the qubits shown in figure \ref{fig:remote_cnot_implemented_in_different_ways}(b), $\alpha$ and $\beta$ are complex numbers such that $|\alpha|^2 + |\beta|^2 = 1$,  and $\ket{\chi}$ is an arbitrary pure state. With this assumption, we find that %
\begin{equation}
\begin{split}
    F_{\rm out} = F_{\rm w} + \frac{1 - F_{\rm w}}{3} \Biggl( & \left( |\alpha|^2 - |\beta|^2 \right)^2 + 2 |\alpha|^4 \left| \bra{\chi} U_{q_2'} \ket{\chi}_{q_2'}\right|^2 \\
    &+ 2 |\beta|^4 \left( \bra{\chi} U^{\dagger}_{q_2'} \ket{\chi}_{q_2'} \right)^2 \Biggr),
\end{split}
\label{eq:cat_analytical_F_out}
\end{equation} %
where $U_{q_2'}$ is the unitary operation applied to $q_2'$ during the remote CU gate. Unlike Eq. \eqref{eq:1TP_analytical_F_out}, this does depend on the type of CU gate applied and the input state of the control and target qubits.

Considering the case where a CNOT gate is applied and adding the further limiting assumption that $\ket{\chi}$ is an eigenstate of the computational basis, Eq. \eqref{eq:cat_analytical_F_out} simplifies to %
\begin{equation}
    \label{eq:cat_analtyical_F_out4CNOT}
    F_{\mathrm{out}} = F_{\mathrm{w}} + \frac{1 - F_{\mathrm{w}}}{3} \left( 2|\alpha|^2 - 1 \right)^2.
\end{equation}%
Although, it applies to a very specific situation, Eq. \eqref{eq:cat_analtyical_F_out4CNOT} is a convenient univariate expression for testing other error analysis tools.

\subsubsection{First-order approximations}

To begin extending beyond the limiting assumptions of the analytical analysis, it is instructive to make first-order approximations to the results. These first-order approximations carry assumptions of their own but allow local multi-qubit gate errors to also be considered. Here, and throughout the remainder of this work, we use the $\mathrm{U}_3$ $+$ CNOT set of gates and so the only multi-qubit gates considered are CNOTs.

For sufficiently small local gate and entanglement error rates, $\epsilon_{\mathrm{cnot}}$ and $\epsilon_{\mathrm{ebit}}$, respectively, one might expect the output fidelity, $F_{\rm out}$, to degrade linearly with the number of imperfect CNOT gates, $n_{\mathrm{cnot}}$, or ebits, $n_{\mathrm{ebit}}$, yielding the equation:%
\begin{equation} \label{eq:first_order_f_out}
    F_{\rm out} \approx  1 - n_{\mathrm{cnot}} \epsilon_{\mathrm{cnot}} - n_{\mathrm{ebit}} \epsilon_{\mathrm{ebit}}.
\end{equation}%

Alternatively, retaining the main desired terms in Eqs. \eqref{eq:werner_state} and \eqref{eq:depol_2_qubit_gate_error}, we can approximate $F_{\mathrm{out}}$ by%
\begin{equation} \label{eq:exp_approx}
    F_{\mathrm{out}} \approx (1-\epsilon_{\mathrm{ebit}})^{n_{\mathrm{ebit}}}(1 - \epsilon_{\mathrm{cnot}})^{n_{\mathrm{cnot}}}.
\end{equation}%
This is consistent with Eq. \eqref{eq:first_order_f_out} in the limit of $n_{\mathrm{ebit}} \epsilon_{\mathrm{ebit}}$ and $n_{\mathrm{cnot}} \epsilon_{\mathrm{cnot}}$ $<<<$ 1 and has been used to estimate the fidelity in the presence of gate errors for larger circuits \cite{Flannigan_2022}.

Both Eqs. \eqref{eq:first_order_f_out} and \eqref{eq:exp_approx} are quite general and can be used to describe local gate errors and entanglement errors for any quantum circuit that adheres to the assumptions used during derivation.

 Memory depolarisation is continuous rather than discrete like local gate errors and entanglement errors. For this reason, memory depolarisation is not well described by an equation with the same form as Eq. \eqref{eq:first_order_f_out} or Eq. \eqref{eq:exp_approx}, and we do not consider a first-order approximation for memory depolarisation in this work.

\subsubsection{Simulation}

In order to move past first-order approximations and enable extension to larger circuits, we generate numerical results using classical simulation. The simulator used is an event-based simulator built using Python 3.9 and the associated libraries: NetSquid \cite{netsquid} and nuqasm2 \cite{nuqasm2}. We modified nuqasm2 for our purposes. 

Our simulation package converts arbitrary monolithic quantum circuits, specified using openQASM 2 \cite{OpenQASM2.0_paper} or as a list of gate tuples, to compiled distributed quantum circuits. The use of communication qubits is handled automatically without explicit user specification and the subroutines associated with each type of remote gate specified in Fig. \ref{fig:remote_cnot_implemented_in_different_ways} can be automatically generated and scheduled. Manual specification of subroutines is also possible. All simulated hardware can have custom or stock noise models specified, and simulated QPUs come with a variety of options, including deciding the size of the communication and processing qubit allocations. Efforts have been made to retain the modularity of the NetSquid package and users can also specify their own compilers, QPUs and connections for use with our simulation package. Further details of our simulation package API and how to access the package will be presented in a separate work.

As the simulator is event-based, anything that happens during the simulation is allocated an amount of time it will take to occur. Time is discretised based on the duration of those events. For example, after a gate or entanglement distribution is carried out, all qubits in the circuit will experience an amount of memory depolarisation correlated to the amount of time that gate or entanglement distribution took. $\Delta t$ in the memory depolarisation model (Eq. \eqref{eq:mem_depol_channel}) is incrementally increased every time such an event occurs in the simulation. The events accounted for are: single-qubit gates, two-qubit gates (CNOTs), measurement, classical communication, and entanglement distribution. For simplicity, the classical communication latency is modeled using NetSquid's fibre delay model, which assumes classical communication occurs in $\frac{d}{2 \times 10^8 \mathrm{ms^{-1}}}$, where $d$ is the inter-QPU distance in metres. Entanglement distribution is carried out using the abstract model described in Sec. \ref{subsec:entanglement_distribution}. 

\subsection{Numerical results}
\label{subsec:single_remote_gate_results}

Armed with the tools described in Sec. \ref{subsec:error_analysis_methods}, we are able to obtain numerical values for the output error. All simulated results are obtained using either a laptop with 16GB of RAM and an AMD Ryzen 7 4700 processor or a desktop with 32GB of RAM and an i7 processor. 

We assume the simulated QDC has two QPUs each possessing just two communication qubits, which is believed to be a realistic limitation for near-term devices \cite{Ferrari, AutoComm}. We also assume that the processing qubits have an input state of \begin{equation}
    \label{eq:remote_CNOT_input_state}
    \frac{1}{\sqrt{2}} \left(  \ket{0} + \ket{1} \right)_{\mathrm{q_2}} \otimes \ket{0}_{\mathrm{q_2'}},
\end{equation}
where the subscripts $q_2$ and $q_2'$ refer to the corresponding qubits in Fig. \ref{fig:remote_cnot_implemented_in_different_ways}. The control state was found to be most susceptible to entanglement errors out of all separable input states that have $q_2'$ in the state  $\ket{0}_{\mathrm{q_2'}}$, as discussed in Appendix \ref{app:variation_input_state}.

The parameter values used for the different types of error and the durations of different operations are based on state-of-the-art values for trapped-ion systems from the literature, and are displayed in Table \ref{tab:state_of_art_params}.\begin{ruledtabular}
\begin{table}
    \centering
    \caption{The nominal state-of-the-art parameter values used in our simulations.}
    \begin{tabular}{ccc}
       Parameter & Value & Source  \\ \colrule
       $F_{\rm w}$ & 0.94 & With $^{88}$Sr$^+$ qubits \cite{ionTrapEntDist94percent2m} \\ 
       $\epsilon_{\mathrm{cnot}}$ & 0.4\% & With $^171$Yb$^+$ qubits \cite{ionQAriaSpecs} \\
       Single-qubit gate time & $135 \mu$s  & \cite{ionQAriaSpecs}  \\
       Two-qubit gate time & $600 \mu$s & \cite{ionQAriaSpecs} \\
       Measurement time & 6 ms & Inferred from \cite{ionQAriaSpecs, AutoComm, metodi2005quantum} \footnote{Refs. \cite{AutoComm, metodi2005quantum} put the measurement time at five to ten times the two-qubit gate time for trapped-ion quantum computers. Therefore, we have estimated the measurement time to be ten times the two-qubit gate time from Ref. \cite{ionQAriaSpecs}.}  \\
       $R$ & 182 Hz & \cite{ionTrapEntDist94percent2m} \\
       $T_1$ \footnote{The qubit lifetime. The memory depolarisation rate, $r$, is given by $r = \frac{1}{T_1}$. When a single state-of-the-art value is required we use the midpoint ($r=0.055$) of the quoted range.} & 10 -- 100s & \cite{ionQAriaSpecs} \\
       $d$ & 2m & \cite{ionTrapEntDist94percent2m}
    \end{tabular}
    \label{tab:state_of_art_params}
\end{table}%
\end{ruledtabular}%
Where possible, parameters for current commercial hardware (IonQ's Aria quantum computer \cite{ionQAriaSpecs}) are used to inform the state-of-the-art-values, so as to make the simulated set-up more realistically implementable in the near-term. 

When we consider a range of parameter values, we refer to the order of magnitude in which the relevant state-of-the-art parameter exists as the state-of-the-art range. To facilitate comparison between entanglement error and local gate error, we also often consider what happens when both error types are varied over the same range. When we do this, we consider a range of entanglement error which is an order of magnitude smaller than the `state-of-the-art' value. As this entanglement error reduction would require one of the methods from Sec. \ref{subsec:entanglement_distribution}, we refer to the resulting range as the distilled range, in reference to entanglement distillation, although entanglement distillation is just one of the possible error reduction methods that could be used.

In the following, we first compare our different error analysis tools. This is done in Sec. \ref{subsubsec:first_order_vs_sim} for the building block modules shown in Fig. \ref{fig:remote_cnot_implemented_in_different_ways}. After that, we compare cat-comm, 1TP, 2TP, and TP-safe in Sec. \ref{subsubsec:remote_gate_comparison}. Finally, we compare the impact of different types of error in Sec. \ref{subsubsec:error_type_comparison}.

\subsubsection{Comparison of error analysis methods}
\label{subsubsec:first_order_vs_sim}

As alluded to in Sec. \ref{subsubsec:analytical_results}, analytical results provide a useful robustness check for simulated ones. In our case, we find that the analytical expressions, Eq. \eqref{eq:1TP_analytical_F_out} and Eq. \eqref{eq:cat_analytical_F_out}, agree exactly with the simulated results for entanglement error, which suggests that the simulated results are likely to be robust.

The agreement is less strong between the analytical/simulation results and the first-order approximations given by Eqs. \eqref{eq:first_order_f_out} and \eqref{eq:exp_approx}. Indeed, we find both first-order expressions to be inadequate, indicating the need for more detailed error analysis, such as the classical simulation of errors conducted in this paper. The failure of the linear approximation is well established for larger monolithic circuits \cite{errorPropagationMonoQC}, however, here we indicate that clear discrepancies between both first-order approximations and the simulated results can occur for individual remote gates, when error parameters within the state-of-the-art or distilled ranges are used.

To demonstrate the inadequacy of the first-order approximation, we consider the percentage difference between simulated and approximate output errors, $\Delta_{\mathrm{oe}}$. This is calculated using the expression: 
\begin{equation}
    \label{eq:percentage_difference}
    \Delta_{\mathrm{oe}} = \frac{\left(1 - (F_{\mathrm{out}})_{\mathrm{approx}} \right) - \left(1 -  (F_{\rm out})_ {\mathrm{ sim}}\right)}{1 - (F_{\rm out})_{\mathrm{sim}}} \times 100.
\end{equation}
 A positive sign to the result indicates that the first-order approximation overestimates the output error, $1-F_{\mathrm{out}}$, relative to the simulated data, and a negative sign means that the first-order approximation underestimates the output error. The subscript `sim' refers to simulated data and `approx' to data calculated using Eq. \eqref{eq:first_order_f_out} or Eq. \eqref{eq:exp_approx}. 

\begin{figure*}
    \centering
    \hspace*{-2em}%
    \begin{overpic}[scale=0.2, trim={0, 1.5cm, 0, 0}, clip]{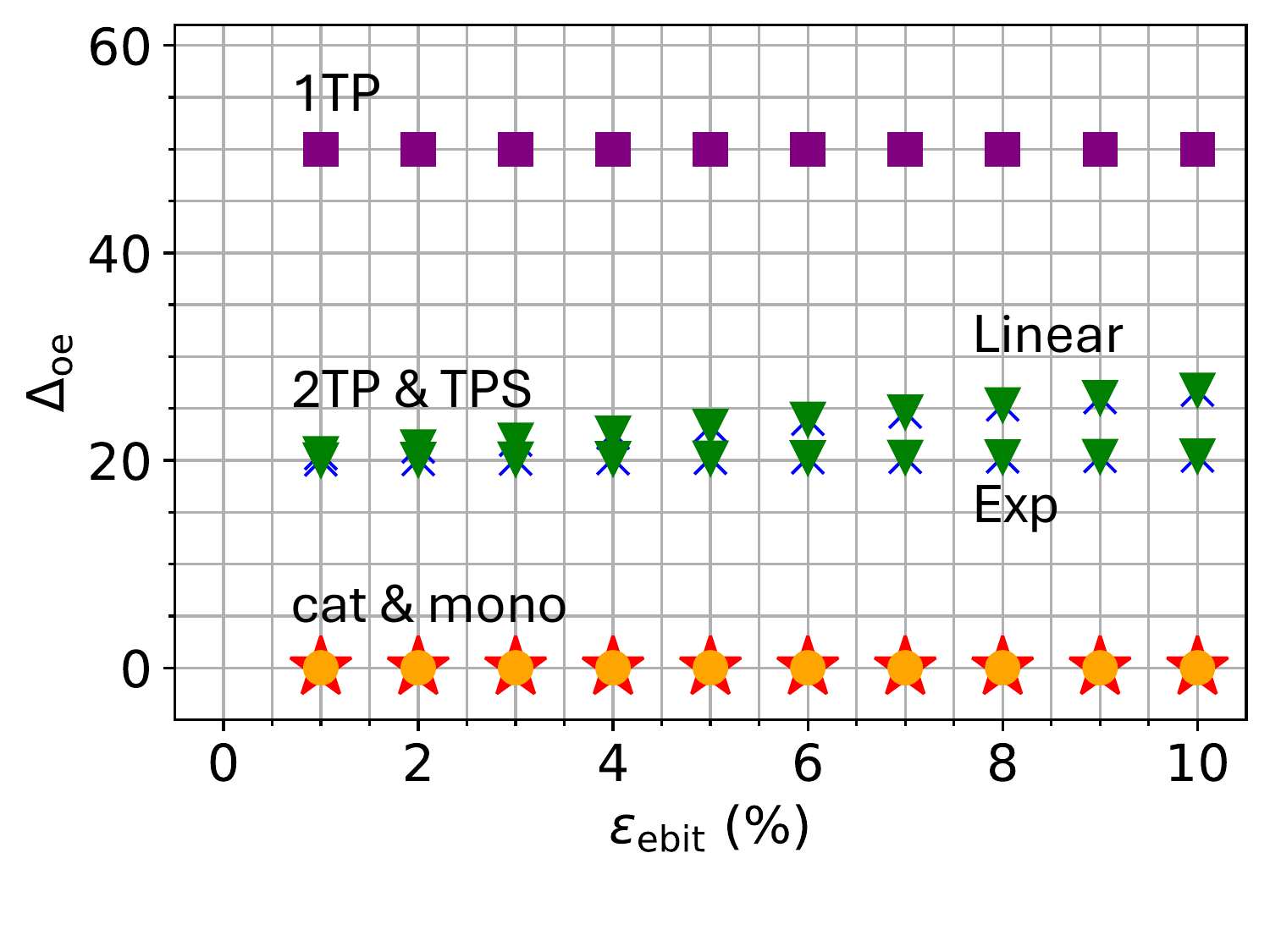}%
        \put(-2, 65){(a)}    
    \end{overpic}\hspace{1em}%
    \begin{overpic}[scale=0.2, trim={0, 1.5cm, 0, 0}, clip]{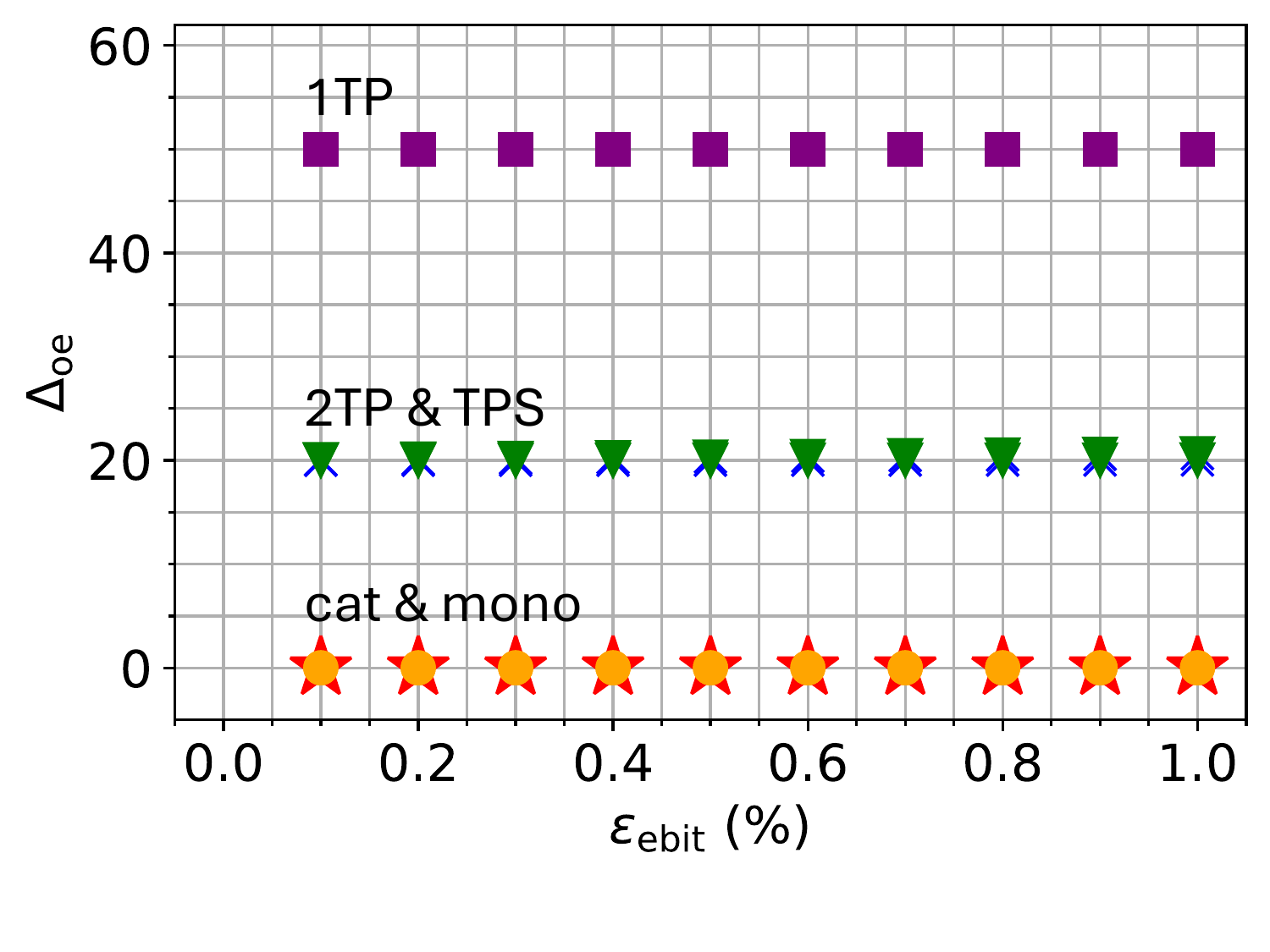}
        \put(-2, 65){(b)}
    \end{overpic}\hspace{1em}%
    \begin{overpic}[scale=0.2, trim={0, 1.5cm, 0, 0}, clip]{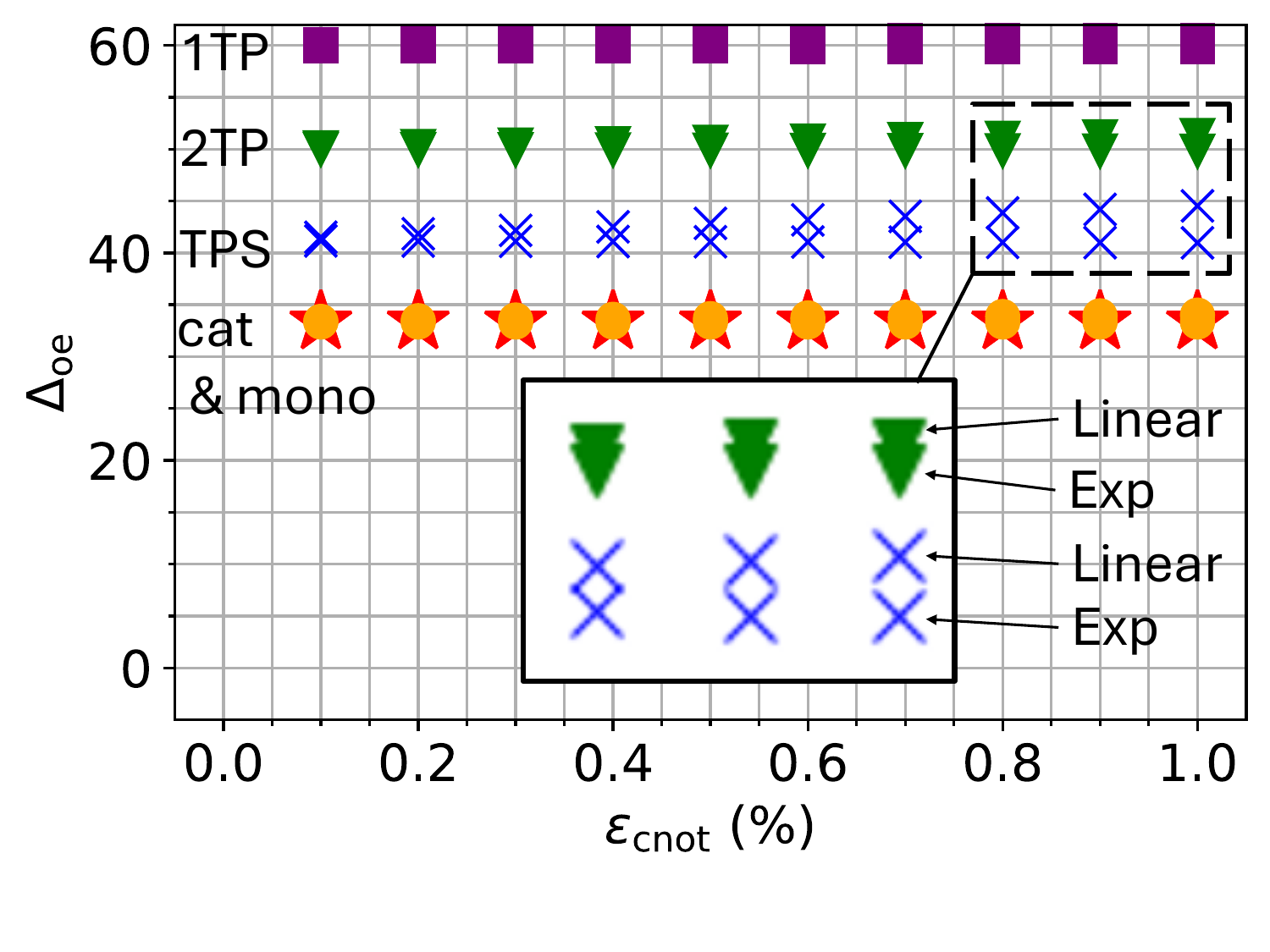}
        \put(-2, 65){(c)}
    \end{overpic}
    \subfloat{%
   \raisebox{0.4\height}{
    \begin{overpic}[scale=0.5, trim={0.5cm, 0.2cm, 0.2cm, 0.2cm}, clip]{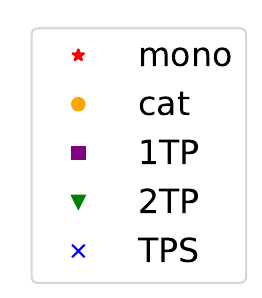}
    \end{overpic}}}
    \caption{The percentage difference between the simulated and approximate results for a remote CNOT gate conducted using each of the schemes depicted in Fig. \ref{fig:remote_cnot_implemented_in_different_ways}: 1TP, cat-comm (cat), 2TP, and TP-safe (TPS). The results for a local CNOT gate conducted on a monolithic processor is also shown for comparison (labelled `mono'). In each case, the percentage difference, calculated using Eq. \eqref{eq:percentage_difference}, is plotted with respect to: (a) the entanglement error, varied over state-of-the-art range, with all other errors set to zero; (b) the entanglement error, varied over the distilled range, with all other errors set to zero; and (c) the local CNOT depolarisation errors, varied over the state-of-the-art-range, with all other errors set to zero. The approximate results are calculated using Eqs. \eqref{eq:first_order_f_out}  and \eqref{eq:exp_approx}, marked as `Linear' and `Exp', respectively, when a difference is discernible.  All simulated results are averaged over ten runs. The standard error is $0$, up to machine precision, for all data points.}
    \label{fig:first_order_diff}
\end{figure*}

Figure \ref{fig:first_order_diff} %
shows $\Delta_{\mathrm{oe}}$ after a remote CNOT gate is implemented using each of the schemes in Fig. \ref{fig:remote_cnot_implemented_in_different_ways}. A local CNOT gate conducted on a single monolithic quantum computer is also considered for comparison. Figures \ref{fig:first_order_diff}(a) and \ref{fig:first_order_diff}(b) show the $\Delta_{\mathrm{oe}}$ with respect to the entanglement error, varied over the state-of-the-art and distilled ranges, respectively. All other errors are set to zero. Figure \ref{fig:first_order_diff}(c) shows $\Delta_{\mathrm{oe}}$ with respect to the local two-qubit gate error. Again all other errors, including entanglement errors, are set to zero.

Using Fig. \ref{fig:first_order_diff}, several observations can be made:
\begin{enumerate}[leftmargin=0pt, listparindent=1.25em, itemindent=20pt, parsep=0pt, label={\roman*.}]
    \item \label{obs:first_order_upper_bound} The first-order approximations provide an upper bound on the output error. All $\Delta_{\mathrm{oe}}$ values are non-negative in Fig. \ref{fig:first_order_diff}. This means that the output errors obtained using the simulation never exceed the approximate ones. This observation is resilient to the choice of input state as discussed in Appendix \ref{app:first_order_approx_robustness}.
    \item \label{obs_1st_order:cat_1TP_distinguishable} The first-order approximations fail to distinguish between cat-comm and 1TP, for which the number of gates, classical communications, measurements, and ebits is the same. By contrast, the markedly different results for cat-comm and 1TP in Fig. \ref{fig:first_order_diff} indicate that such a discrepancy does exist. This is one case where the numerical simulation reveals behaviour that is not apparent from first-order analysis alone.
    \item The discrepancy between simulated and approximate results can be large. For the input state chosen, this discrepancy is especially significant for 1TP, 2TP, and TP-safe. Figures \ref{fig:first_order_diff}(a) and \ref{fig:first_order_diff}(b) indicate that the discrepancy is approximately 20-50\% when only entanglement error is considered and Fig. \ref{fig:first_order_diff}(c) indicates that it is 40-60\% when local two-qubit gate errors are considered. Therefore, even with just one remote gate, the difference in the output error predicted by the first-order approximations and the simulation can be significant.
    \item \label{obs:good_agreement_possible_btwn_approx_and_sim} Good agreement between simulated and first-order results is possible for some schemes. For cat-comm and the monolithic case, the simulated and first-order results agree when only entanglement error is considered. This can be seen from Figs. \ref{fig:first_order_diff}a and \ref{fig:first_order_diff}b. This agreement is highly dependent on input state as discussed in Appendix \ref{app:first_order_approx_robustness}.
    \item The linear and exponential approximations, given by Eqs. \eqref{eq:first_order_f_out} and \eqref{eq:exp_approx}, respectively, perform similarly over the investigated error ranges, but, even with just one remote gate, the exponential approximation can be seen to perform slightly better than the linear one for 2TP and TP-safe. This holds for both entanglement error and local two-qubit gate error, within the state-of-the-art range, as can be seen from Figs. \ref{fig:first_order_diff}a and \ref{fig:first_order_diff}c.
\end{enumerate}

A key takeaway from these observations is that we cannot rely on first-order approximations alone for a quantitative error analysis of a QDC. Given the complexity of closed-form analytical analysis of arbitrary distributed quantum circuits, numerical simulation is therefore a key tool for understanding the role of errors in the QDC context.

\subsubsection{Comparison of remote gate implementations}
\label{subsubsec:remote_gate_comparison}

Using numerical simulation, we are able to compare different ways of implementing remote gates. The remote gate varieties we consider are shown in Fig. \ref{fig:remote_cnot_implemented_in_different_ways}. The distributed quantum computing literature \cite{heightIncreaseNonLocalGates, DQCassistedByRemoteToffoli, NickersonDQCfaultTolerantThesis, GeneralProtocolForDistributedGates, catCommDistributedShor, generalizedGHZandDQC,
OptimisingTeleportaionCost, reordingPartitionDQCcircuitsTPcommOnly, ConnectivityMatrixCompiler, OptimisedQuantumCircuitPartitioning, TimeSlicedPartitioning, EfficientDistributionQuantumCircuits, davis2023dqcPartitioning, Ferrari, AutoComm, QuComm} typically bases such comparisons on the number of EPR pairs required by each scheme in various contexts, which falls short of the more detailed analysis possible using numerical simulation.

\begin{figure}
    \centering
    \hspace*{-2em}%
    \begin{overpic}[scale=0.2, trim={0, 2cm, 0, 0}, clip]{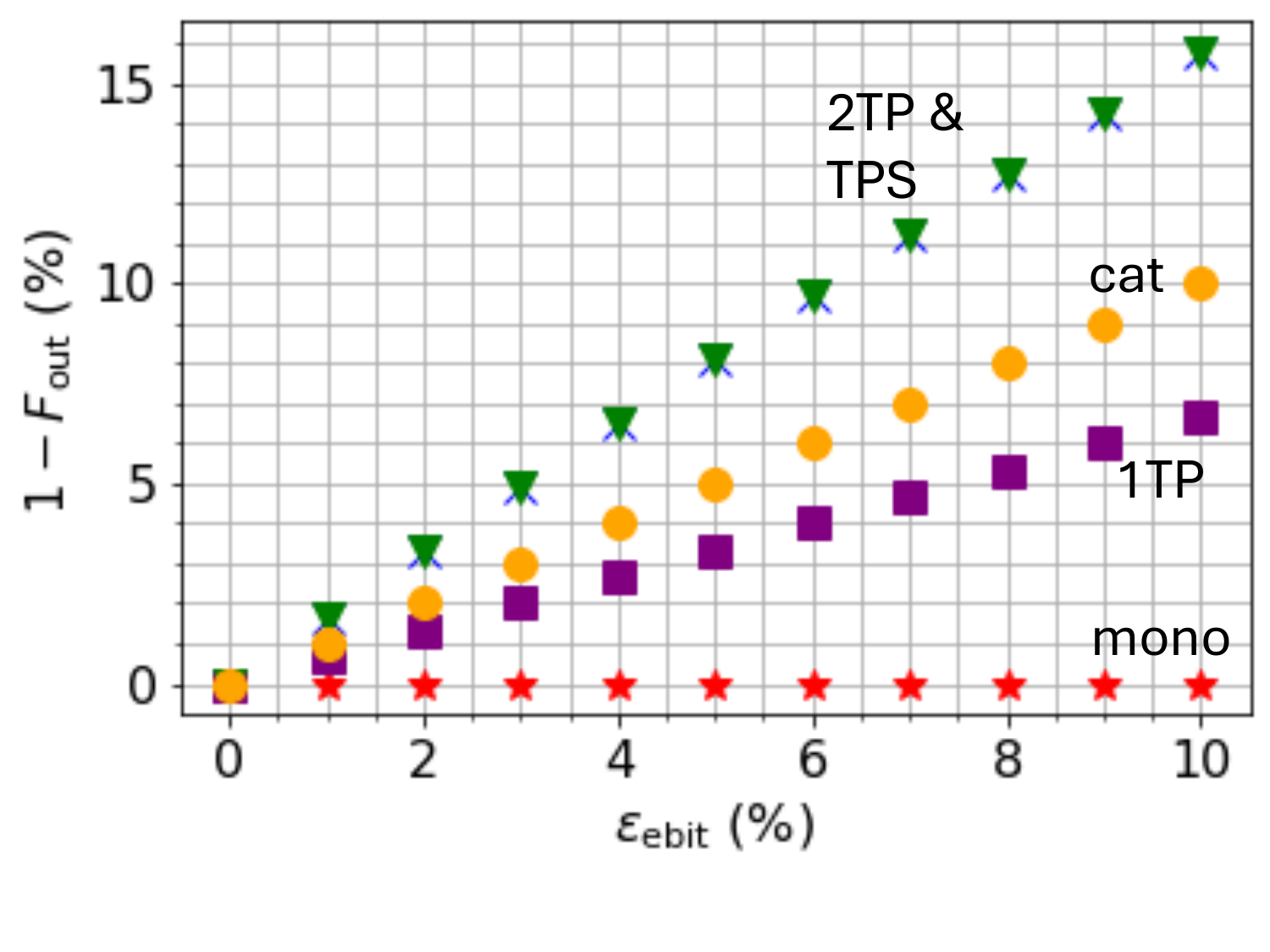}
        \put(-4, 60){(a)}        
    \end{overpic}\hspace{2em}%
    \begin{overpic}[scale=0.2, trim={0, 2cm, 0, 0}, clip]{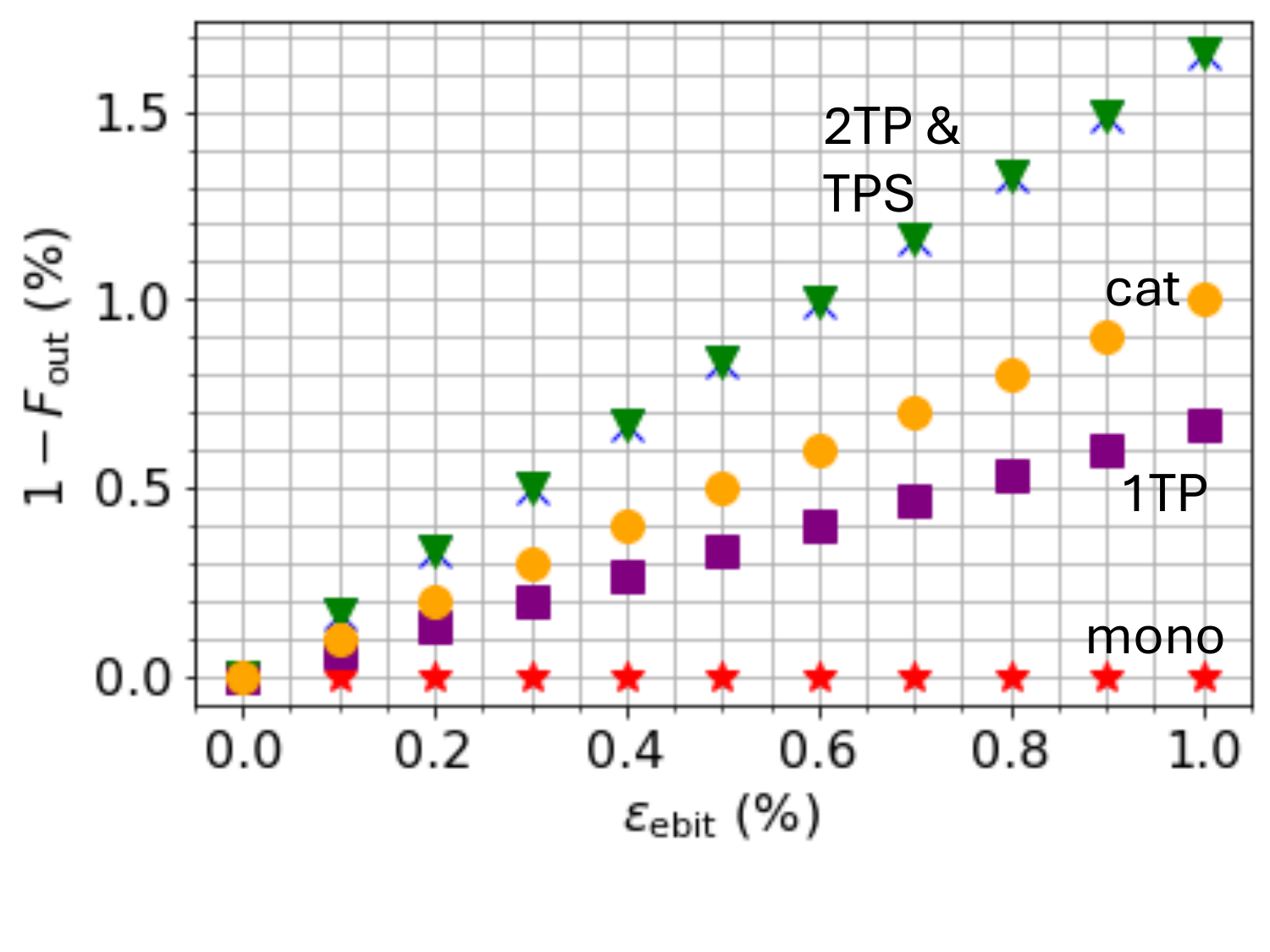}
        \put(-10, 64){(b)}
    \end{overpic}\hspace{2em}%
    \begin{overpic}[scale=0.2, trim={0, 1.5cm, 0, 0}, clip]{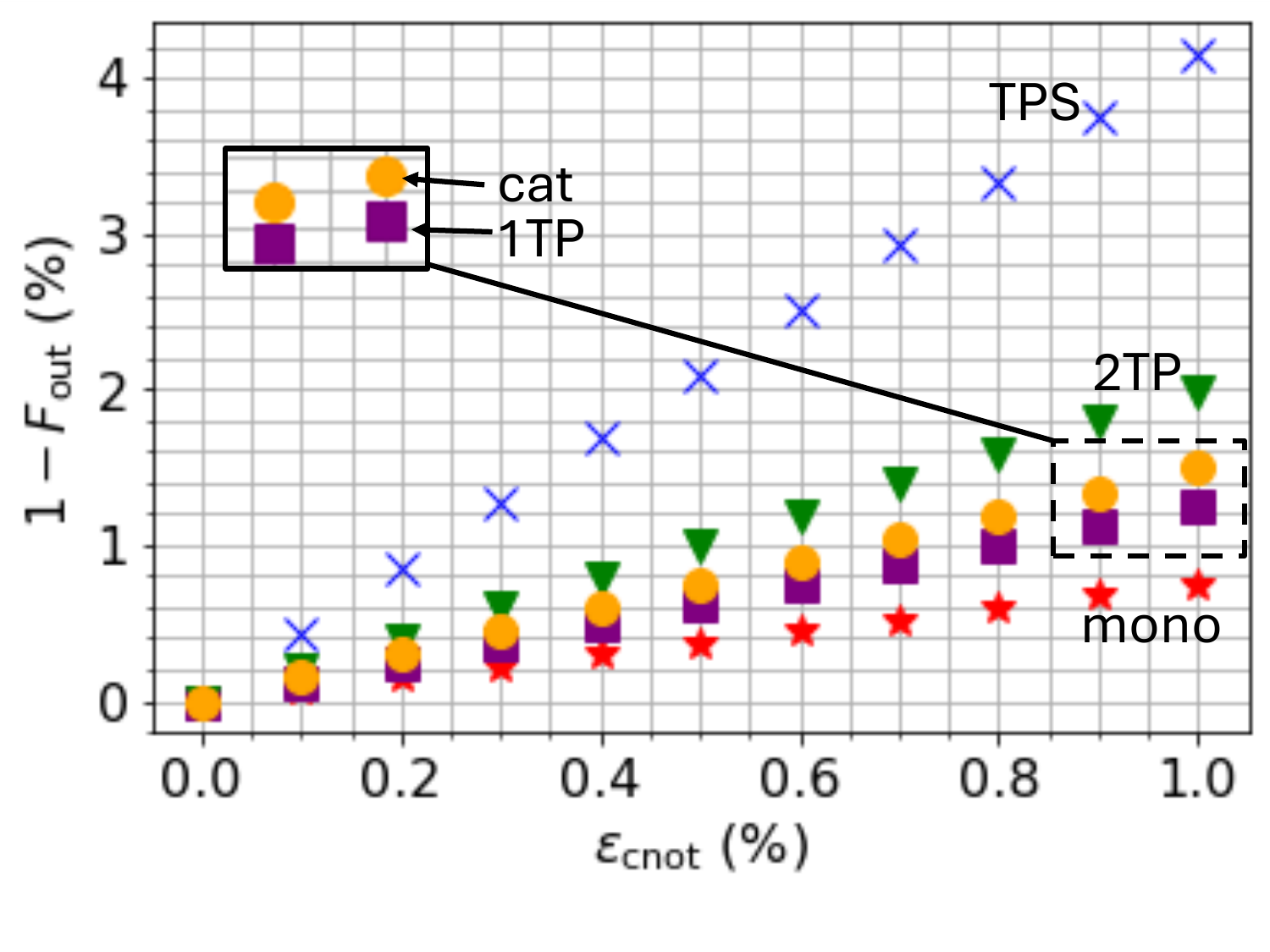}
        \put(-10, 64){(c)}        
    \end{overpic}
    \begin{overpic}[scale=0.2, trim={0, 2.5cm, 0, 0}, clip]{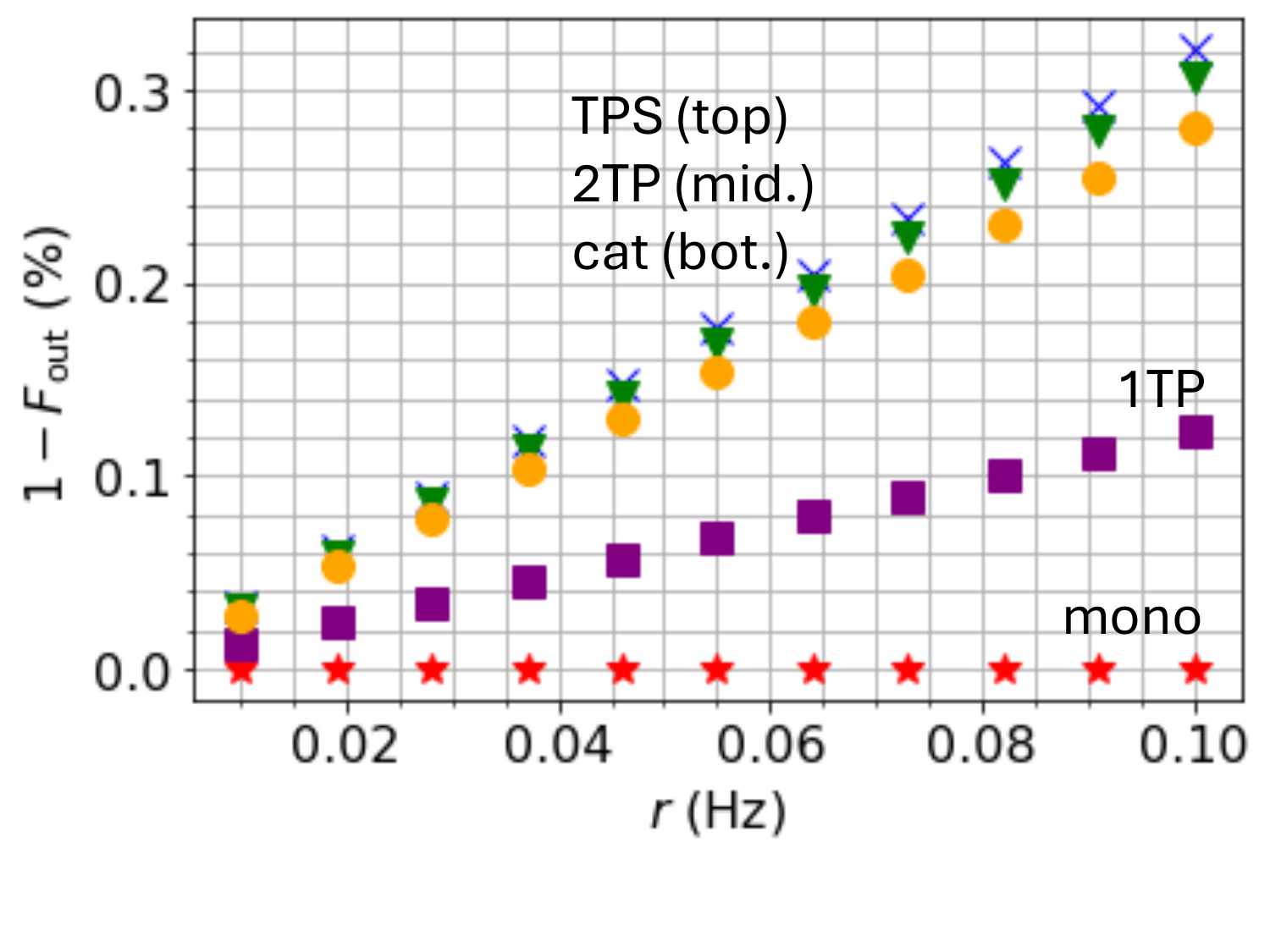}
        \put(-10, 64){(d)}
    \end{overpic}\hspace{2em}%
    \subfloat{%
    \raisebox{0.4\height}{
    \begin{overpic}[scale=0.5, trim={0.5cm, 0.2cm, 0.2cm, 0.2cm}, clip]{comparing_schemes_legend.pdf}
    \end{overpic}}}
    \caption{The output error, $1 - F_{\rm out}$, for an individual remote CNOT gate with increasing: (a) entanglement error (state-of-the-art range); (b) entanglement error (distilled range); (c) local two-qubit gate error; and (d) memory depolarisation, for cat-comm (cat), 1TP, 2TP, and TP-safe (TPS). For the monolithic case (mono) a single local CNOT gate is considered. The input state is given by Eq. \eqref{eq:remote_CNOT_input_state}. All simulated results are averaged over ten runs. The standard error is zero, up to machine precision, for all data points.}
    \label{fig:single_CNOT_scheme_comparison}
\end{figure}

Figure \ref{fig:single_CNOT_scheme_comparison} %
 shows the output error, $1-F_{\mathrm{out}}$, after a remote CNOT gate. We vary one error at a time, respectively varying:  entanglement error, over the state-of-the-art range in Fig. \ref{fig:single_CNOT_scheme_comparison}(a) and the distilled range in Fig. \ref{fig:single_CNOT_scheme_comparison}(b); local two-qubit gate error in Fig. \ref{fig:single_CNOT_scheme_comparison}(c); and memory depolarisation rate in Fig. \ref{fig:single_CNOT_scheme_comparison}(d). Again, all error parameters not explicitly varied in a given subfigure are set to zero. 

Two key observations can be made:
\begin{enumerate}[leftmargin=0pt, listparindent=1.25em, itemindent=20pt, parsep=0pt, label={\roman*.}]
    \item For the input state given by Eq. \eqref{eq:remote_CNOT_input_state}, the remote gate schemes are ordered from lowest to highest by their impact on output error, as: 1TP, cat-comm, 2TP, and then TP-safe, where 2TP and TP-safe are equally damaging to the output when only entanglement error is considered. This ordering leads to the next more general observation.
    \item \label{obs_scheme_order:1TP_beats_cat} Cat-comm and 1TP can be distinguishable. This is consistent with observation \ref{obs_1st_order:cat_1TP_distinguishable} from Sec. \ref{subsubsec:first_order_vs_sim}, but is easier to see using Fig. \ref{fig:single_CNOT_scheme_comparison}. As discussed previously, it is not obvious a priori that cat-comm and 1TP could lead to different output errors, as they use exactly the same gates, measurements, ebits and classical communication. The difference between the output error produced by 1TP and cat-comm may offer an optimisation opportunity if either scheme is found to perform better when averaged over all input states. Such averaging is beyond the scope of the current work.
\end{enumerate}

\subsubsection{Comparison of error types}
\label{subsubsec:error_type_comparison}

Although the relative qualitative performance of different schemes is unchanged by the type of error considered, another important algorithmic consideration for any quantum computer is how best to mitigate or, ideally, correct errors. A greater understanding of which errors have the most impact may help focus such efforts on where they are most needed and assist the efficient deployment of hardware resources to get the best possible output with the least possible effort. To this end, we investigate the impact of the different types of error on the output error.

\begin{figure*}
    \centering
    \begin{overpic}[scale=0.2, trim={0, 0, 0, 0}, clip]{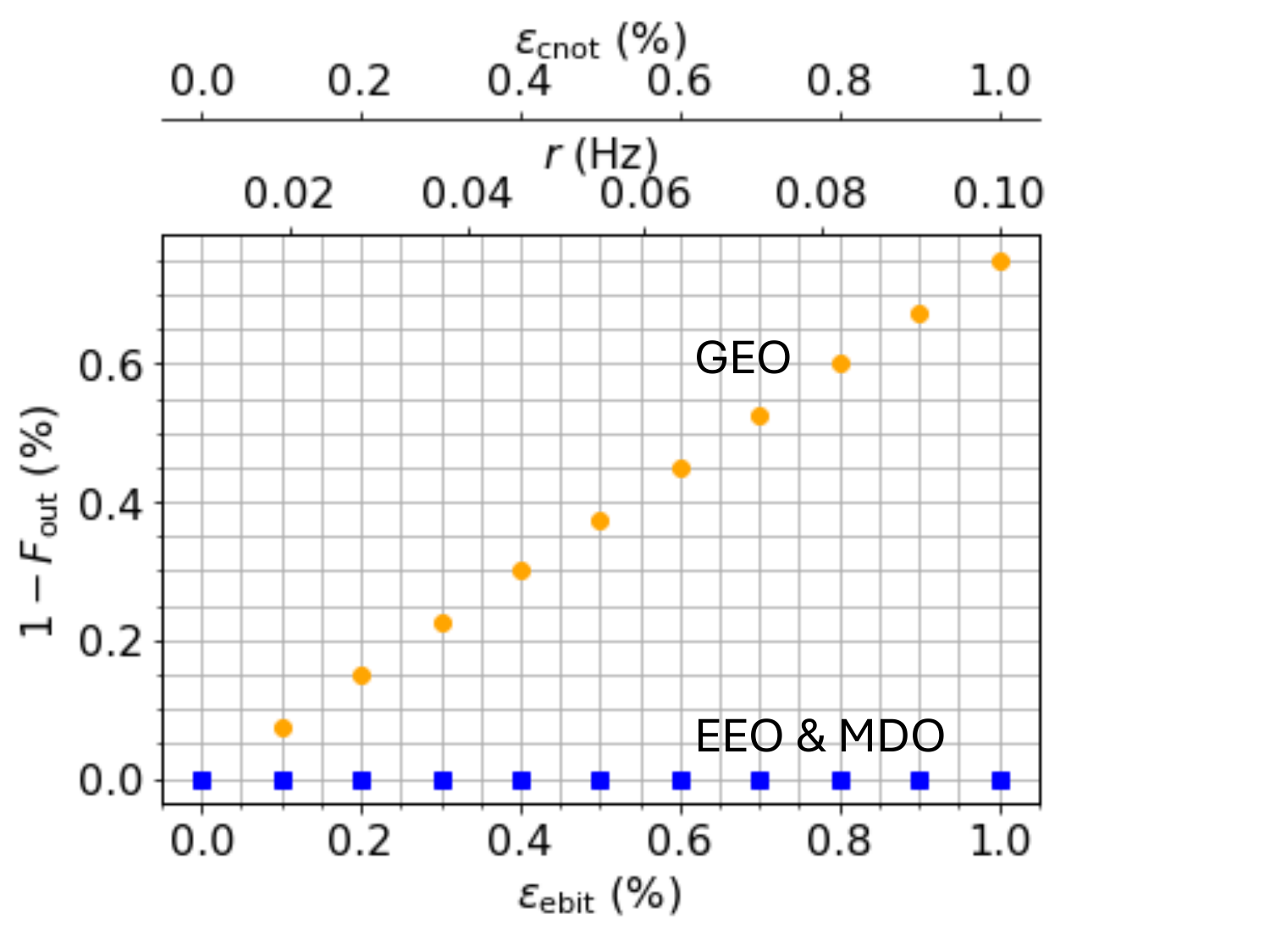}
        \put(20, 45){(a)}
    \end{overpic}
    \begin{overpic}[scale=0.2, trim={0, 0, 0, 0}, clip]{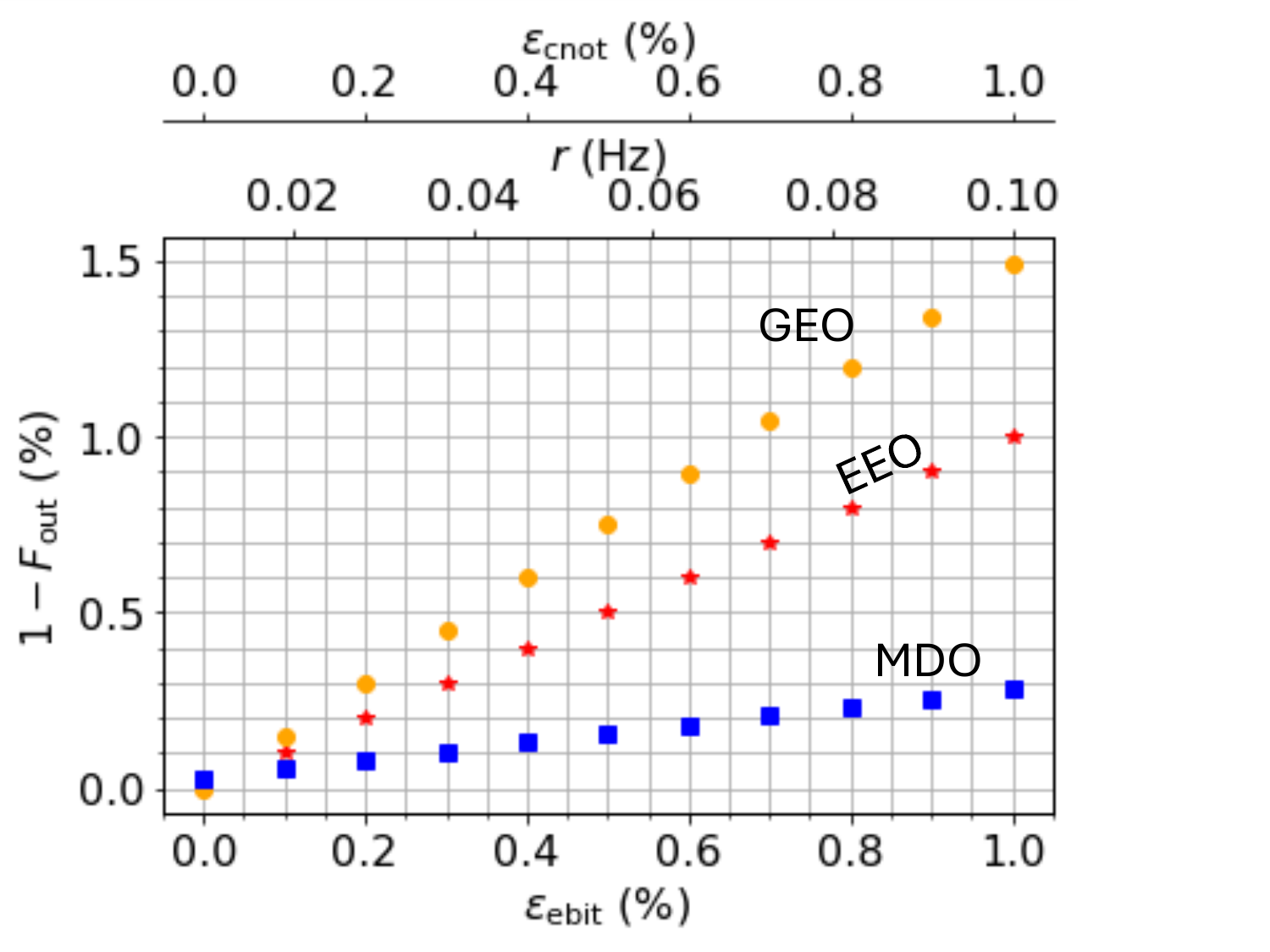}
        \put(20, 45){(b)}
    \end{overpic}
    \begin{overpic}[scale=0.2, trim={0, 0, 0, 0}, clip]{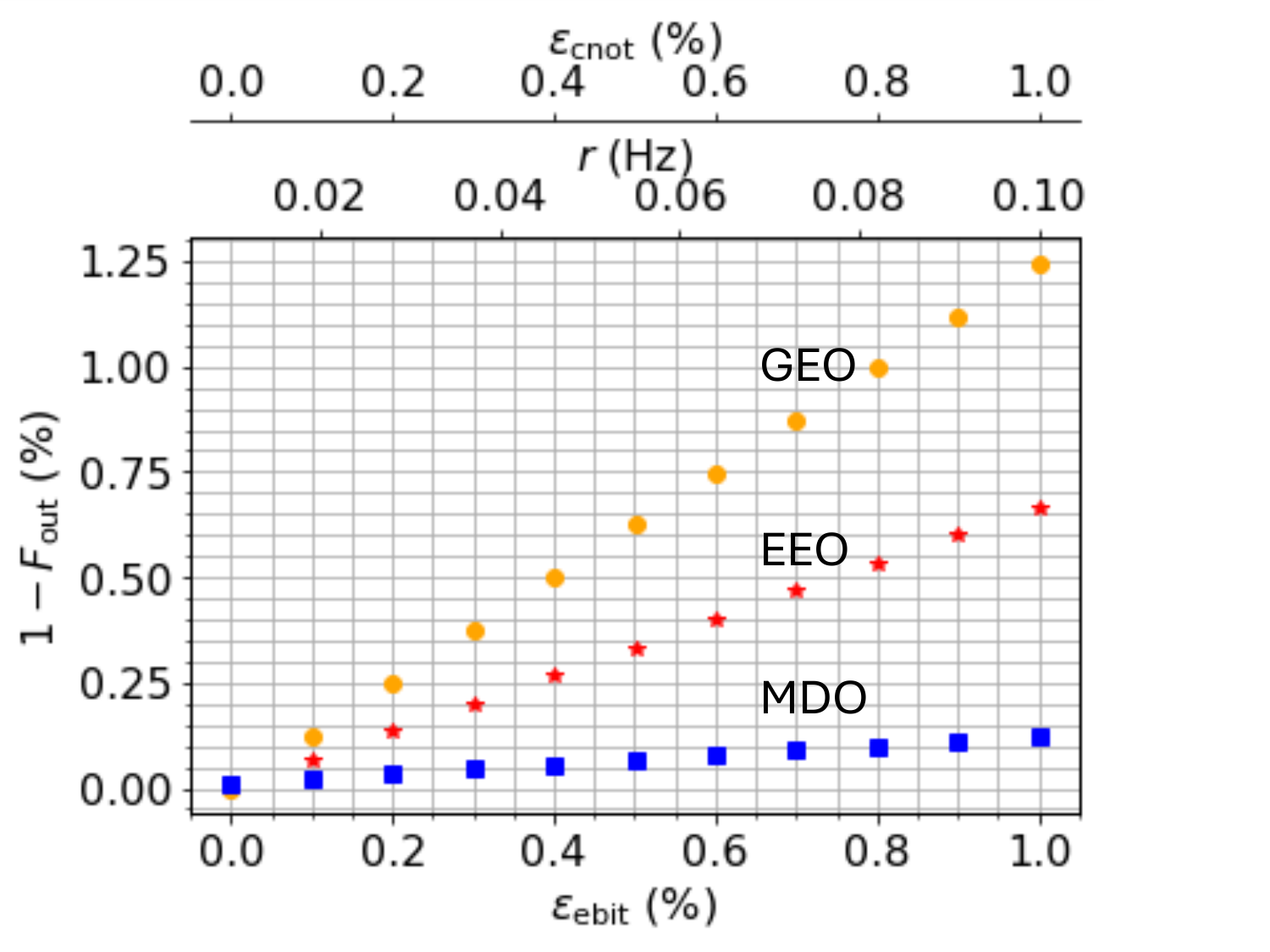}
        \put(20, 45){(c)}
    \end{overpic}
    \begin{overpic}[scale=0.2]{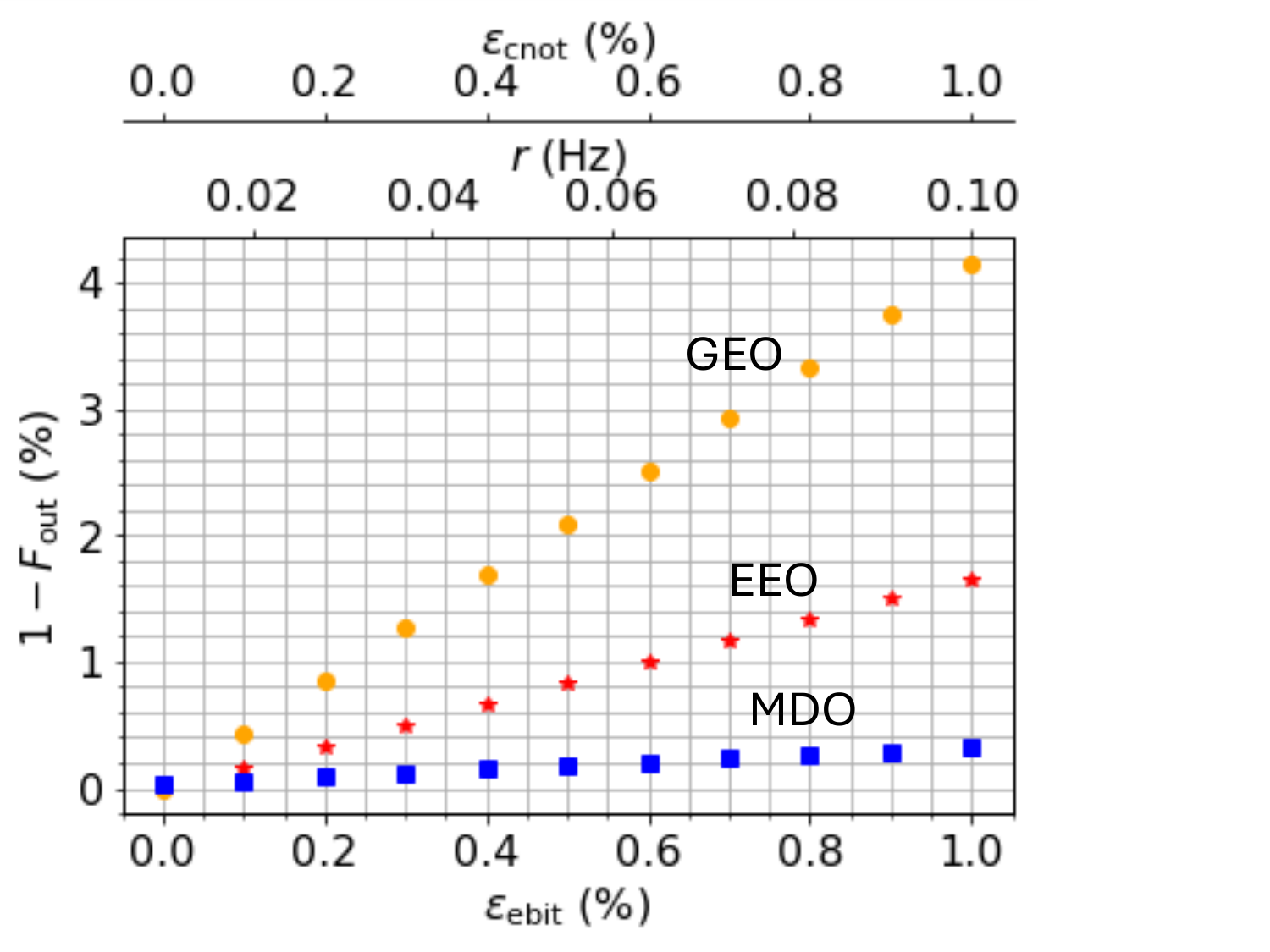}
        \put(20, 45){(d)}
    \end{overpic}
    \begin{overpic}[scale=0.2, trim={0, 0, 0, 0}, clip]{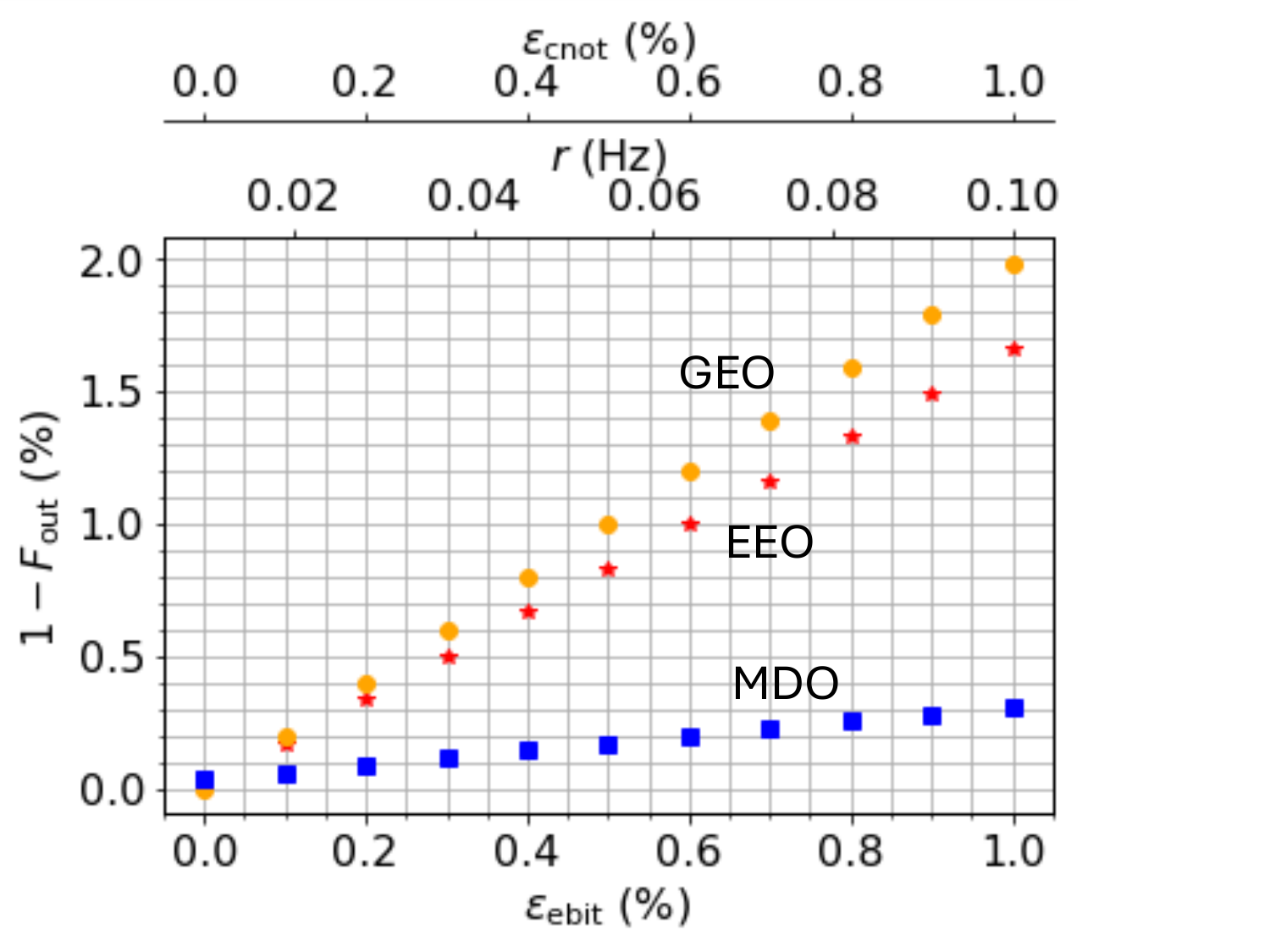}
        \put(20, 45){(e)}
    \end{overpic}
    \subfloat{%
    \raisebox{0.6\height}{
    \begin{overpic}[scale=0.5, trim={0.5cm, 0.2cm, 0.2cm, 0.2cm}, clip]{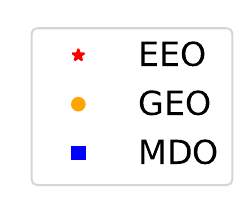}
    \end{overpic}}}
    \caption{The output error for a single remote (local for monolithic case) CNOT gate, implemented using: (a) the monolithic case, (b) cat-comm, (c) 1TP, (d) 2TP, (e) TP-safe.  The input state is given by Eq. \eqref{eq:remote_CNOT_input_state}. Each set of markers corresponds to the results obtained by varying a single error parameter and setting all other errors to zero. EEO indicates entanglement error only, GEO indicates gate error only, and MDO indicates memory depolarisation only. Entanglement errors are varied over the distilled range. All other errors are varied over the state-of-the-art range.}
    \label{fig:single_cnot_error_comparison_distilled}
\end{figure*}

We gain insight into the relative impact of different error types in two different ways. Firstly, we compare the quantitative values of output error displayed in the different subfigures of Fig. \ref{fig:single_CNOT_scheme_comparison}. This allows us to re-use Fig. \ref{fig:single_CNOT_scheme_comparison} to gain insight into the relative impact of errors when entanglement error is varied within the state-of-the-art range. However, this analysis is insufficiently precise to compare errors when entanglement error is varied within the distilled range,  where the magnitudes of output error are much smaller. To perform comparison between entanglement error in the distilled range and other types of error, it is useful to directly plot the output error caused by each type of error on the same figure. For this reason, in Fig. \ref{fig:single_cnot_error_comparison_distilled}, %
we collect the output fidelity data used in Figs. \ref{fig:single_CNOT_scheme_comparison}(b)-(d), and re-express it all on one plot per remote gate scheme. In each plot, the loss of output fidelity due to entanglement error only (EEO), two-qubit gate error only (GEO), and memory depolarisation only (MDO) are all shown. 

Using Figs. \ref{fig:single_CNOT_scheme_comparison} and \ref{fig:single_cnot_error_comparison_distilled}, we find that:
\begin{enumerate}[leftmargin=0pt, listparindent=1.25em, itemindent=20pt, parsep=0pt, label={\roman*.}]
    \item \label{obs:state_of_art_ordering} When entanglement error is varied over the state-of-the-art range, in our simulation, the error types are ranked from least to most detrimental to output error as: memory depolarisation, local two-qubit gate error, and then entanglement error.

    Memory depolarisation is relatively negligible. When only memory depolarisation is non-zero, as in the bottom curve of Figs. \ref{fig:single_cnot_error_comparison_distilled}(a)-(e), we can see that memory depolarisation leads to significantly less output error than local CNOT error for all considered depolarisation rates.

    Local two-qubit gate error is in turn dominated by entanglement error. This can be seen by comparing Fig. \ref{fig:single_CNOT_scheme_comparison}(a), in which only entanglement error is non-zero, with Fig. \ref{fig:single_CNOT_scheme_comparison}(c), in which only local two-qubit gate error is non-zero. In Fig. \ref{fig:single_CNOT_scheme_comparison}(c), only TP-safe ever has an output error exceeding 2\% for any local gate error in the considered range and there is no data point where the output error exceeds 4.2\%. In Fig. \ref{fig:single_CNOT_scheme_comparison}(a), all schemes have an output error exceeding 2\% as soon as the entanglement error reaches 4\% and at 8\% entanglement error, the output error of all schemes is greater than 4.2\%. 
    \item \label{obs:distilled_ordering} When entanglement error is varied over the distilled range, in our simulation, the ranking of the error types from least to most detrimental to output error changes to: memory depolarisation, entanglement error, and then local two-qubit gate error. 
    
    The ordering of entanglement error and local two-qubit gate error flips now that their magnitude is varied over the same range. This is easily seen from Figs. \ref{fig:single_cnot_error_comparison_distilled}(a)-(e). This finding is not immediately obvious a priori, as one might expect the correlations intrinsic to entanglement to propagate errors more rapidly through the system.

    Again, memory depolarisation is negligible and its impact is significantly lower than the other types of error shown in \ref{fig:single_cnot_error_comparison_distilled}(a)-(e) for the parameter values considered here.
\end{enumerate}

Overall, the findings suggest that entanglement error will be the most urgent optimisation consideration in the near term, however if the entanglement error is reduced, for example by using one of the methods from Sec. \ref{subsec:entanglement_distribution}, the local gate errors may well be more significant. The trade-off between improving entanglement errors and introducing additional local errors and/or latency via the methods discussed in Sec. \ref{subsec:entanglement_distribution} would be an interesting and potentially very important point of future study.

\section{Larger quantum circuits}
\label{sec:larger_quantum_circuits}

The behaviour of individual remote gates can tell us many things, however, when many such gates operate in tandem, any errors in one gate will impact the rest of the system. How this happens may depend on the structure of the circuits in question and so it is not automatically obvious that the trends seen for individual remote gates will also hold for larger circuits. Here, we verify that the observations on the relative impact of different errors made in Sec. \ref{subsubsec:error_type_comparison} do in fact hold for larger quantum circuits and discuss some differences between single remote gates and larger circuits. We describe the problem in more detail in Sec. \ref{subsec:larger_circuits_problem_description} and then give the results of our numerical simulations in Sec. \ref{subsec:larger_circuits_results}.

\subsection{Problem description}
\label{subsec:larger_circuits_problem_description}

We consider 22 different quantum algorithms taken from the MQT bench library \cite{MQTBench}. We predominantly consider the five-qubit implementations of these algorithms but do also consider the impact of varying qubit number on each algorithm. These algorithms represent a variety of applications and range from algorithms of historical interest, such as the Deutsch-Joza algorithm, to common quantum subroutines, like the quantum Fourier transform, and NISQ applications, such as QAOA and VQE applied to different problems. 

Compilation is kept simple. All quantum circuits are distributed by simple bipartitioning between two QPUs. Half of the processing qubits are given to one QPU and half to the other. If there is an odd number of processing qubits, then the additional qubit is arbitrarily allocated to a QPU based on the indices supplied for the original monolithic circuits by MQT bench \cite{MQTBench}---the QPU whose qubits have the lower indices are given the extra qubit. Scheduling of operations is done greedily, but, for simplicity, ebits are requested only when needed.  No attempt is made to merge remote gates together or otherwise optimise compilation. In this way, we avoid making the results specific to any one distributed quantum computing compiler, which may have imposed unforseen structural bias on the circuits. The results represent a lower bound on what can be achieved with QDC.

The reasons for considering multiple implementations for TP-comm become apparent for these larger quantum circuits, as it becomes impossible to implement 1TP in many cases and even 2TP is often inapplicable with the simple compilation strategy employed. Therefore, we only consider TP-safe and cat-comm, which are compatible with the simple compilation strategy used. For each circuit considered, results are taken for when all remote gates are conducted with cat-comm and independently when all remote gates are conducted with TP-safe. 

Of the error analysis tools discussed in Sec. \ref{subsec:error_analysis_methods}, we focus primarily on numerical simulation.

\subsection{Numerical results}
\label{subsec:larger_circuits_results}

\begin{figure}
    \begin{overpic}[scale=0.2, trim={0, 2.5cm, 0, 0}, clip]{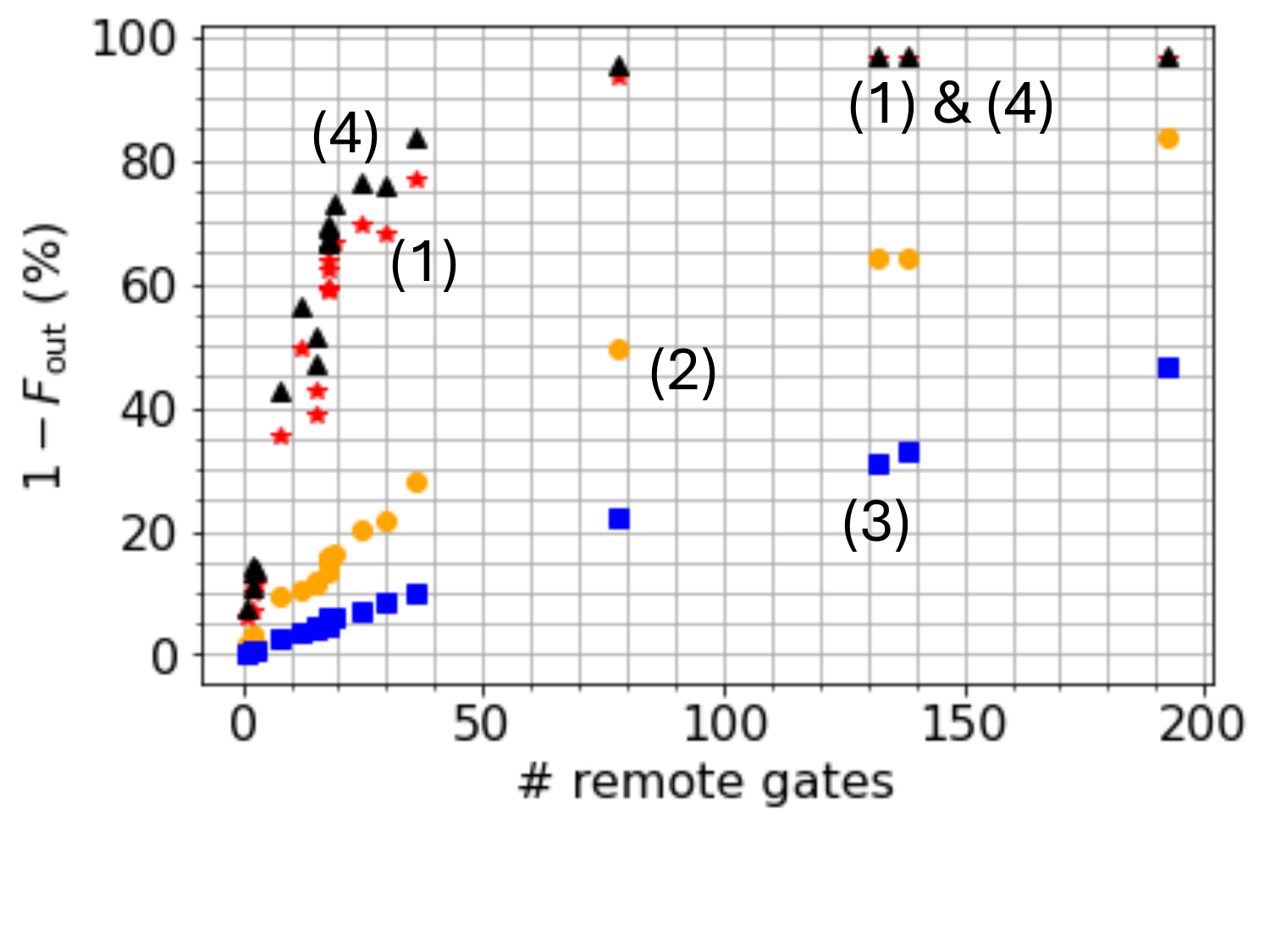}
        \put(-4, 60){(a)}
    \end{overpic}\hspace{4em}%
    \begin{overpic}[scale=0.2, trim={0, 2.5cm, 0, 0}, clip]{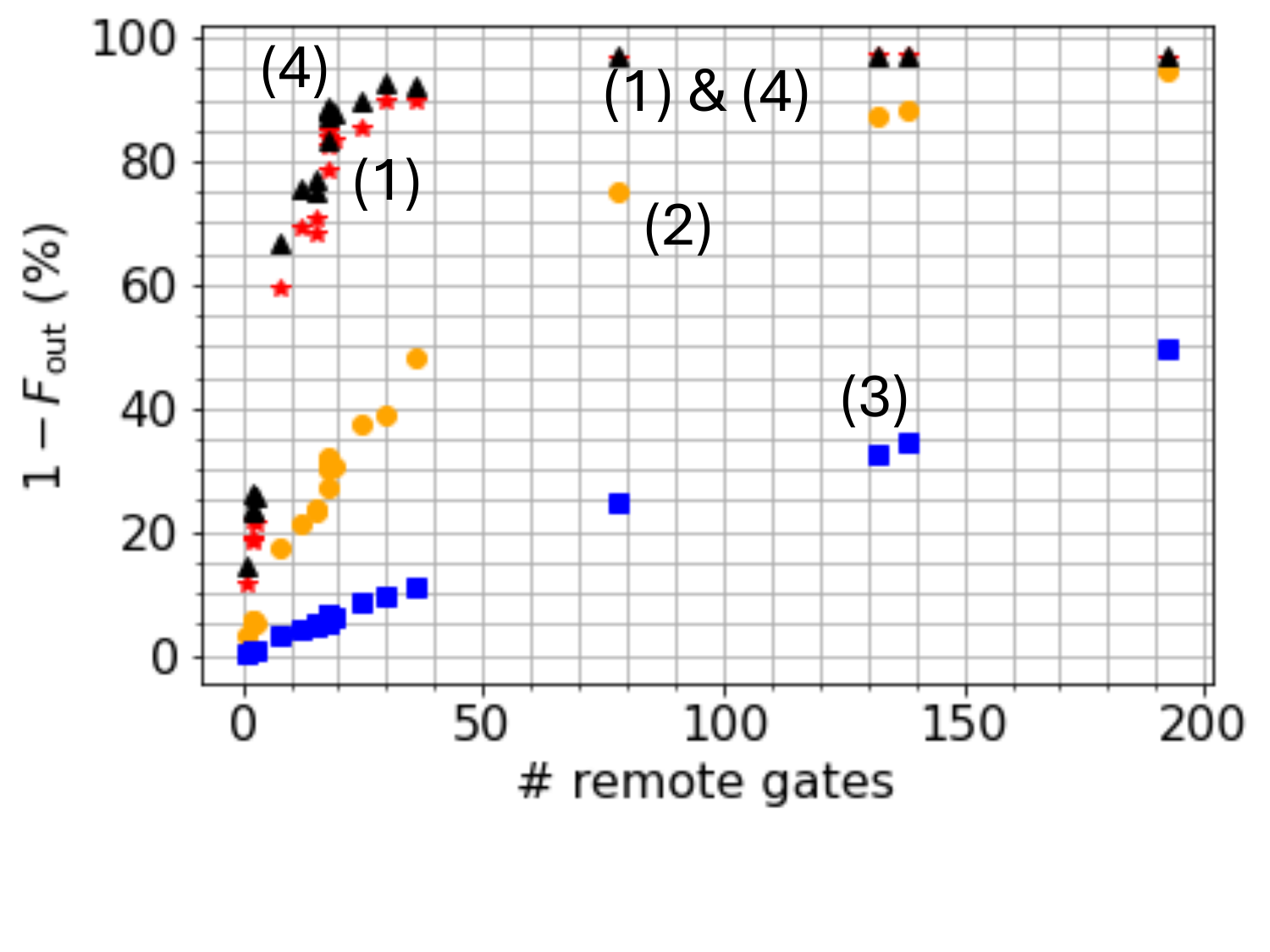}
        \put(-4, 60){(b)}
    \end{overpic}
    \subfloat{%
    \raisebox{0.6\height}{
    \begin{overpic}[scale=0.5, trim={0.5cm, 0.2cm, 0.2cm, 0.2cm}, clip]{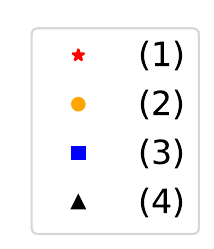}
    \end{overpic}}}
    \caption{Output error, $1 - F_{\rm out}$, as a function of the number of remote gates for 22 five-qubit MQT bench \cite{MQTBench} quantum circuits implemented using (a) cat-comm, (b) TP-safe. The markers correspond to the ouptut fidelity when: (1) $\epsilon_{\mathrm{ebit}} = 6\%$ is the only source of error; (2) $\epsilon_{\mathrm{cnot}} = 0.4\%$ is the only source of error; (3) $r=0.055$ Hz is the only source of error; and (4) all three types of errors are present and have the values quoted for (1)-(3). In some cases, multiple quantum circuits share the same number of remote gates and a separate data point is plotted for each. This occurs for three of the remote gate values.}
    \label{fig:mqt_per_num_remote_gates_state_of_art}
\end{figure}

It is convenient to concisely express the behaviour of all 22 of the circuits investigated in one place. Doing this also allows the very different quantitative values obtained for different circuits to be directly compared. To this end, in Fig. \ref{fig:mqt_per_num_remote_gates_state_of_art}, %
we consider the output error as a function of the number of remote gates. When multiple circuits share the same number of remote gates, all data points are shown---one for each circuit. Once again, unless otherwise noted, each type of error discussed in Sec. \ref{subsec:error_models} is considered separately, with the other errors set to zero. When a given error is being considered, the relevant error parameter is set to the state-of-the-art value stated in Table \ref{tab:state_of_art_params}. Here, we also include an additional curve corresponding to the output error when all of the errors are set to their state-of-the-art value at once. 

\begin{figure*}
    \begin{overpic}[scale=0.2, trim={0, 2.5cm, 0, 0}, clip]{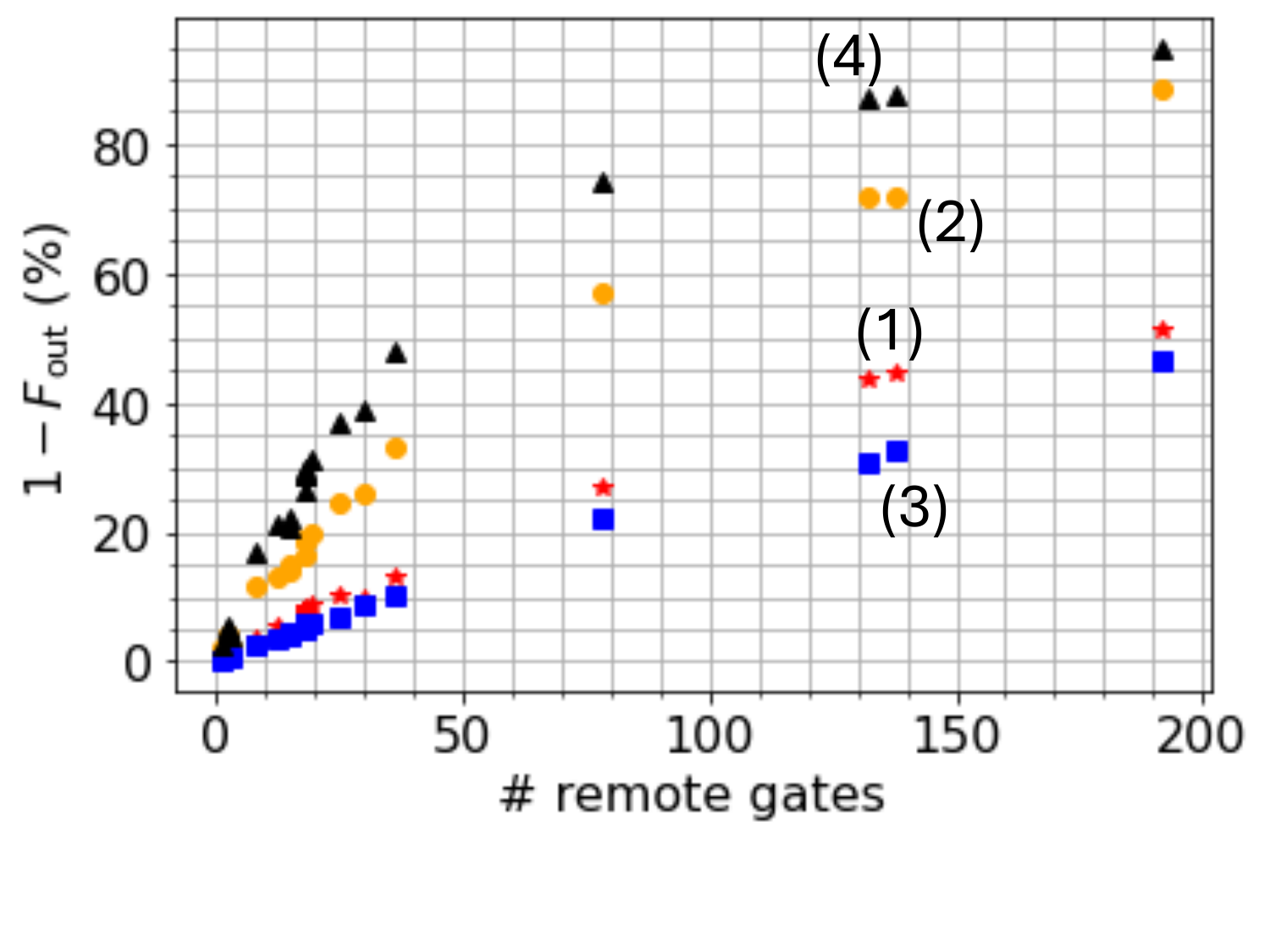}
        \put(-4, 60){(a)}
    \end{overpic}\hspace{2em}%
    \begin{overpic}[scale=0.2, trim={0, 2.5cm, 0, 0}, clip]{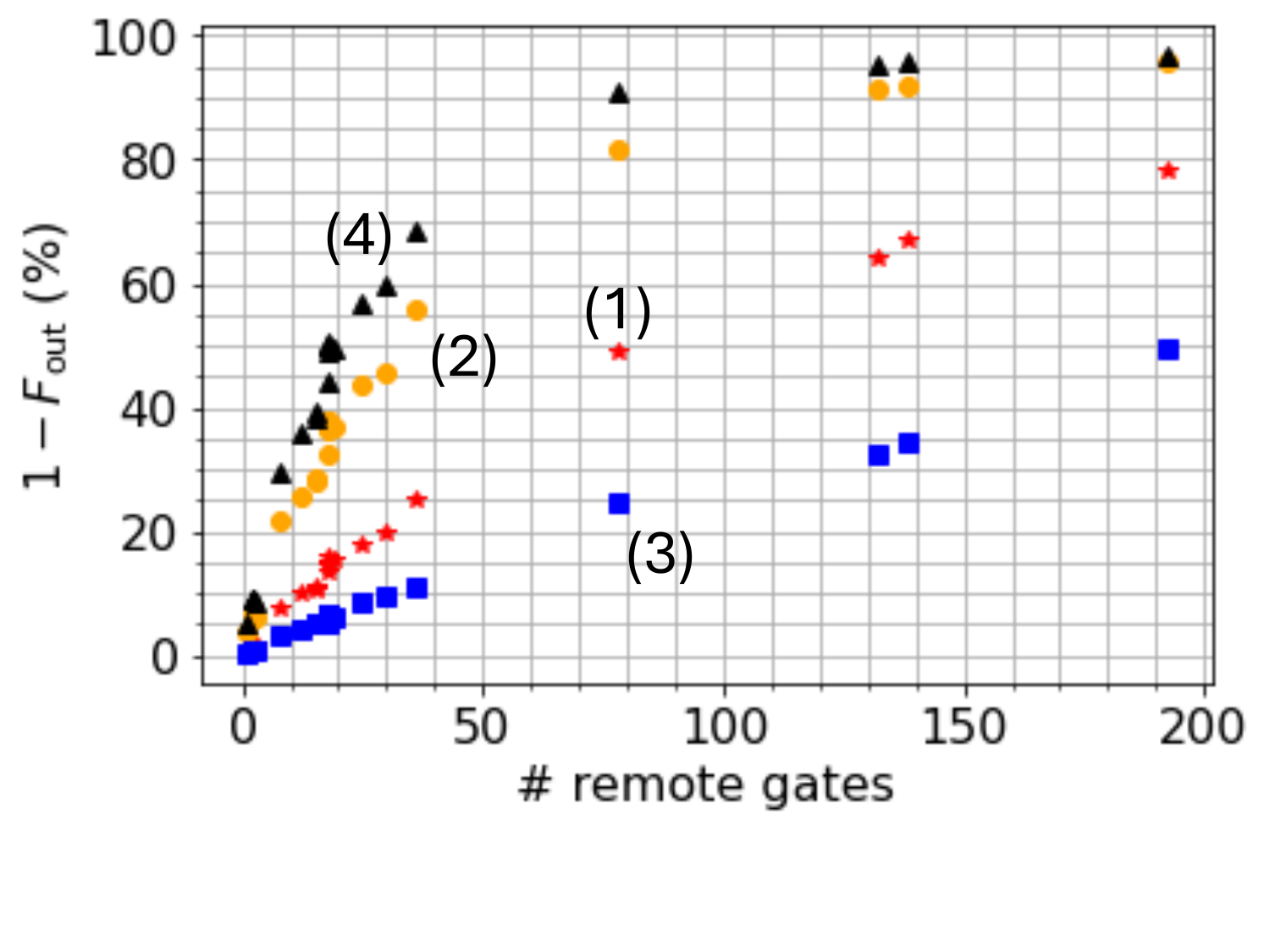}
        \put(-4, 60){(b)}
    \end{overpic}
    \subfloat{%
    \raisebox{0.6\height}{
    \begin{overpic}[scale=0.5, trim={0.5cm, 0.2cm, 0.2cm, 0.2cm}, clip]{comparing_errors_legend_including_all_errors.pdf}
    \end{overpic}}}
    \caption{The output error as a function of the number of remote gates for 22 five-qubit MQT bench \cite{MQTBench} quantum circuits implemented using (a) cat-comm, (b) TP-safe. The markers correspond to the output error when: (1) $\epsilon_{\mathrm{ebit}} = 0.5\%$ is the only source of error; (2) $\epsilon_{\mathrm{cnot}}=0.5\%$ is the only source of error; (3) $r=0.055$ Hz is the only source of error; and (4) all three types of errors are present and have the values quoted for (1)-(3). In some cases, multiple quantum circuits share the same number of remote gates and a separate data point is plotted for each. This occurs for three of the remote gate values.}
    \label{fig:mqt_per_num_remote_gates_distilled}
\end{figure*}

Figure \ref{fig:mqt_per_num_remote_gates_distilled} %
shows the same thing as Fig. \ref{fig:mqt_per_num_remote_gates_state_of_art} except that the entanglement and local gate errors are set to $\epsilon_{\mathrm{ebit}} = 0.5\%$ and $\epsilon_{\mathrm{cnot}} = 0.5\%$, respectively, for comparison. For memory depolarisation, $r=0.055$Hz, as before.

Figure \ref{fig:mqt_per_num_remote_gates_state_of_art} shows that observation \ref{obs:state_of_art_ordering} from Sec. \ref{subsubsec:error_type_comparison} also holds for larger circuits. Once more, the impact of entanglement error dominates when the state-of-the-art error probability is used, followed by the impact of local gate errors, and then that of memory depolarisation. The one caveat is that, unlike for single remote gates, the impact of memory depolarisation is not negligible for some of the circuits considered. Nonetheless, memory depolarisation is still the least impactful type of error. In most cases, the difference in the impact of memory depolarisation from that of the other types of error considered remains quite large.

The magnitudes of the output error observed in Fig. \ref{fig:mqt_per_num_remote_gates_state_of_art} also offer some insight into what will be achievable with QDCs in the near-term. With as few as 10-20 remote gates in a circuit, the three errors considered can drive the output error to approximately 50\% when cat-comm is used. An output error as high as 50\% makes quantum advantage very unlikely. Even with entanglement error alone, under 20 remote gates can be tolerated before the output error exceeds 50\%. For TP-safe, around five remote gates can be tolerated before the output error exceeds 50\%.

When the magnitude of the entanglement error is reduced to fall within the distilled range and we impose that $\epsilon_{\mathrm{ebit}}=\epsilon_{\mathrm{cnot}} =0.5\%$ (see Fig. \ref{fig:mqt_per_num_remote_gates_distilled}), observation \ref{obs:distilled_ordering} from Sec. \ref{subsubsec:error_type_comparison} is also found to be true for larger circuits. As with the single remote gates, local errors have the greatest adverse effect on the output error, followed by entanglement error and then memory depolarisation. Unlike for single remote gates, the impact of entanglement error and memory depolarisation is similar when cat-comm is used to implement all remote gates. This can be seen from the proximity of curves (1) and (3) in Fig. \ref{fig:mqt_per_num_remote_gates_distilled}(a).  For  TP-safe, the impacts of entanglement error and memory depolarisation on output error are less similar to each other than for cat-comm, but they are more similar to each other than is observed for single remote gates. This can be seen using curves (1) and (3) of Fig. \ref{fig:mqt_per_num_remote_gates_distilled}(b).

Additionally, reducing entanglement error to this extent has a significant quantitative impact on the output error. If cat-comm is used, around 30-40 remote gates could be tolerated before the output error observed in curve (4) of Fig. \ref{fig:mqt_per_num_remote_gates_distilled}(a) exceeds 50\%. With TP-safe, around 10-20 remote gates could be tolerated. Thus, reducing the entanglement error by an order of magnitude is likely to dramatically improve the number of quantum circuits with which potentially meaningful experiments could be done using a QDC.

\begin{figure*}
    \begin{overpic}[scale=0.2, trim={0, 2.5cm, 0, 0.1em}, clip]{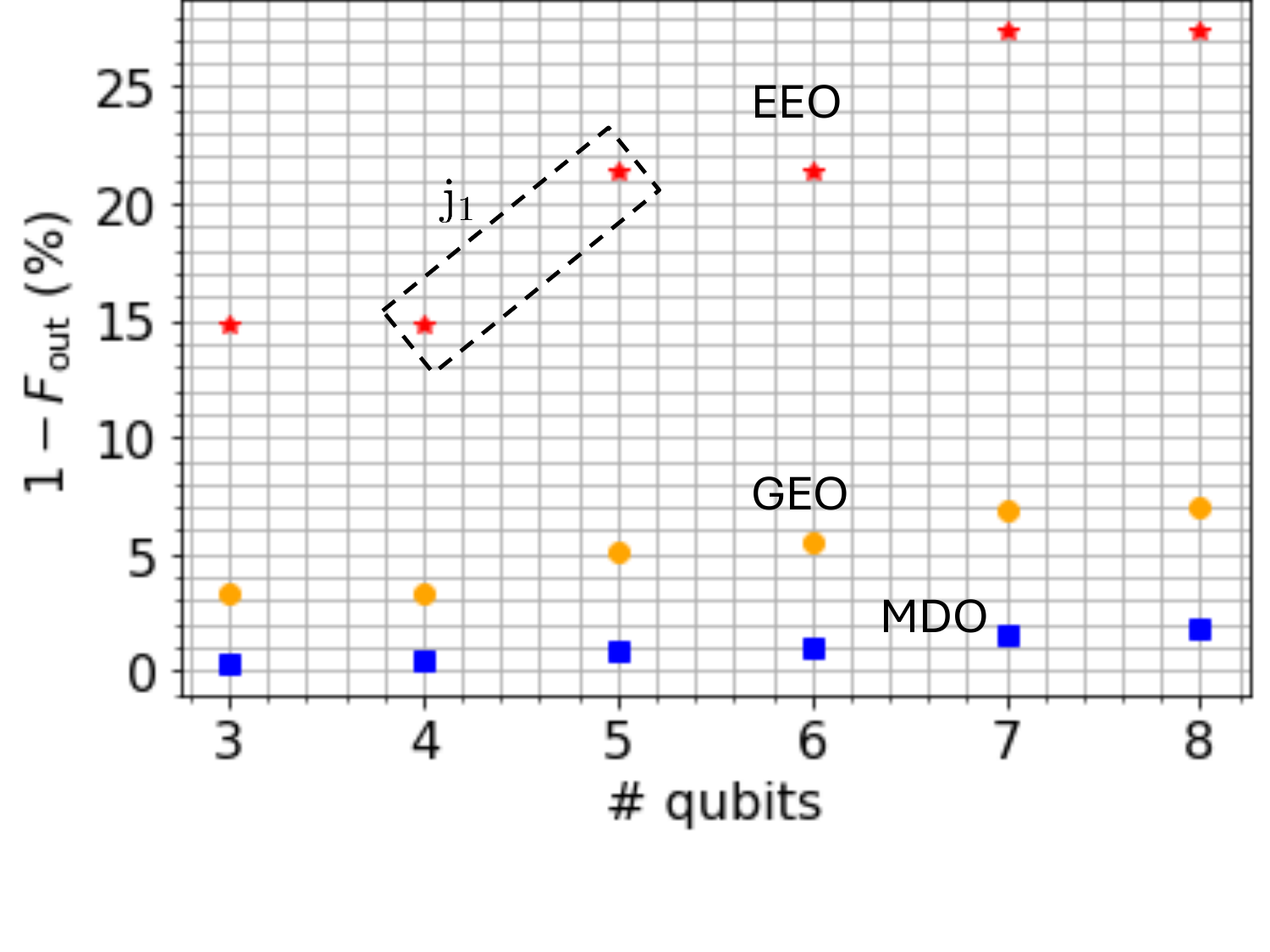}
        \put(23, 58){(a)}
    \end{overpic}
    \begin{overpic}[scale=0.2, trim={0, 2.5cm, 0, 0.1em}, clip]{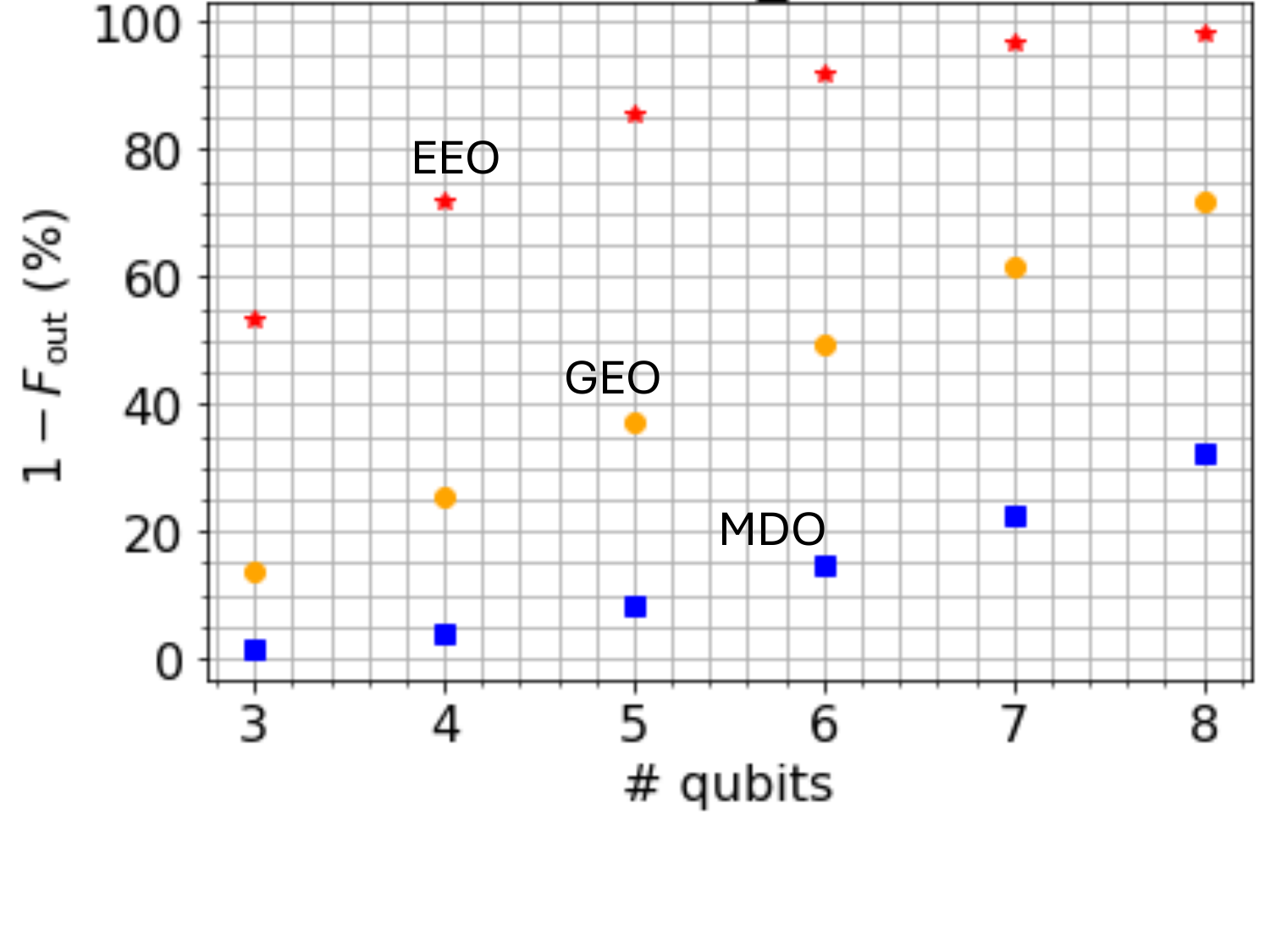}
       \put(23, 58){(b)} 
    \end{overpic}
    \begin{overpic}[scale=0.2, trim={0, 2.5cm, 0, 0}, clip]{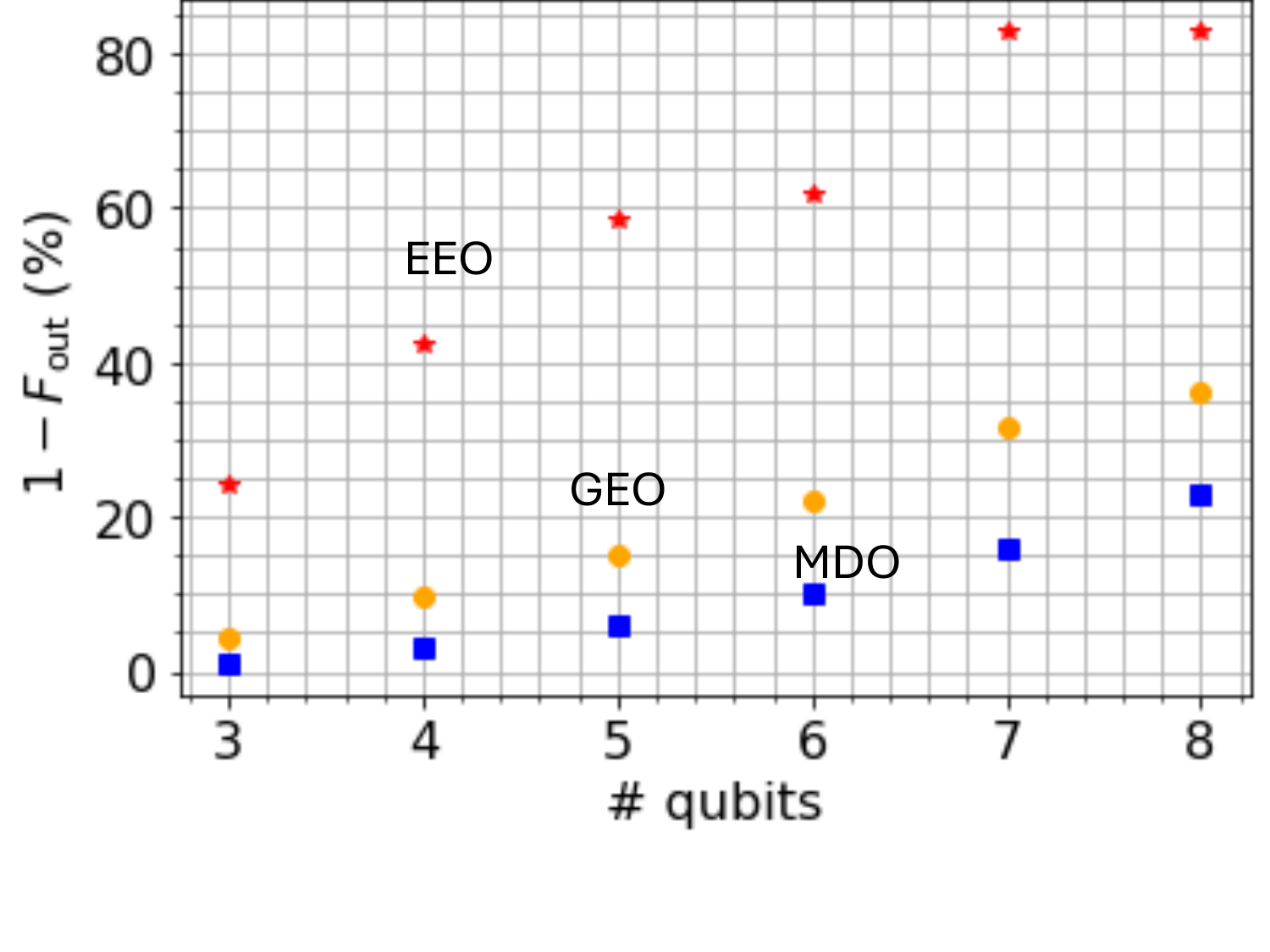}
        \put(23, 58){(c)}
    \end{overpic}
    \begin{overpic}[scale=0.2, trim={0, 2.5cm, 0, 0}, clip]{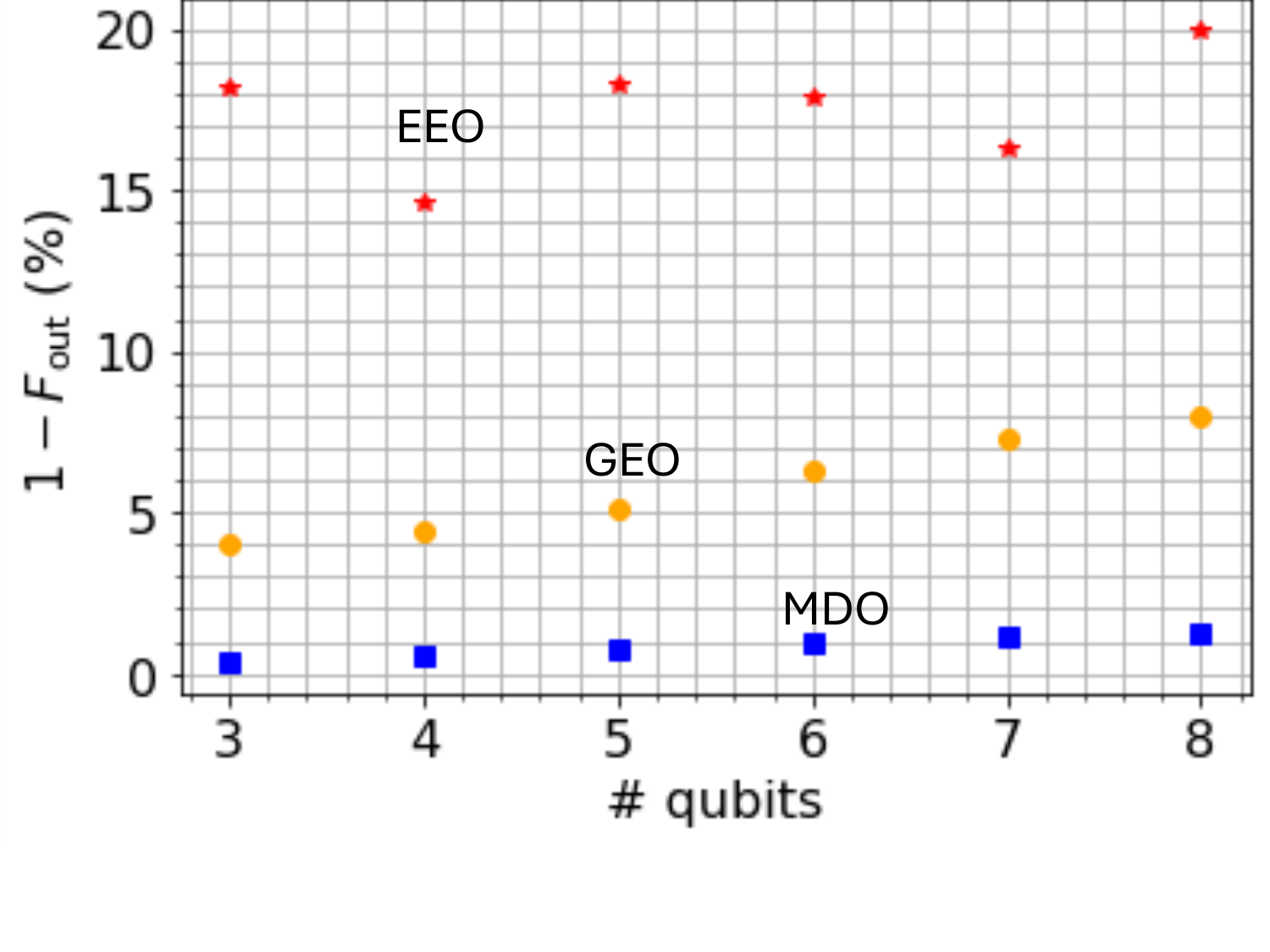}
        \put(23, 58){(d)}
    \end{overpic}
    \subfloat{%
    \raisebox{0.6\height}{
    \begin{overpic}[scale=0.5, trim={0.5cm, 0.2cm, 0.2cm, 0.2cm}, clip]{comparing_errors_legend.pdf}
    \end{overpic}}}
    \caption{The output error, $1-F{\mathrm{out}}$, as a function of the number of qubits used to implement: (a) the Deutsch-Joza circuit; (b) a quantum neural network; (c) the variational quantum eigensolver (VQE) applied to portfolio optimisation; and (d) VQE with a TwoLocal ansatz applied to the max-cut problem. All circuits are taken from Ref. \cite{MQTBench}. In plot (c) cat-comm is used to implement all remote gates and in all other case TP-safe is used. Each set of data points shows the results when one of the error parameters (see the annotations) is non-zero, and all other errors are set to zero. The non-zero error values had the state-of-the-art values given in Table \ref{tab:state_of_art_params}.}
    \label{fig:num_qubits_state_of_art}
\end{figure*}

\begin{figure*}[t]
    \begin{overpic}[scale=0.2, trim={0, 2cm, 0, 0}, clip]{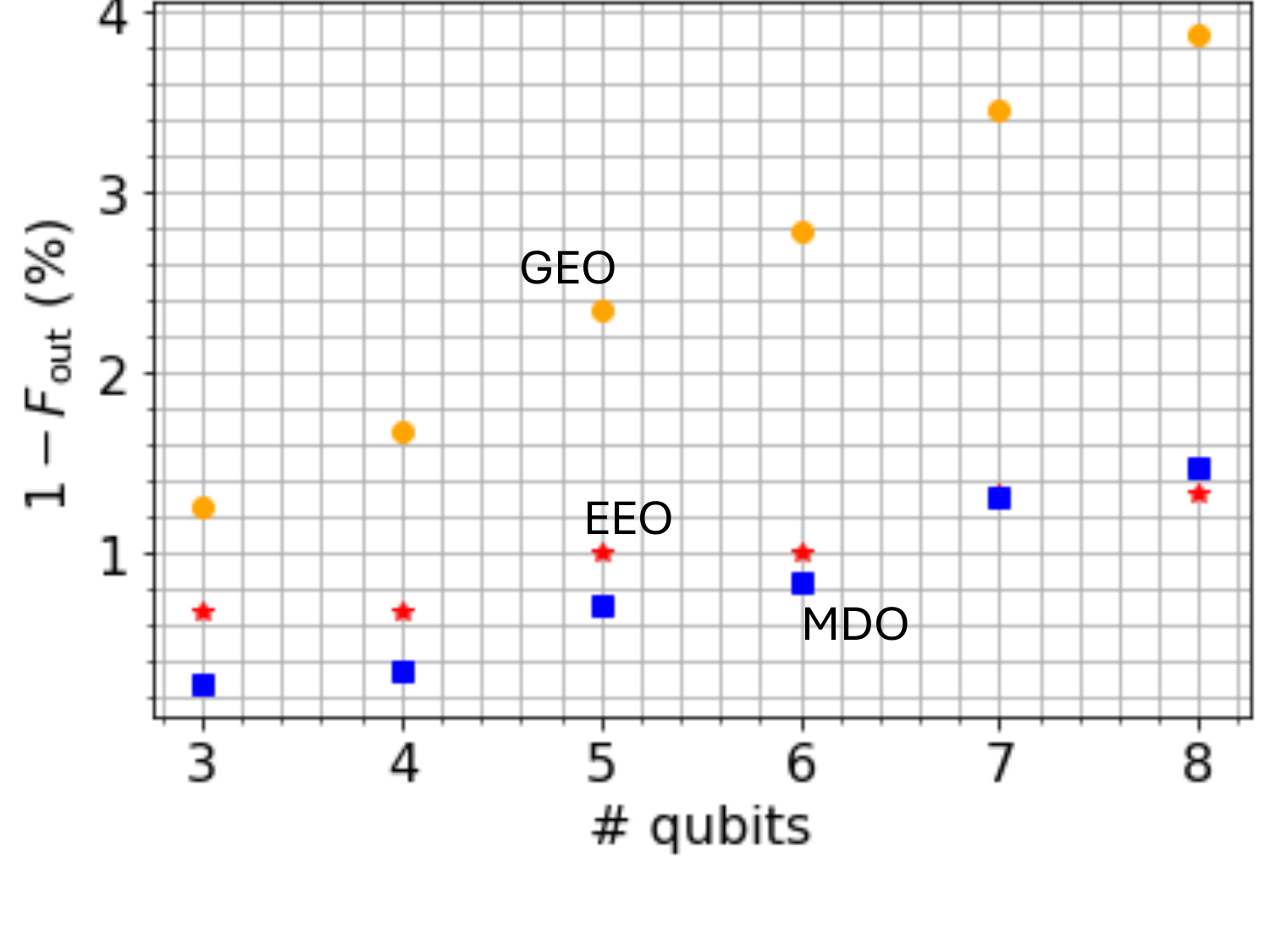}
        \put(-4, 60){(a)}
    \end{overpic}\hspace{2em}%
    \begin{overpic}[scale=0.2, trim={0, 2cm, 0, 0}, clip]{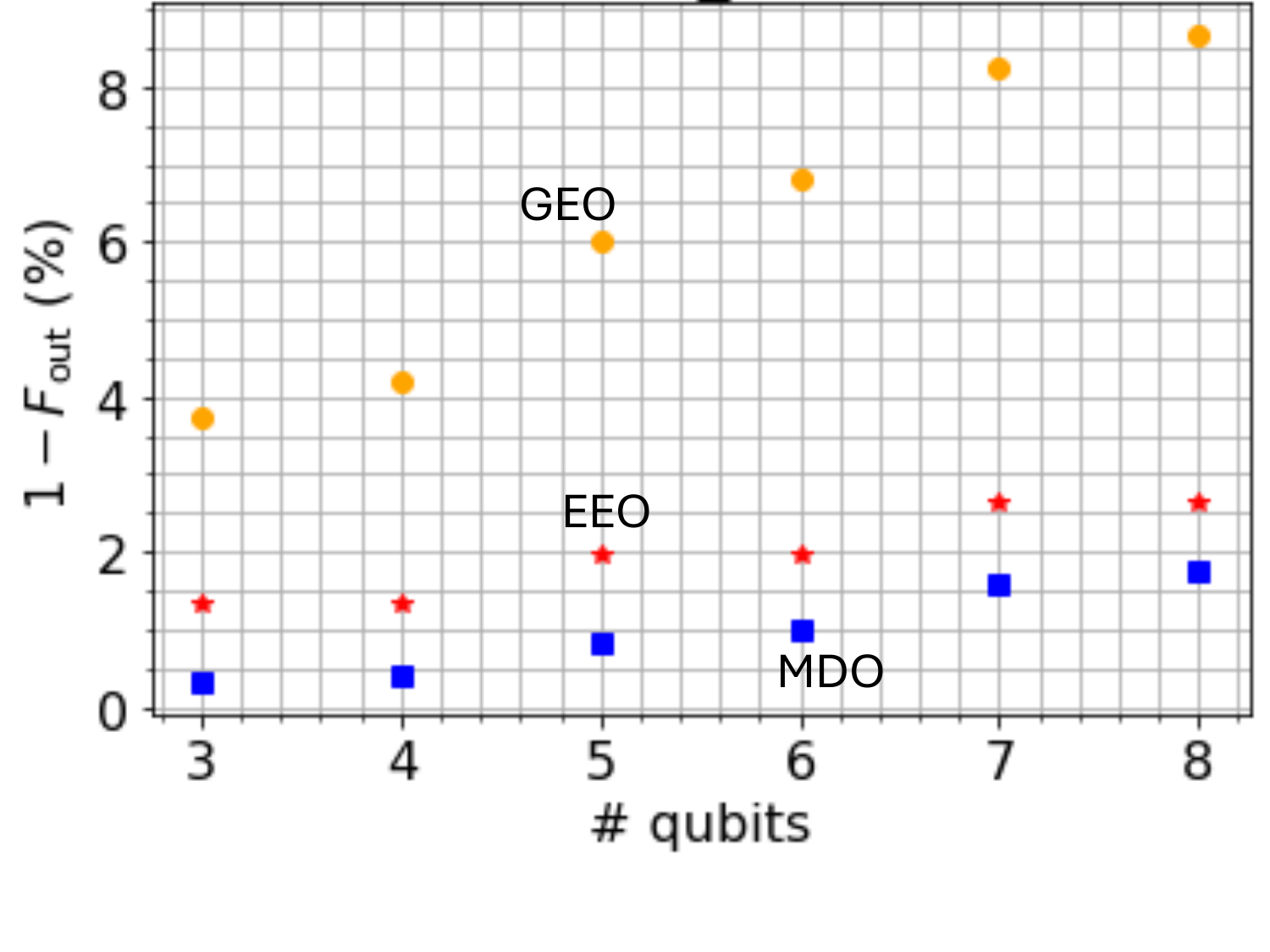}
        \put(-8, 60){(b)}
    \end{overpic}\hspace{2em}%
    \begin{overpic}[scale=0.2, trim={0, 2cm, 0, 0}, clip]{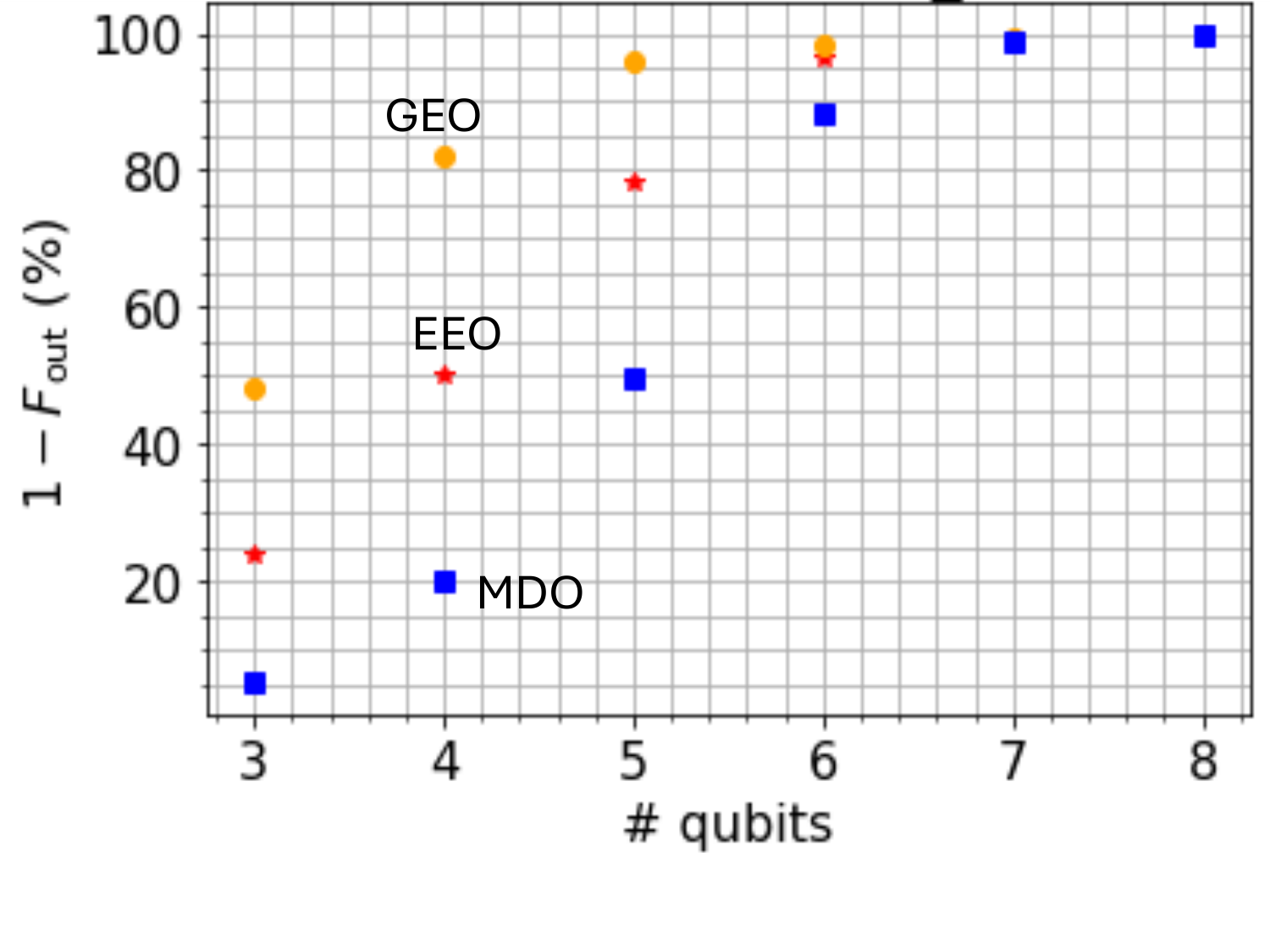}
        \put(-8, 60){(c)}
        \put(60, 20){\includegraphics[scale=0.5, trim={0.7cm, 0.2cm, 0.2cm, 0.2cm}, clip]{comparing_errors_legend.pdf}}
    \end{overpic}
    \caption{The output error, $1-F{\rm out}$, as a function of the number of qubits used to implement: (a) the Deutsch-Joza circuit with cat-comm; (b) the Deutsch-Joza algorithm with TP-safe; (c) a quantum walk without ancilla qubits using TP-safe. Each set of data points shows the results when one of the error parameters (see the annotations) is non-zero, and all other errors are set to zero. The non-zero error parameters used are: $\epsilon_{\mathrm{ebit}} = 0.5\%$ for entanglement error only (EEO); $\epsilon_{\mathrm{cnot}} = 0.5\%$ for two-qubit gate errors only (GEO); and $r=0.055$Hz for memory depolarisation only (MDO). For $\epsilon_{\mathrm{ebit}}$, this means the distilled range is used.}
    \label{fig:num_qubits_distilled}
\end{figure*}

Another instructive resource to consider is the number of qubits. The impact of varying the number of qubits is considered for all of the circuits in the MQT bench suite by implementing versions of the same circuits but with a different quantum register size in each case. The effect of this on the output error is shown in Figs. \ref{fig:num_qubits_state_of_art} %
and \ref{fig:num_qubits_distilled} %
for a selection of the MQT Bench circuits with fixed error parameters in the state-of-the-art and distilled ranges, respectively. For each set of data points, one type of error is non-zero and the other types of error are set to zero. The results for the remainder of the circuits considered are shown in the supplementary information. As the standard error is negligible for the five-qubit circuits, which are averaged over ten runs, no such averaging is done for the other circuits, with different numbers of qubits, to avoid incurring the significant increase in runtime associated with repeating the simulation.

Figures \ref{fig:num_qubits_state_of_art} and \ref{fig:num_qubits_distilled} provide further evidence for the previous discussion on the relative impact of each error type. Using Fig. \ref{fig:num_qubits_state_of_art}, we again see that, with the state-of-the-art error parameters, entanglement error has the largest adverse impact on output error, followed by local gate errors and then memory depolarisation. When entanglement error is reduced to be within the distilled range (see Fig. \ref{fig:num_qubits_distilled}), we again find that the impact of local gate errors dominates. However, if all remote gates are implemented with cat-comm, we find that for many of the circuits considered, the impact of memory depolarisation can exceed that of entanglement error when the number of qubits used is high enough. Figure \ref{fig:num_qubits_distilled}(a) shows this happening for the Deutsch-Joza algorithm. Further examples can be found in the supplementary information.  This is intuitive, as increasing the number of qubits will increase the latency of the circuit, meaning qubits are more likely to undergo memory decoherence. 

A less intuitive feature of Figs. \ref{fig:num_qubits_state_of_art} and \ref{fig:num_qubits_distilled} is the existence of discrete jumps and plateaus in the output error.  In some cases, the jumps and plateaus can be easily understood. For example, consider jump $\mathrm{j}_1$ in Fig. \ref{fig:num_qubits_state_of_art}(a), which shows $1-F_{\mathrm{out}}$ as a function of the number of qubits for Deutsch-Josza circuits implemented with TP-safe. Jump $\mathrm{j}_1$ occurs when increasing the number of qubits from four to five because the number of ebits increases from two to three when doing so. As such, it is natural for the entanglement error to increase and cause an increase in $1-F_{\mathrm{out}}$. By contrast, when we consider the five and six qubit circuits, both have the same number of ebits (three) and so we see a corresponding plateau in $1-F_{\mathrm{out}}$ for the curve in which only entanglement error, $\epsilon_{ebit}$, is non-zero. Similar arguments can be used to explain the other jumps observed in Fig. \ref{fig:num_qubits_state_of_art}(a) and Figs. \ref{fig:num_qubits_distilled}(a)-(b). On the other hand, in some cases, what causes jumps and plateaus remains an open question. Consider, for example, Fig. \ref{fig:num_qubits_state_of_art}(d), which shows $1-F_{\mathrm{out}}$ for circuits implementing the variational quantum eigensolver (VQE) algorithm using TP-safe with increasingly many qubits. The curve for which only the entanglement error is non-zero fluctuates quite a bit, but the number of ebits remains constant at two for all data points. Therefore, unlike for the Deutsch-Josza algorithm, variations in $1-F_{\mathrm{out}}$ cannot be explained by changes in the number of ebits. What causes these fluctuations in $1-F_{\mathrm{out}}$ for VQE remains an open question. These features of the results allude to the fact that knowing the quantity of resources such as the number of ebits or local CNOT gates is not always enough to predict the error propagation in distributed quantum circuits. We infer that the exact structure of the compiled quantum circuit is also important to the error propagation and suggest caution against overreliance on heuristic optimisation that considers only resource quantities when compiling distributed quantum circuits, although this remains a useful starting point. There is still much to learn about error propagation in QDCs and distributed quantum computers more generally.

\section{Conclusion and future outlook}
\label{sec:conclusion}

In this work, we have used classical simulation to emulate operations on a quantum data centre. We studied the behaviour of individual remote gates and a variety of larger distributed quantum circuits, in the presence of various types of error.

We found that first-order approximations to error propagation fail to accurately capture the behaviour of even the smallest systems for any finite error. 

We also found that for error values obtainable with current trapped-ion hardware, the detrimental impact of imperfect inter-nodal entanglement dominates that of local errors, due to the much higher magnitude of entanglement error. The next most impactful of the errors considered is imperfect implementation of local gates, and time-dependent memory decoherence is negligible in comparison to other forms of error. However, if high-quality entanglement could be generated, so that the magnitudes of the gate and entanglement errors are comparable, then gate errors become more impactful than entanglement errors.

Additionally, we discovered that, despite having the same quantities and types of ebits, gates, classical communications, and measurements, cat-comm and 1TP can have different output errors for the same error types and parameter values. The exact discrepancy in output error is input state dependent.

Finally, we found that if current quantum technology could be successfully integrated to form a single QDC, a circuit containing around 10 remote gates could be implemented with cat-comm before the output error exceeded 50\%. If the magnitude of the entanglement error could be reduced to the same value as local gate error, a circuit containing at least 30 remote gates could be implemented. These numbers could most likely be significantly improved by using more optimal compilation strategies.

Our results are likely to be most applicable to matter-based QPUs with photonic interconnects, although, it is possible that they could be relevant to other platforms in the future, if a viable method of photon to communication qubit transduction is discovered for those platforms. Viewed collectively, our results indicate that small to medium-sized experimental demonstrations of a QDC are feasible in the near to mid-term. This has been recently corroborated by an experimental implementation of a small two-QPU QDC \cite{firstDeterministicQDC}, albeit with a very low entanglement distribution rate (of 9.7Hz)---to keep the entanglement error low. It may take longer before QDCs can compete with monolithic quantum computers in terms of output error. Our results also indicate which errors should be focused on for QDC hardware and circuit optimisation, and introduce some additional factors that should be considered when developing QDC compilation heuristics. 

The further development of QDC compilation heuristics represents a possible avenue for future research, and there remains tremendous scope for future QDC optimisation. It would also be interesting to extend our work to larger QDCs with more QPUs and explore more complicated multipartite entanglement within a QDC network. Further work could include more detailed and specific analysis of certain hardware platforms, such as the one implemented in Ref. \cite{firstDeterministicQDC}. For now, comparisons to such work are challenging as they would require more bespoke simulation to mirror the idiosyncrasies of each platform. For example, each platform has its own specific set of native gates, which may differ from ours, and it may be necessary to implement specific error mitigation strategies, such as dynamical decoupling, if we wish to emulate specific devices as they are being implemented in experiments. We defer such investigations to future work.

\section{Acknowledgements}

We acknowledge funding from the  Leeds Doctoral Scholarship and the UK EPSRC grants EP/Y037421/1 and EP/X040518/1. 

\section{Data availability}

The data that support the findings of this study are openly available at the following URL/DOI: 10.5281/zenodo.13773229.

\appendix

\section{Analytical derivations of the output fidelity for 1TP and cat-comm}
\label{app:analytical_derivations}

In the main text, the working for Eqs. \eqref{eq:1TP_analytical_F_out} and \eqref{eq:cat_analtyical_F_out4CNOT}, the analytical expressions of $F_{\mathrm{out}}$ for 1TP and cat-comm, respectively, are omitted for brevity. Here, we provide the working for both expressions.

For both calculations, we model entanglement error as the distribution of the Werner state (see Eq. \eqref{eq:werner_state}) rather than the ideal $\ket{\Phi^+}$ state. The Werner state is a classical mixture of the four Bell states. Therefore, it is instructive to consider what happens to  1TP and cat-comm when each of the Bell states is distributed between QPUs.

The 1TP scheme depicted in Fig. \ref{fig:remote_cnot_implemented_in_different_ways}(b) assumes that the $\ket{\Phi^+}$ state, specifically, was distributed between QPUs during the teleportation process. If one of the other Bell states is distributed instead then a Pauli error occurs in the teleported state. We represent this error as a fictitious gate $R_i \in \{\mathcal{I}, \sigma_x, \sigma_z, \sigma_z \sigma_x\}$, where $\sigma_x$ and $\sigma_z$ are the Pauli $x$ and $y$ operators, respectively.

\begin{figure}
    \centering
    \includegraphics{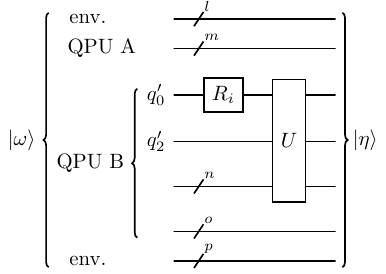}
    \caption{A simplified version of the 1TP quantum circuit shown in Fig. \ref{fig:remote_cnot_implemented_in_different_ways}(b). The circuit shown is derived from Fig. \ref{fig:remote_cnot_implemented_in_different_ways}(b) by assuming a measurement result of `$0$' for all measurements in the circuit and generalising to an arbitrary remote gate, rather than a remote CNOT gate. $\ket{\omega}$ is the purified ideal state produced after an ideal teleportation from QPU A to QPU B. To produce $\ket{\omega}$, the system has been combined with a possibly fictitious environment (labelled env. in the figure), so that it can be considered to be pure, as discussed in Sec. 2.5 of Ref. \cite{NielsenChuang}. $\ket{\eta}$ is the ideal output from the circuit. $l$, $m$, and $n$ $\in \mathcal{N}$ are arbitrary, possibly infinite numbers of qubits, which could be in the QDC itself or an external environment, and may be real or fictitious. $R_i$ is a rotation of the ideal teleported state imposed by the possible non-ideality of the entangled state distributed, and $U$ is a  quantum gate local to QPU B's qubits. The qubit labels $q_0'$ and $q_2'$ refer to the same qubits as in the original circuit shown in Fig. \ref{fig:remote_cnot_implemented_in_different_ways}(b).}
    \label{fig:simplified_1TP_comm}
\end{figure}

Further simplifications can be made using the fact that when any of the Bell states is distributed between QPUs, the output is independent of the BSM results used during the teleportation process, provided that there are no local, intra-QPU, errors of any kind. This is simply verified by calculating the output for each measurement result and can be intuitively understood by noting that any measurement result of qubits on QPU A in Fig. \ref{fig:remote_cnot_implemented_in_different_ways}(b) is equally likely to be obtained, regardless of which Bell state is distributed between QPU A and QPU B. Consequently, the circuit shown in Fig. \ref{fig:remote_cnot_implemented_in_different_ways}(b) can be greatly simplified by assuming a specific BSM result has occurred, without loss of generality. For simplicity, we assume here that the BSM result is `$00$', and obtain a simplified circuit diagram for implementing an arbitrary remote gate using 1TP, as shown in Fig. \ref{fig:simplified_1TP_comm}. %

Combining the results for each of the Bell states, the overall output state, $\rho_{\rm out}$, can be represented in terms of the ideal input state, $\rho_{\omega}$, as \begin{equation}
\label{eq:single_tele_ouput_dm}
    \rho_{\rm out} = U (\sum^4_{i=1} p_i R_i \rho_{\omega} R_{i}^{\dag}) U^{\dag},
\end{equation}
where $p_i$ is the probability of being in one of the Bell states (corresponding to the coefficients in the Werner state), and $U$ is a unitary gate acting on QPU B from Fig. \ref{fig:simplified_1TP_comm}. For example, if a remote CNOT gate were being considered, as in Fig. \ref{fig:remote_cnot_implemented_in_different_ways}(b), $U$ would be the X (or NOT) gate. Here, we consider the more general case in which $U$ is arbitrary and the remote gate need not be a controlled-unitary (CU) gate.

Purifying $\rho_{\omega}$, by combining it with a, potentially fictitious, environment, $E$, as discussed in Sec. $2.5$ of Ref. \cite{NielsenChuang}, we obtain:%
\begin{equation} \label{eq:simplified_tp_comm_output}
    \rho_{\rm out} = U (\sum^4_{i=1} p_i R_i \ket{\omega} \bra{\omega} R_{i}^{\dag}) U^{\dag},
\end{equation}%
where $\ket{\omega}$ is the purified ideal input obtained when $\ket{\Phi^+}$ is distributed between QPU A and QPU B. The environment, $E$, may include qubits within the quantum data centre and fictitious environmental qubits external to it.

Again using purification, the purified ideal output from the circuit, when the correct Bell state, $\ket{\Phi^+}$, is distributed initially, is given by:%
\begin{equation} \label{eq:simplified_tp_comm_ideal_output}
    \ket{\eta} = U \ket{\omega}.
\end{equation}%

For an arbitrary pure state, $\ket{\psi}$, the fidelity definition given by Eq. \eqref{eq:fidelity_definition} simplifies to \cite{NielsenChuang}:%
\begin{equation} \label{eq:pure_state_fidelity_def}
    F(\ket{\psi} \bra{\psi}, \rho_{\rm noisy}) =
   \bra{\psi} \rho_{\rm noisy} \ket{\psi}.
\end{equation}%
Inserting Eqs. \eqref{eq:simplified_tp_comm_output} and \eqref{eq:simplified_tp_comm_ideal_output} into  \eqref{eq:pure_state_fidelity_def} gives an output fidelity $F_{\rm out}$ of%
\begin{equation} \label{eq:first_output_fidelity_tp_comm}
\begin{split}
    F_{\rm out} &= \bra{\omega} U^{\dag} U (\sum^4_{i=1} p_i R_i \ket{\omega} \bra{\omega} R_{i}^{\dag}) U^{\dag} U \ket{\omega} \\
    &=\sum^4_{i=1} p_i |\bra{\omega} U^{\dag} U  R_i \ket{\omega}|^2,
\end{split}
\end{equation}
which by the unitarity of $U$ simplifies to \begin{equation}\label{eq:unitarity_simplified_output_fidelity_tp_comm}
    F_{\rm out} =  \sum^4_{i=1} p_i |\bra{\omega} R_i \ket{\omega} |^2.
\end{equation}

Using the Schmidt decomposition \cite{NielsenChuang}, \\%
$\ket{\omega} = \sum_k \lambda_k \ket{k_{q_0'} } \ket{k_{q_2', E}}$:
\begin{equation}
    F_{\rm out} =  \sum^4_{i=1} p_i \sum_{j, k} \lambda_j \lambda_k | \bra{j_{q_0'}} \bra{j_{q_2', E}} (R_i)_{q_0'} \ket{k_{q_0'}} \ket{k_{q_2', E}} |^2,
\end{equation}%
where $\lambda_j$ and $\lambda_k$ are non-negative real numbers satisfying $\sum_{j} \lambda_{j}^2 = 1$, and $\ket{k_{q_0'}}, \ket{k_{q_2', E}}$ are orthonormal states for the $q_0'$ and $\{q_2', E\}$ systems, respectively (see Fig. \ref{fig:simplified_1TP_comm}).

By the orthonormality of $\ket{j_{q_2', E}}$ and $\ket{k_{q_2', E}}$ and the commutativity of the basis states for the $q_0'$ and $\{q_2', E\}$ subsystems: \begin{equation}
\label{eq:penultimate_result_for_single_tele_TP_fidelity}
    \begin{split}
        F_{\rm out} & = 
\sum_{j, k} \lambda_j \lambda_k \sum^4_{i=1} p_i | \bra{j_{q_0'}} (R_i)_{q_0'} \ket{k_{q_0'}} \delta_{j, k} |^2 \\
        & =  \sum_{j} \lambda_j^2 \sum^4_{i=1} p_i | \bra{j_{q_0'}} (R_i)_{q_0'} \ket{j_{q_0'}}|^2,
    \end{split}
\end{equation}
where $\delta_{j, k}$ is the kronecker delta.

From there, we individually consider each term in the sum over index $i$.

For $i=1$, $p_i = F_{\rm w}$, and %
\begin{equation}
\label{eq:tp_single_tele_i_is1}
        p_1 \left| \bra{j_{q_0'}} (R_1)_{q_0'} \ket{j_{q_0'}}\right|^2 = F_{\rm w} \left| \bra{j_{q_0'}}  \1 \ket{j_{q_0'}}\right|^2 
        = F_{\rm w}.
\end{equation}

For the remaining three terms, $p_i = \frac{1-F_{\rm w}}{3}$. For $i=2$, using the Schmidt decomposition, $ \ket{\omega} = \sum_m \lambda_m \ket{m_{q_0'}} \ket{m_{q_2', E}} $:%
\begin{equation}
\label{eq:tp_analytical_fidelity_one_tele_i_is2}
\begin{split}
    p_2 & \left| \bra{j_{q_0'}} (R_2)_{q_0'} \ket{j_{q_0'}}\right|^2 \\
    &= \frac{1-F_{\rm w}}{3} \sum_{m, n} \lambda_m \lambda_n \left( \bra{m_{q_0'}}  \bra{m_{q_2', E}} \sigma_{x_{q_0'}} \ket{n_{q_0'} }\ket{n_{q_2', E}} \right)^2.
\end{split}
\end{equation}%
By the orthonormality of $\ket{m}_{q_2', E}$ and $\ket{n}_{q_2', E}$, and the independence of the $q_0'$ and $\{q_2', E \}$ subsystems from each other: \begin{equation}
    \label{eq:1tp_analytical_fidelity_i_is2_penultimate}
    \begin{split}
    \eqref{eq:tp_analytical_fidelity_one_tele_i_is2} &= \frac{1-F_{\rm w}}{3} \sum_{m, n} \lambda_m \lambda_n \left|\bra{m_{q_0'}}(\sigma_x)_{q_0'} \ket{m_{q_0'}}\delta_{m, n}\right|^2 \\
    &= \frac{1-F_{\rm w}}{3} \sum_m \lambda_m^2 \left|\bra{m_{q_0'}}(\sigma_x)_{q_0'}  \ket{m_{q_0'}}\right|^2
    \end{split}
\end{equation}

Any pure, single-qubit state can be represented in the form $\alpha \ket{0} + \beta \ket{1}$ and so we can say $\ket{m_{q_0'}}= \alpha \ket{0}_{q_0'} + \beta \ket{1}_{q_0'}$ without loss of generality. Using this, Eq. \eqref{eq:1tp_analytical_fidelity_i_is2_penultimate} becomes%
\begin{equation}
\label{eq:tp_single_tele_i_is2}
\begin{split}
        & \frac{1-F_{\rm w}}{3} \sum_m \lambda_m^2 \left| (\alpha^* \bra{0} + \beta^* \bra{1}) (\alpha \ket{1} + \beta \ket{0}\right|^2 \\
        & = \frac{1-F_{\rm w}}{3} 
        \sum_m \lambda_m^2 \left| \alpha^* \beta + \beta^* \alpha \right|^2 \\
        & = \frac{1-F_{\rm w}}{3} \sum_m \lambda_m^2 \left( \left| \alpha^* \beta \right|^2 + \left| \beta^* \alpha \right|^2 + \alpha^* \alpha^* \beta \beta + \alpha \alpha \beta^* \beta^* \right) \\
        & = \frac{1-F_{\rm w}}{3}  \left( 2 |\alpha|^2 |\beta|^2 + \alpha^* \alpha^* \beta \beta + \alpha \alpha \beta^* \beta^* \right),
\end{split}
\end{equation}%
where the last line follows from the fact that $\sum_m \lambda_m^2 = 1$, as per the definition of Schmidt decomposition.

The calculations for $i=3$ and $i=4$ proceed similarly, except that we replace $\sigma_{x_{q_0'}}$ with $\sigma_{z_{q_0'}}$ and $\sigma_{z_{q_0'}} \sigma_{x_{q_0'}}$, respectively. We obtain:%
\begin{equation}
\label{eq:tp_single_tele_i_is3}
        \frac{1-F_{\rm w}}{3} 
        \left| \left| \alpha \right|^2 - \left| \beta \right|^2 \right|^2,
\end{equation}
for $i=3$, and%
\begin{equation}
\label{eq:tp_single_tele_i_is4}
        \frac{1-F_{\rm w}}{3} \left| (\alpha^* \bra{0} + \beta^* \bra{1}) (\alpha \ket{1} - \beta \ket{0}\right|^2
\end{equation}%
for $i=4$.

Substituting Eqs. \eqref{eq:tp_single_tele_i_is1}, \eqref{eq:tp_single_tele_i_is2}, \eqref{eq:tp_single_tele_i_is3}, and \eqref{eq:tp_single_tele_i_is4} into Eq. \eqref{eq:penultimate_result_for_single_tele_TP_fidelity} gives
\begin{equation}
    F_{\rm out} = \frac{1 + 2F_{\rm w}}{3}.
\end{equation}

Remarkably, this simple result has no dependence on the local operation, $U$, applied to QPU B's qubits, provided that $U$ is unitary. The result is also independent of the input state to the processing qubits.

For cat-comm, things are complicated by the measurement on QPU B in the cat-disentanglement subroutine. This means that, although a similar simplified circuit to that depicted in Fig. \ref{fig:simplified_1TP_comm} can be concocted, the operation $U$ would no longer be unitary and so the cancellations possible when working with the Schmidt decomposition for 1TP no longer apply. Consequently, it is more convenient to use the density matrix formalism throughout the calculation. 

The circuit output remains independent of the measurement outcomes, as before, and so there again exist four possible pure output states, one for each of the terms in the Werner state. To make the problem tractable, we make the limiting assumption that the processing qubits are initially separable. This reduces the generality of the results greatly but does allow some initial investigations to be made and means that simulated results can be checked. With these assumptions made, the possible output states, $\ket{\mathrm{out}}$, to the circuit when the ebits used are in the state $\ket{\Phi^{\pm}}$ or $\ket{\Psi^{\pm}}$ are:%
\begin{equation} \label{eq:phi_plus_minus_output_from_remote_CU}
    |\Phi^{\textcolor{red}{\pm}}\rangle_{q_0, q_0'} \rightarrow |\rm out\rangle_{q_2,q_2'} = \alpha |0\rangle_{q_2} |\chi \rangle_{q_2'} \textcolor{red}{\pm} \beta |1\rangle_{q_2} \textcolor{red}{U_{q_2'}}|\chi \rangle_{q_2'},
\end{equation}%
\begin{equation} \label{eq:psi_plus_minus_output_from_remote_CU}
        |\Psi^{\textcolor{red}{\pm}}\rangle_{q_0,q_0'} \rightarrow |\rm out\rangle_{q_2,q_2'} = \alpha |0\rangle_{q_2} \textcolor{red}{U_{q_2'}}|\chi \rangle_{q_2'} \textcolor{red}{\pm} \beta |1\rangle_{q_2}|\chi \rangle_{q_2'},
\end{equation}
where here and hereafter we have highlighted discrepancies between the states in red. $\ket{\chi}_{q_2'}$ is the arbitrary pure state of qubit $q_2'$ from Fig. \ref{fig:simplified_1TP_comm}.

The output states from Eqs. \eqref{eq:phi_plus_minus_output_from_remote_CU} and \eqref{eq:psi_plus_minus_output_from_remote_CU} correspond to the density matrices:
\begin{equation}
\begin{split}
    \rho_{\ket{\Phi^{\textcolor{red}{\pm}}}} =&\, |\alpha|^2 \ket{0} \bra{0}_{q_2} \ket{\chi}\bra{\chi}_{q_2'} \, \textcolor{red}{\pm} \, \alpha \beta^* \ket{0}\bra{1}_{q_2} \ket{\chi}\bra{\chi}_{q_2'} \textcolor{red}{U_{q_2'}^{\dagger}} \\
    &\textcolor{red}{\pm} \, \beta \alpha^* \ket{1}\bra{0}_{q_2} \textcolor{red}{U_{q_2'}} \ket{\chi} \bra{\chi}_{q_2'} \\ &+ 
    |\beta|^2 \ket{1} \bra{1}_{q_2} \textcolor{red}{U_{q_2'}} \ket{\chi}\bra{\chi}_{q_2'} \textcolor{red}{U_{q_2'}^{\dagger}},
\end{split}
\label{eq:rho4phi_plus_minus_output_from_remote_CU}
\end{equation}%
and%
\begin{equation}
\begin{split}
    \rho_{\ket{\Psi^{\textcolor{red}{\pm}}}} =& \, |\alpha|^2 \ket{0} \bra{0}_{q_2} \textcolor{red}{U_{q_2'}} \ket{\chi}\bra{\chi}_{q_2'} \textcolor{red}{U_{q_2'}^{\dagger}} \textcolor{red}{ \pm } \, \alpha \beta^* \ket{0}\bra{1}_{q_2} \textcolor{red}{U_{q_2'}} \ket{\chi}\bra{\chi}_{q_2'} \\
    &\textcolor{red}{ \pm } \, \beta \alpha^* \ket{1}\bra{0}_{q_2} \ket{\chi} \bra{\chi}_{q_2'} \textcolor{red}{U_{q_2'}^{\dagger}} + 
    |\beta|^2 \ket{1} \bra{1}_{q_2} \ket{\chi}\bra{\chi}_{q_2'} ,
\end{split}
\label{eq:rho4psi_plus_minus_output_from_remote_CU}
\end{equation}
respectively.

Moreover, the ideal input state, $\ket{\omega}$, from before now has the state%
\begin{equation}
\label{eq:seperable_gamma}
    \ket{\omega} = (\alpha \ket{0} + \beta \ket{1})_{q_2} (a \ket{0} + b\ket{1})_{q_2'},
\end{equation}%
where $a$ and $b$ are complex numbers such that $|a|^2 + |b|^2 = 1$.

Inserting Eq. \eqref{eq:seperable_gamma} and Eq. into \eqref{eq:pure_state_fidelity_def} gives 
\begin{equation}
\begin{split}
    F_{\rm out} =& \sum_{i = 1}^4 p_i \bra{\omega}_{q_2, q_2'} (\rho_i)_{q_2, q_2'} \ket{\omega}_{q_2, q_2'} \\
    =& \sum_{i = 1}^4 p_i ( |\alpha|^2 \bra{\chi}_{q_2'}  \bra{0}_{q_2} (\rho_i)_{q_2, q_2'} \ket{0}_{q_2} \ket{\chi}_{q_2'} \\
    &+ \alpha^* \beta \bra{\chi}_{q_2'}  \bra{0}_{q_2}  (\rho_i)_{q_2, q_2'} \ket{1}_{q_2} U_{q_2'} \ket{\chi}_{q_2'} \\
    &+ \beta^* \alpha \bra{\chi}_{q_2'} U_{q_2'}^{\dagger} \bra{1}_{q_2} (\rho_i)_{q_2, q_2'} \ket{0}_{q_2} \ket{\chi}_{q_2'} \\
    &+ |\beta|^2 \bra{\chi}_{q_2'} U_{q_2'}^{\dagger} \bra{1}_{q_2} (\rho_i)_{q_2, q_2'} \ket{1}_{q_2} \ket{\chi}_{q_2'})
\end{split}
\label{eq:cat_comm_fidelity_in_terms_of_DMs}
\end{equation}
with $\rho_i$ being the density matrix for one of the states in Eqs. \eqref{eq:rho4phi_plus_minus_output_from_remote_CU} to \eqref{eq:rho4psi_plus_minus_output_from_remote_CU}.

Inserting \eqref{eq:rho4phi_plus_minus_output_from_remote_CU} and \eqref{eq:rho4psi_plus_minus_output_from_remote_CU} into \eqref{eq:cat_comm_fidelity_in_terms_of_DMs} for each $i$, and using the orthonormality of $\{\ket{0}, \ket{1}\}$ to cancel terms yields%
\begin{equation}
    \begin{split}
         & p_{1, 2} \bra{\omega}_{q_2, q_2'} (\sigma_{1, 2})_{q_2, q_2'} \ket{\omega}_{q_2, q_2'} \\
         =& \, \left( |\alpha|^4 \left| \langle \chi |  \chi\rangle_{q_2'} \right|^2 \textcolor{red}{\pm} \, |\alpha|^2 |\beta|^2 \langle \chi |  \chi\rangle_{q_2'} \bra{\chi} \textcolor{red}{U_{q_2'}^{\dagger}} U_{q_2'}\ket{\chi}_{q_2'} \right.\\
         &\textcolor{red}{\pm} \,  |\alpha|^2 |\beta|^2 \bra{\chi} \textcolor{red}{U_{q_2'}^{\dagger}} U_{q_2'}\ket{\chi}_{q_2'} \langle \chi |  \chi\rangle_{q_2'} \\
         & \left. +  |\beta|^4 \left|\bra{\chi}_{q_2'} \textcolor{red}{U_{q_2'}^{\dagger}} U_{q_2'} \ket{\chi}_{q_2'} \right|^2\right) p_{1, 2}, 
    \end{split}
\end{equation}%
for $i=1$ and $i=2$, where $p_{1, 2} = p_1$ or $p_2$.

Using $UU^+ = U^+U = \1$ and $\langle \chi| \chi \rangle = 1$,  then this simplifies to %
\begin{equation}
    \begin{split}
         &\left( |\alpha|^4 \, \textcolor{red}{\pm} \, 2|\alpha|^2 |\beta|^2 + |\beta|^4 \right) p_{1, 2} \\
         &= \left( |\alpha|^2 \, \textcolor{red}{\pm} \, |\beta|^2 \right)^2 p_{1, 2}.
    \end{split} 
\end{equation}%
For $i=1$ ($\textcolor{red}{+}$ case), the expression further simplifies to $p_1 = F_{\rm w}$ because $|\alpha|^2 + |\beta|^2 = 1$, by the normalisation condition of quantum states.

Proceeding in a similar vein, for $i=3$ and $i=4$:%
\begin{equation}
    \begin{split}
         & p_{3, 4} \bra{\omega}_{q_2, q_2'} (\sigma_{3, 4})_{q_2, q_2'} \ket{\omega}_{q_2, q_2'} \\
         =& \, |\alpha|^4 \left| \bra{\chi} \textcolor{red}{U_{q_2'}^{\dagger}}\ket{\chi}_{q_2'} \right|^2 \textcolor{red}{\pm} \, |\alpha|^2 |\beta|^2 \bra{\chi} U_{q_2'}\ket{\chi}_{q_2'} \\
         &\textcolor{red}{\pm} \, |\alpha|^2 |\beta|^2 \left(\bra{\chi}_{q_2'} \textcolor{red}{U_{q_2'}^{\dagger}} \ket{\chi}_{q_2'} \right)^2  \\
         & +  |\beta|^4 \left(\bra{\chi}_{q_2'} \textcolor{red}{U_{q_2'}^{\dagger}} \ket{\chi}_{q_2'}\right)^2. 
    \end{split}
\end{equation}

Inserting these terms back into Eq. \eqref{eq:cat_comm_fidelity_in_terms_of_DMs}, we find that%
\begin{equation}
\label{eq:cat_analtyical_F_out4CNOT_appendix_version}
\begin{split}
    F_{\rm out} = F_{\rm w} + \frac{1 - F_{\rm w}}{3} (& \left( |\alpha|^2 - |\beta|^2 \right)^2 + 2 |\alpha|^4 \left|\bra{\chi} U \ket{\chi}_{q_2'}\right|^2 \\
    &+ 2 |\beta|^4 \left( \bra{\chi} U^{\dagger} \ket{\chi}_{q_2'} \right)^2),
\end{split}
\end{equation}%
which is equivalent to Eq. \eqref{eq:cat_analtyical_F_out4CNOT} in the main text. Unlike the expression for 1TP, Eq. \eqref{eq:cat_analtyical_F_out4CNOT_appendix_version} depends on the exact local operations conducted on QPU B's qubits and the input state of QPU A and QPU B's processing qubits.

\section{Variation with respect to input state}
\label{app:variation_input_state}

Throughout Sec. \ref{subsec:single_remote_gate_results} in the main text, we assume that the input state to the remote CNOT gate considered is given by Eq. \eqref{eq:remote_CNOT_input_state}. Here, we demonstrate that the output error after a remote CNOT gate depends on its input, discuss the implications of our choice of input state and indicate why cat-comm and 1TP differ.

For 1TP, it is clear from Eq. \eqref{eq:1TP_analytical_F_out} that the output fidelity, and therefore output error, is unaffected by the input to the remote gate when only entanglement error is considered. However, the same cannot be said for cat-comm, 2TP, and TP-safe. Moreover, for cat-comm, some input state dependence can be seen in the analytical expression for output state fidelity given by Eq. \eqref{eq:cat_analytical_F_out}. To fully understand the role of input state would require a non-trivial averaging over all possible quantum states for the control and target qubits, which correspond to $q_2$ and $q_2'$ respectively in Fig. \ref{fig:remote_cnot_implemented_in_different_ways}. Such an averaging would most likely require a detailed analytical description of the problem and multivariate integration or Monte Carlo analysis. This is beyond the scope of the current work. 

Instead, for convenience, we constrain the problem by assuming that the input state has the form:%
\begin{equation}
\label{eq:remote_CNOT_arbitrary_input_state}
\ket{\mathrm{input}}_{q_2, \, q_2'} = (\alpha \ket{0}_{q_2} + \beta \ket{1}_{q_2}) \ket{0}_{q_2'},
\end{equation}%
where $q_2$ and $q_2'$ are the processing qubits depicted in Fig. \ref{fig:remote_cnot_implemented_in_different_ways}; $\ket{\mathrm{input}}_{q_2, \, q_2'}$ is the input state of qubits $q_2$ and $q_2'$, prior to the remote CNOT gate taking place; and $\alpha$ and $\beta$ are complex numbers such that $|\alpha|^2 + |\beta|^2 = 1$. 

To identify the input state, with the form of Eq. \eqref{eq:remote_CNOT_arbitrary_input_state}, that gives the lowest output fidelity, in Fig. %
\begin{figure*}
    \centering
    \hspace*{-3em}
    \begin{overpic}[scale=0.4]{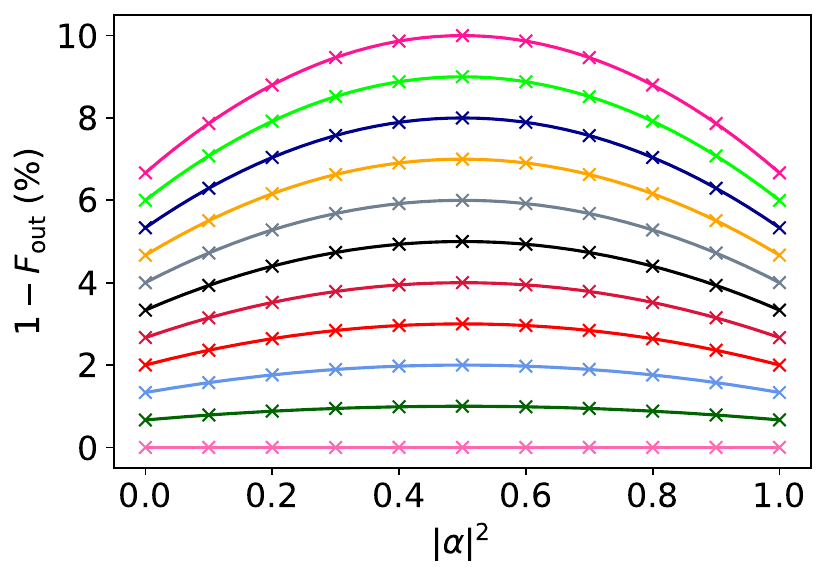}
        \put(-4, 60){(a)}
    \end{overpic}\hspace{1em}%
    \begin{overpic}[scale=0.4]{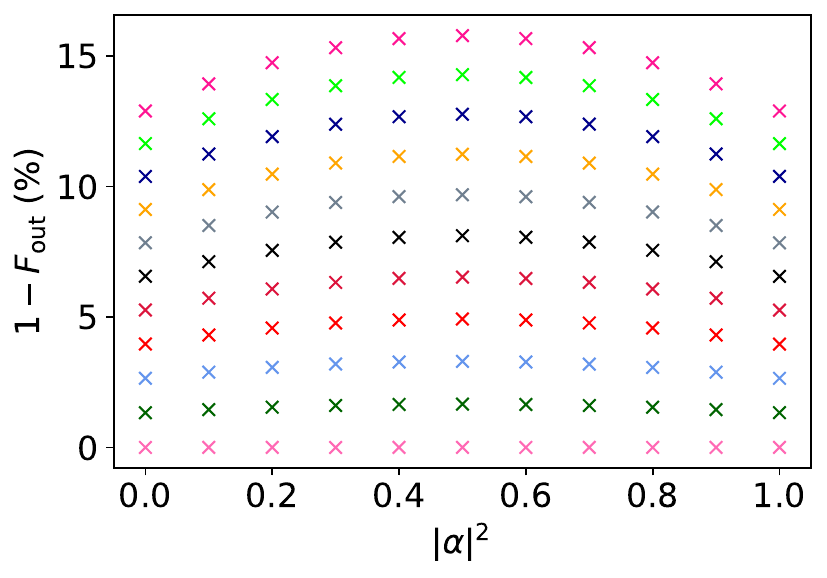}
    \put(-4, 60){(b)}
\end{overpic}\hspace{1em}%
 \subfloat{%
  \raisebox{1.6\height}{
\begin{overpic}[scale=0.4]{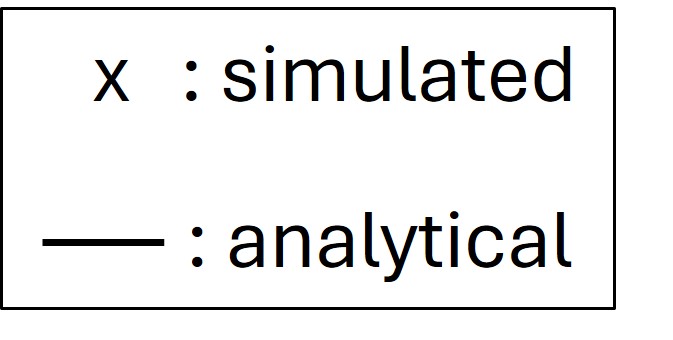}
\end{overpic}}}
\hspace{1em}
    \subfloat{%
    \raisebox{0.1\height}{
    \begin{overpic}[scale=0.2, trim={10cm, 1.0cm, 10cm, 1.5cm}, clip]{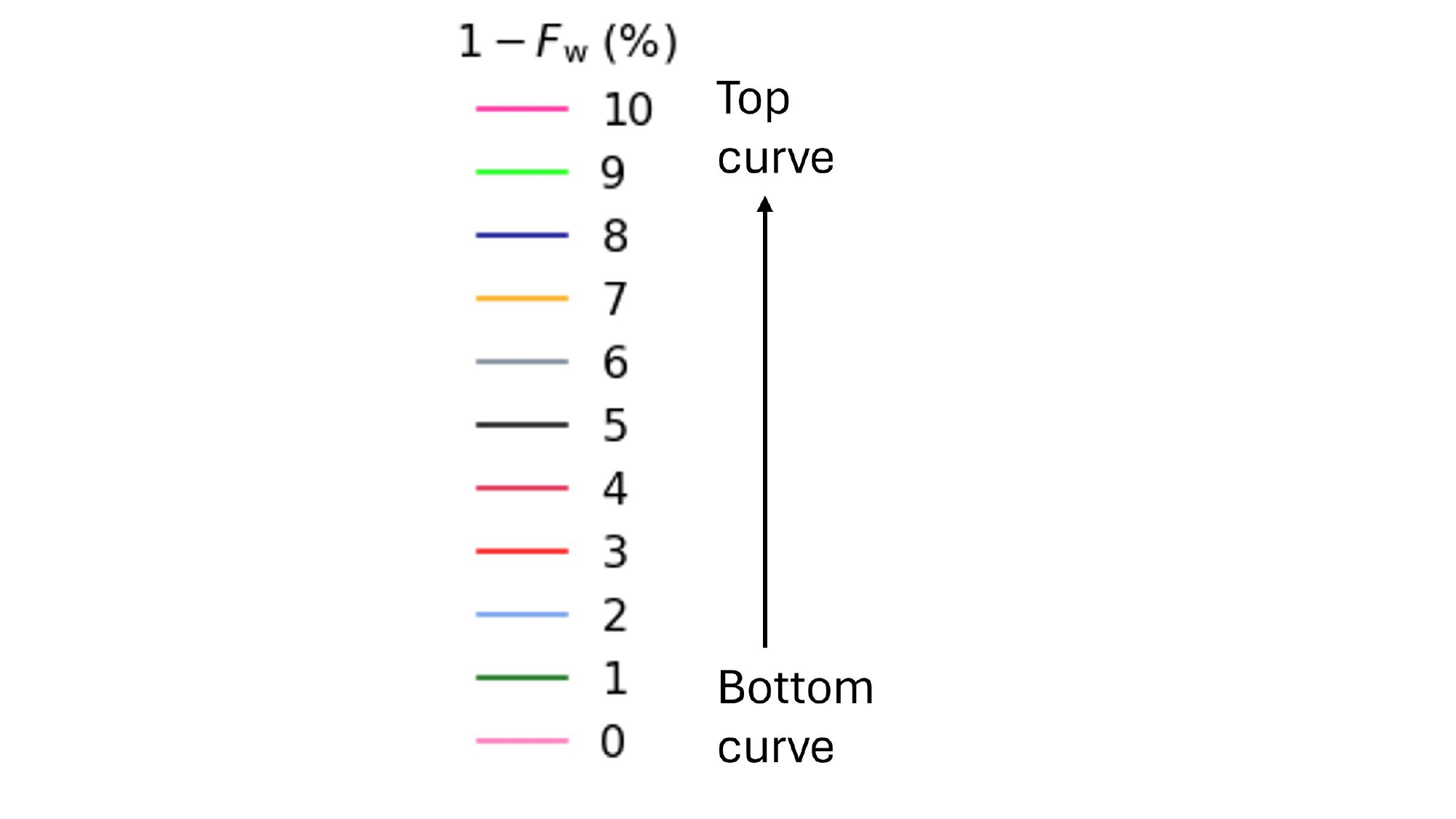}
    \end{overpic}}}
    \caption{Output fidelity as a function of input state for a remote CNOT implemented using: (a) cat-comm, (b) 2TP. The input state of the control qubit for the remote CNOT gate is changed by varying $|\alpha|^2$, where $\alpha$ is the coefficient from Eq. \eqref{eq:remote_CNOT_arbitrary_input_state}. Markers represent simulated data and solid lines represent analytical data created using Eq. \eqref{eq:cat_analtyical_F_out4CNOT}.}
    \label{fig:output_error_vs_mod_alpha_squared}
\end{figure*}\ref{fig:output_error_vs_mod_alpha_squared}, we show the output error as a function of the input state of $q_2$. As noted in the main text, 2TP and TP-safe give identical results when only entanglement error is considered, and so only cat-comm and 2TP are considered in Fig. 
\ref{fig:output_error_vs_mod_alpha_squared}.

From Fig. 
\ref{fig:output_error_vs_mod_alpha_squared}, it is apparent that there is a clear maximum in the output error when $|\alpha|^2 = \frac{1}{2}$. This means that the magnitude squared of the coefficients, $\alpha = \beta = \frac{1}{\sqrt{2}}$, in Eq. \eqref{eq:remote_CNOT_input_state} correspond to the highest output error of any input state of the form given by Eq. \eqref{eq:remote_CNOT_arbitrary_input_state}.

\begin{figure}
    \centering
    \includegraphics[scale=0.4]{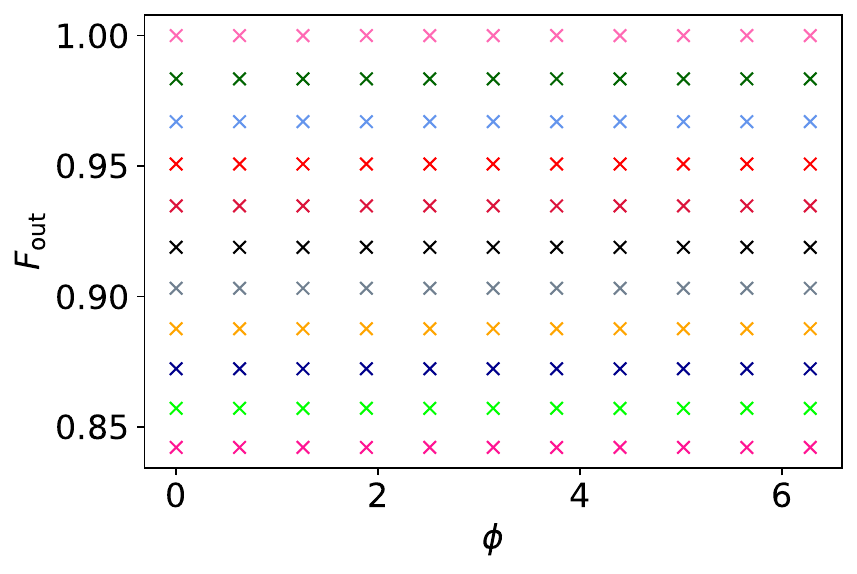}%
    \subfloat{%
    \raisebox{0.1\height}{
    \begin{overpic}[scale=0.2, trim={10cm, 1.0cm, 10cm, 1.5cm}, clip]{srg_input_state_state_of_art_range_legend.pdf}
    \end{overpic}}}
    \caption{Simulated output fidelity as a function of the relative phase, $\phi$, in the input state for a remote CNOT implemented using 2TP. $\phi$ is defined as in Eq. \eqref{eq:remote_CNOT_arbitrary_phase_input}. $\alpha = \frac{1}{\sqrt{2}}$, where $\alpha$ is the coefficient from Eq. \eqref{eq:remote_CNOT_arbitrary_phase_input}.}
    \label{fig:output_error_wrt_phase}
\end{figure}

The output error is independent of the relative phase between $\alpha$ and $\beta$. For cat-comm, the phase independence of the output error can be seen from Eq. \ref{eq:cat_analytical_F_out}, which has no terms that depend on the relative phases of $\alpha$ and $\beta$. For 2TP, we can see the phase independence of the output error by assuming $\beta$ is real and re-writing Eq. \eqref{eq:remote_CNOT_arbitrary_input_state} as: %
\begin{equation}
\label{eq:remote_CNOT_arbitrary_phase_input}
\ket{\mathrm{input}}_{q_2, \, q_2'} = (\alpha \ket{0}_{q_2} + e^{i\phi} \sqrt{1 - |\alpha|^2} \ket{1}_{q_2}) \ket{0}_{q_2'},
\end{equation}
where $0\leq \phi < 2\pi$ is a real number. We plot the simulated output error as a function of $\phi$ in Fig. \ref{fig:output_error_wrt_phase}.%
The flatness of the curves in Fig. \ref{fig:output_error_wrt_phase}, indicates that the results are independent of the phase, $\phi$. 

The independence of the output error to phase terms in the input, indicates that any input state with $|\alpha|^2 = \frac{1}{2}$ can yield the maximum output error for input states with the form given by Eq. \eqref{eq:remote_CNOT_arbitrary_input_state}. We choose $\alpha = \beta = \frac{1}{\sqrt{2}}$ for convenience, recovering Eq. \eqref{eq:remote_CNOT_input_state}.

To get some insight into the ramifications of our choice of input state, we consider the slightly more general case where only entanglement error is varied and the input state of $q_2$ and $q_2'$ is separable. In such a scenario, the input state of $q_2$ and $q_2'$ is given by Eq. \eqref{eq:general_separable_input_state}, which can be re-written, up to a global phase, more explicitly as:
\begin{equation}
\label{eq:general_separable_2qubit_input}
\begin{split}
    \ket{\mathrm{input}}_{q_2, \, q_2'} =& (\alpha \ket{0}_{q_2} + e^{i\phi} \sqrt{1-|\alpha|^2} \ket{1}_{q_2}) \otimes \\
    &  (\gamma \ket{0}_{q_2'} + e^{i\theta} \sqrt{1-|\gamma|^2} \ket{1}_{q_2'}),
\end{split}
\end{equation}
where $\alpha$, $\gamma$ are real numbers with magnitudes $0\leq |\alpha| \leq 1$ and $0\leq |\gamma| \leq 1$, and $0 \leq \phi < 2\pi$ and $0 \leq \theta < 2\pi$ are real numbers.

\begin{figure*}
    \centering
    \begin{overpic}[scale=0.8]{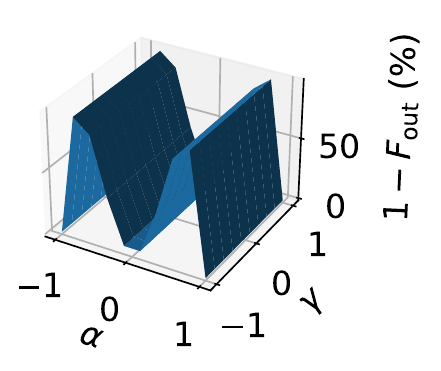}%
        \put(0, 70){(a)}    
    \end{overpic}
    \begin{overpic}[scale=0.8, trim={0, 0, 1.5cm, 0}, clip]{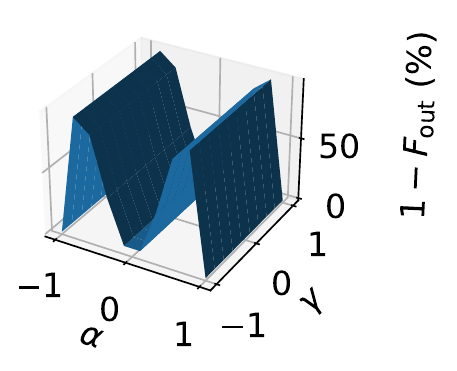}%
        \put(0, 75){(b)}    
    \end{overpic}
    \begin{overpic}[scale=0.8, trim={0, 0, 1.3cm, 0}, clip]{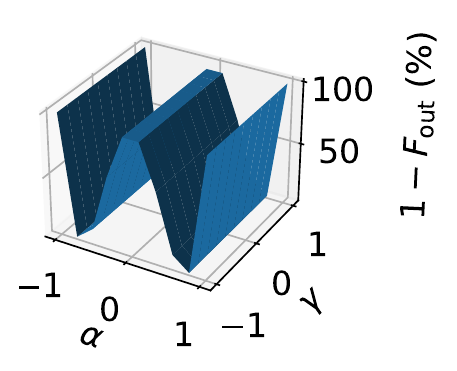}%
        \put(0, 75){(c)}    
    \end{overpic}
    \begin{overpic}[scale=0.8, trim={0, 0, 1.3cm, 0}, clip]{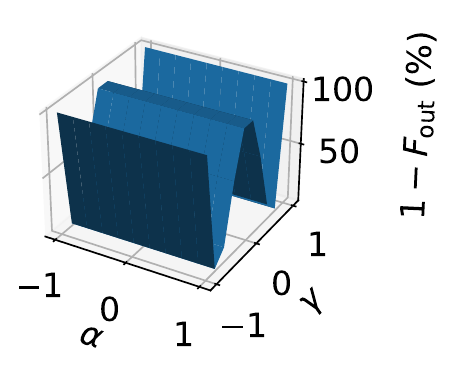}%
        \put(0, 75){(d)}    
    \end{overpic}
    \begin{overpic}[scale=0.8, trim={0, 0, 1.3cm, 0}, clip]{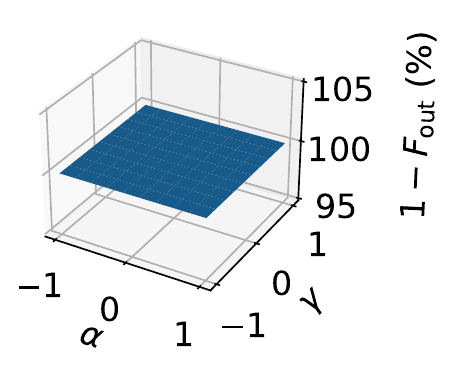}%
        \put(0, 75){(e)}    
    \end{overpic}
    \begin{overpic}[scale=0.8, trim={0, 0, 1.3cm, 0}, clip]{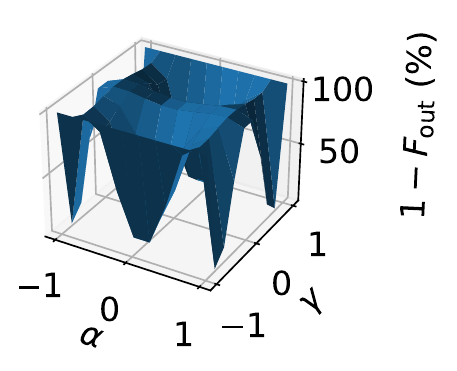}%
        \put(0, 75){(f)}    
    \end{overpic}
    \caption{The output error between the state outputted from a remote CNOT gate when: $\ket{\Phi^+}$ is distributed as the ebit and when: (a)-(b) $\ket{\Phi^-}$ is distributed as the ebit for 1TP and cat-comm, respectively; (c)-(d) $\ket{\Psi^+}$ is distributed as the ebit for 1TP and cat-comm, respectively; and when (e)-(f) $\ket{\Psi^-}$ is distributed as the ebit for 1TP and cat-comm, respectively. All other forms of error are set to zero. $\alpha$ and $\gamma$ are the coefficients from Eq. \eqref{eq:general_separable_2qubit_input}.}
    \label{fig:output_error_for_different_input_bell_states_wrt_alpha}
\end{figure*}

\begin{figure*}
    \centering
    \begin{overpic}[scale=0.8]{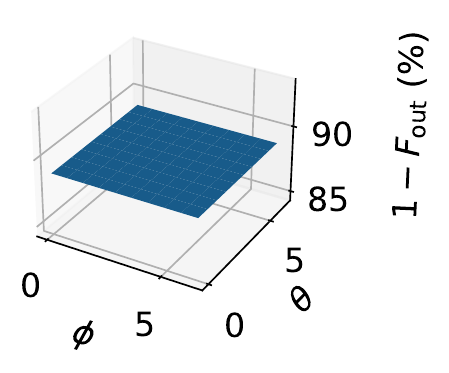}%
        \put(0, 70){(a)}    
    \end{overpic}
    \begin{overpic}[scale=0.8, trim={0, 0, 1.3cm, 0}, clip]{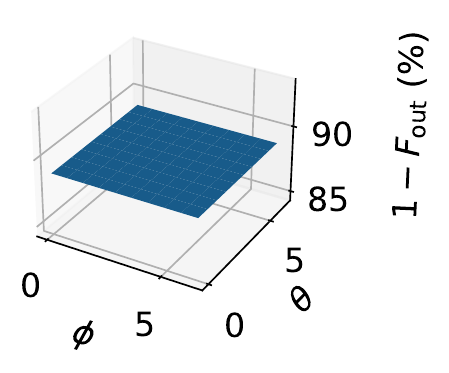}%
        \put(0, 75){(b)}    
    \end{overpic}
    \begin{overpic}[scale=0.8, trim={0, 0, 1.3cm, 0}, clip]{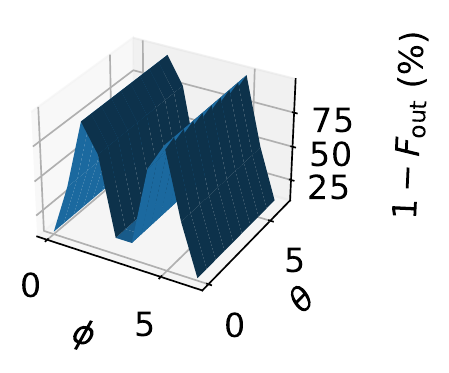}%
        \put(0, 75){(c)}    
    \end{overpic}
    \begin{overpic}[scale=0.8, trim={0, 0, 1.3cm, 0}, clip]{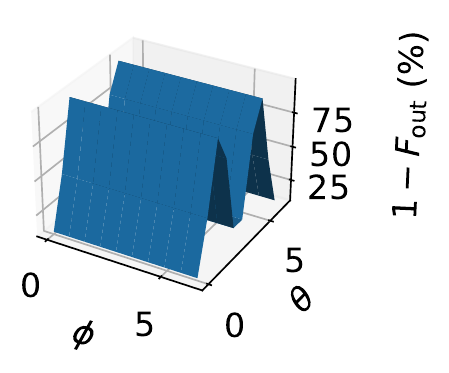}%
        \put(0, 75){(d)}    
    \end{overpic}
    \begin{overpic}[scale=0.8, trim={0, 0, 1.3cm, 0}, clip]{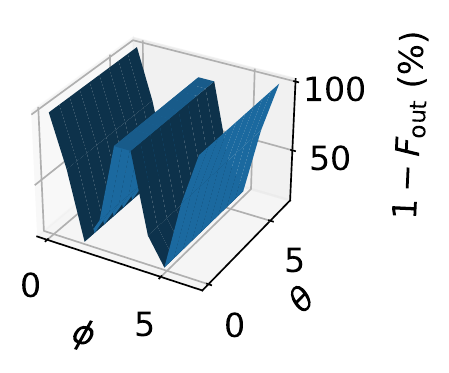}%
        \put(0, 75){(e)}    
    \end{overpic}
    \begin{overpic}[scale=0.8, trim={0, 0, 1.3cm, 0}, clip]{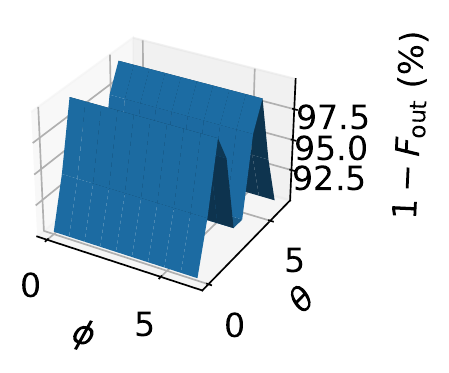}%
        \put(0, 75){(f)}    
    \end{overpic}
    \caption{The output error between the state outputted from a remote CNOT gate when $\ket{\Phi^+}$ is distributed as the ebit and when: (a)-(b) $\ket{\Phi^-}$ is distributed for 1TP and cat-comm, respectively; (c)-(d) $\ket{\Psi^+}$ is distributed for 1TP and cat-comm, respectively; and  (e)-(f) $\ket{\Psi^-}$ is distributed for 1TP and cat-comm, respectively. All other forms of error are set to zero. We fix $\alpha=\gamma=\frac{1}{\sqrt{3}}$ and vary $\phi$ and $\theta$, the relative phases of the input state for $q_2$ and $q_2'$, respectively, as defined in Eq. \eqref{eq:general_separable_2qubit_input}.}
    \label{fig:output_error_for_different_input_bell_states_wrt_phase}
\end{figure*}

We model the entanglement error by assuming that ebits are in the Werner state, see Eq. \eqref{eq:werner_state},  rather than $\ket{\Phi^+}$,  as discussed in Sec. \ref{subsec:error_models}. As the Werner state is just a mixture of the Bell states, we separately consider the output error when each of the Bell states is distributed as an ebit. Figures \ref{fig:output_error_for_different_input_bell_states_wrt_alpha}-\ref{fig:output_error_for_different_input_bell_states_wrt_phase} %
show surface plots of the output error between the state outputted from a remote CNOT gate when the ideal state, $\ket{\Phi^+}$, is distributed as an ebit and the state outputted when one of the other Bell states is erroneously distributed as an ebit for different values of $\alpha$ and $\gamma$, and $\phi$ and $\theta$, respectively. 

Several observations can be made from Figs. \ref{fig:output_error_for_different_input_bell_states_wrt_alpha}-\ref{fig:output_error_for_different_input_bell_states_wrt_phase}:%
\begin{enumerate}
    \item \label{obs:chosen_input_near_local_max} Eq. \eqref{eq:remote_CNOT_input_state} is at or near a local maximum in output error for cat-comm. Figures \ref{fig:output_error_for_different_input_bell_states_wrt_alpha}(b), \ref{fig:output_error_for_different_input_bell_states_wrt_alpha}(f), which show the output error with varying $\alpha$ and $\gamma$ for cat-comm, have maxima at approximately $\alpha = \frac{1}{\sqrt{2}} \approx 0.707$, while Fig. \ref{fig:output_error_for_different_input_bell_states_wrt_alpha}(d), which is also for cat-comm, is constant with respect to $\alpha$ but has maxima at $\gamma = \pm 1$, corresponding to the state $\ket{0}$. Therefore, although the choice of Eq. \eqref{eq:remote_CNOT_input_state} is somewhat arbitrary, it does correspond to an especially high entanglement error. That said, Eq. \eqref{eq:remote_CNOT_input_state} cannot be said to be a true upper bound on the entanglement error without knowing the global maximum.
    \item Cat-comm and 1TP differ due to discrepancies in the circuit output produced by the $\ket{\Psi^{\pm}}$ terms only. The $\ket{\Phi^{\pm}}$ terms yield identical outputs for cat-comm and 1TP. 
    \item Cat-comm is constant with respect to $\phi$ but varies with respect to $\theta$. This can be seen from Figs. \ref{fig:output_error_for_different_input_bell_states_wrt_phase}(b), \ref{fig:output_error_for_different_input_bell_states_wrt_phase}(d) and \ref{fig:output_error_for_different_input_bell_states_wrt_phase}(f). 
    \item In the presence of entanglement error only, 1TP gives a constant output error for any input state, but cat-comm does not. This could be seen from Eqs. \eqref{eq:1TP_analytical_F_out} and \eqref{eq:cat_analytical_F_out}, but the reason behind it is made clearer by Fig. \ref{fig:output_error_for_different_input_bell_states_wrt_alpha} and Fig. \ref{fig:output_error_for_different_input_bell_states_wrt_phase}. For 1TP, the different output errors resemble out of phase oscillations. When the output error caused by the distribution of one Bell state is low, it is high for a different Bell state. For example, Fig. \ref{fig:output_error_for_different_input_bell_states_wrt_alpha}(a) has maxima at the points where Fig. \ref{fig:output_error_for_different_input_bell_states_wrt_alpha} (c) has minima and slopes up at the $\alpha$ values where Fig. \ref{fig:output_error_for_different_input_bell_states_wrt_alpha} (c) slopes down. The only other term, shown in Fig. \ref{fig:output_error_for_different_input_bell_states_wrt_alpha} (e) is constant with respect to $\alpha$ and $\gamma$ and so does not disrupt this trend. Similar observations can be made about phase by comparing Fig. \ref{fig:output_error_for_different_input_bell_states_wrt_phase}(a), Fig. \ref{fig:output_error_for_different_input_bell_states_wrt_phase}(c), and Fig. \ref{fig:output_error_for_different_input_bell_states_wrt_phase}(e). By contrast, for cat-comm the output errors caused by the distribution of each Bell state are often rotated by $\frac{\pi}{2}$ radians relative to each other, as in  Figs. \ref{fig:output_error_for_different_input_bell_states_wrt_alpha}(b) and \ref{fig:output_error_for_different_input_bell_states_wrt_alpha}(d), and Figs. \ref{fig:output_error_for_different_input_bell_states_wrt_phase}(d) and \ref{fig:output_error_for_different_input_bell_states_wrt_phase}(f). This means that the different error terms do not balance out in the same way as they do for 1TP. This accounts for the difference between cat-comm and 1TP.
\end{enumerate}

\section{Verification that results from main text are robust to input state}

While observation \ref{obs:chosen_input_near_local_max} from Appendix \ref{app:variation_input_state} indicates that Eq. \eqref{eq:remote_CNOT_input_state} corresponds to an input state of particular interest, it is also clear from Appendix \ref{app:variation_input_state} that cat-comm, 2TP, and, by extension, TP-safe will vary with respect to input state. As such, it is important to understand if certain claims made in the main text are robust to input state variation. In particular, it is not clear from the figures shown in the main text alone that observation \ref{obs:first_order_upper_bound} from Sec. \ref{subsubsec:first_order_vs_sim} and any of the observations made in Sec. \ref{subsubsec:error_type_comparison} hold for arbitrary input states. Although a true proof is not possible without more detailed input state analysis, here we demonstrate that the claims from the main text hold for a variety of input states. In Appendix \ref{app:first_order_approx_robustness}, we demonstrate that observation \ref{obs:first_order_upper_bound} from Sec. \ref{subsubsec:first_order_vs_sim} is robust to variation in input state and make some further comments on the impact of input state variation on the relative difference between first-order and simulated output errors values. In Appendix \ref{app:error_type_comparison_robustness}, we verify the observations about the relative impact of different error types made in Sec. \ref{subsubsec:error_type_comparison} of the main text. Finally, in Sec. \ref{app:discussion_of_remote_gate_scheme_comparison}, we make some further comments on the observations made about the relative impact of different remote gate schemes in Sec.  \ref{subsubsec:remote_gate_comparison} of the main text.

\subsection{Verification of observations  from Sec. \ref{subsubsec:first_order_vs_sim}}
\label{app:first_order_approx_robustness}

\begin{figure}
    \centering
    \begin{overpic}[scale=0.39]{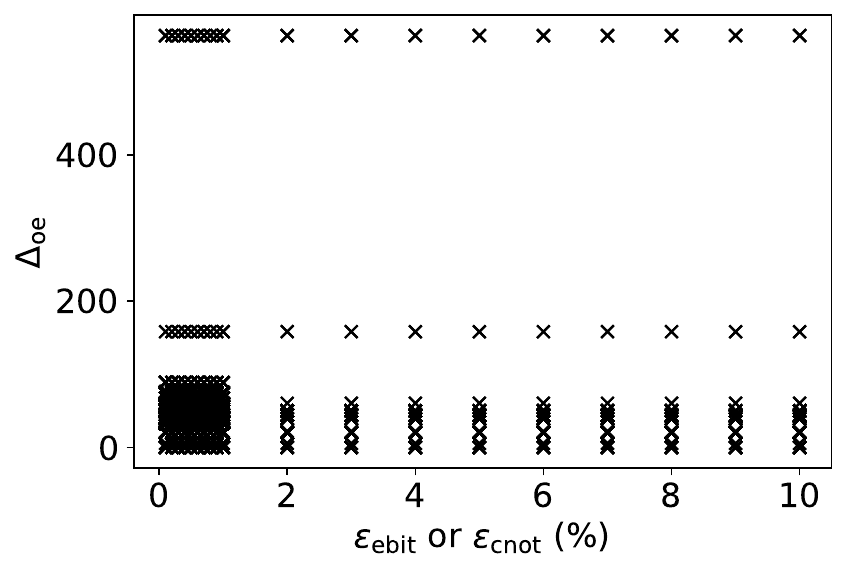}
        \put(0, 65){(a)}    
    \end{overpic}
    \begin{overpic}[scale=0.22, trim={0, 2cm, 0, 0}, clip]{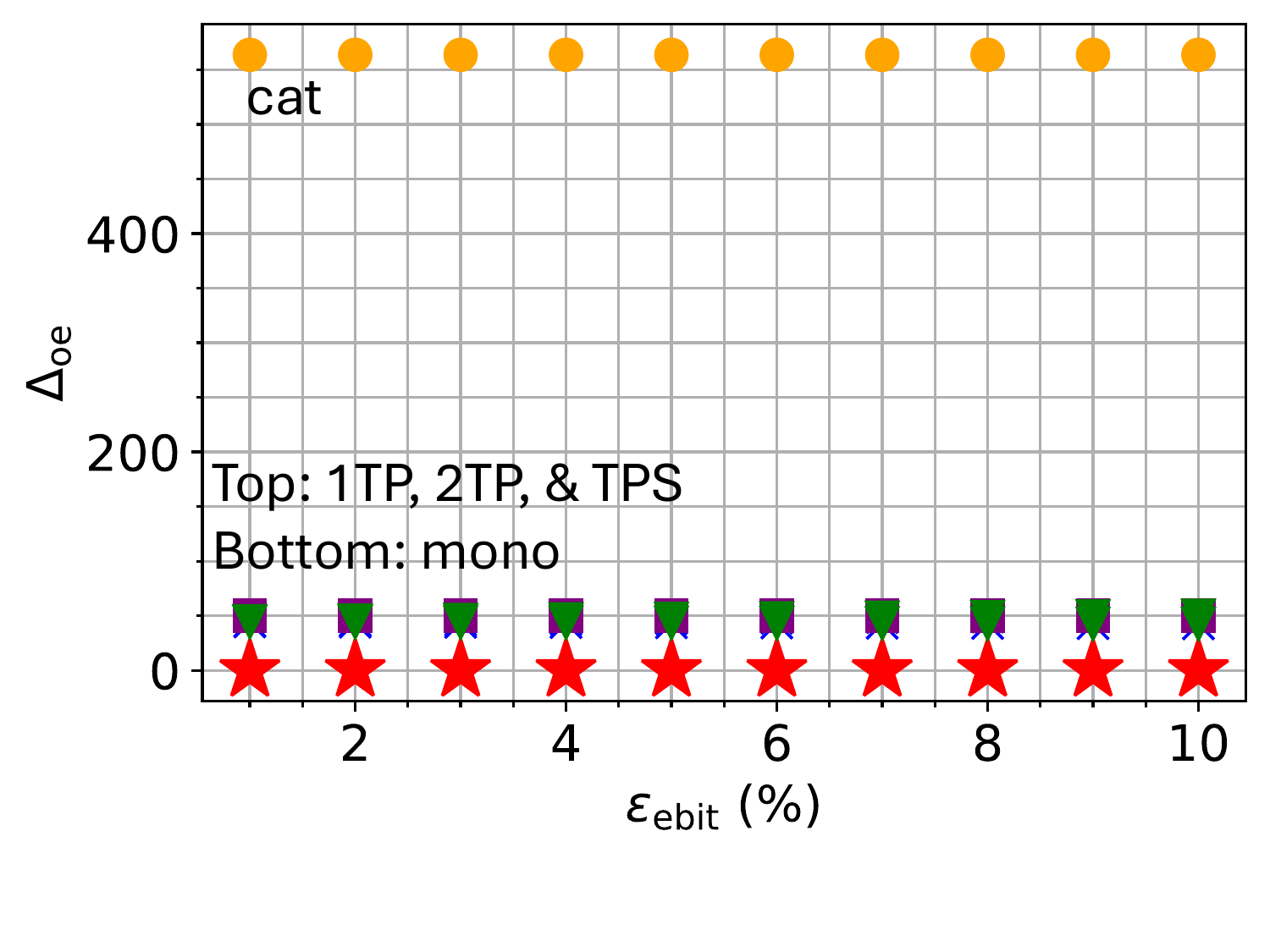}
        \put(0, 65){(b)}   
        \put(60, 22){\includegraphics[scale=0.5, trim={0.5cm, 0.2cm, 0.2cm, 0.4cm}, clip]{comparing_schemes_legend.pdf}}
    \end{overpic}\hspace{1.5em}%
    \begin{overpic}[scale=0.22, trim={0, 2cm, 0, 0}, clip]{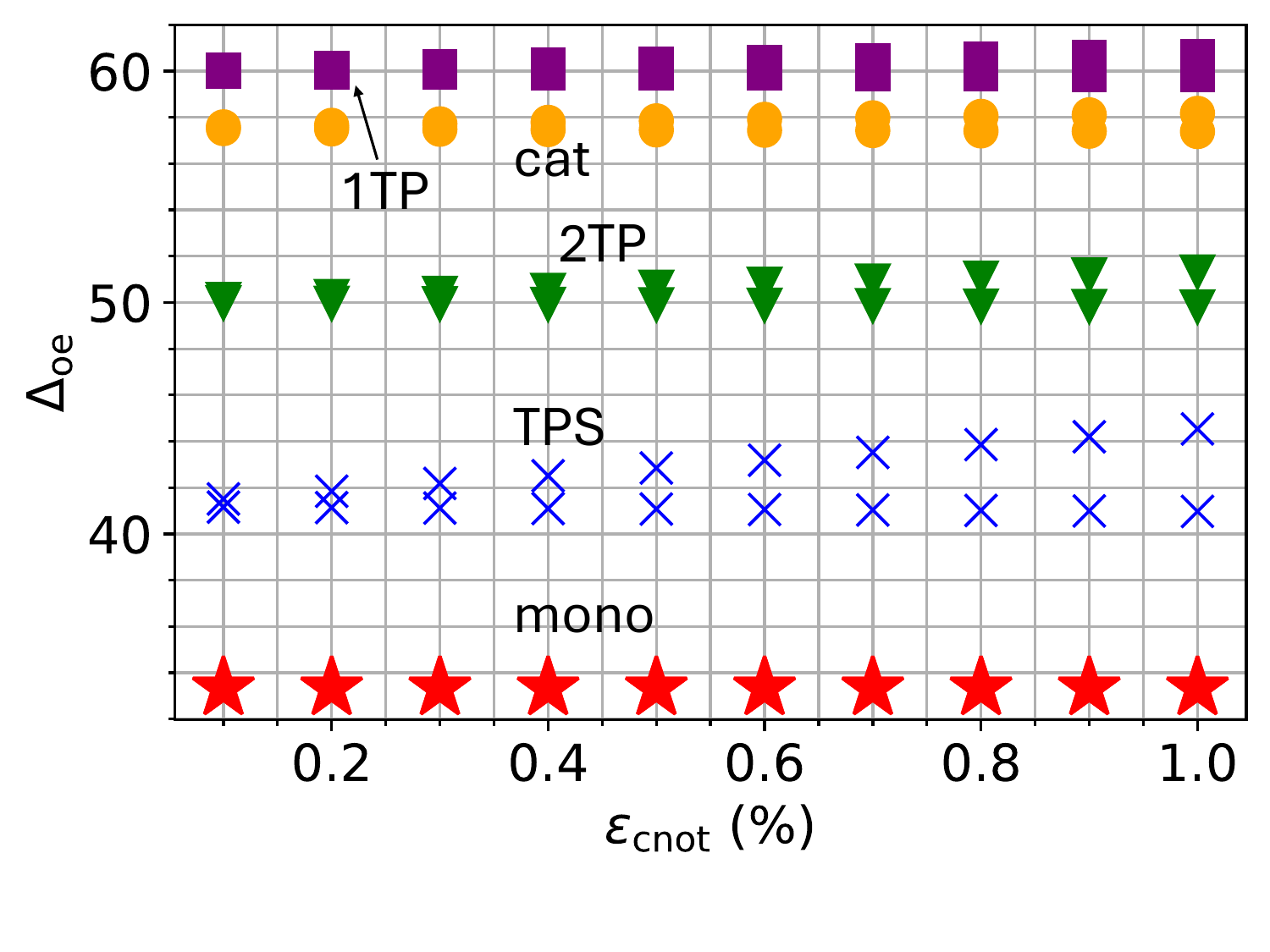}
        \put(0, 65){(c)}    
    \end{overpic}
    \caption{The percentage difference, $\Delta_{\mathrm{oe}}$, in the output error calculated using the first-order approximations, given by Eqs. \eqref{eq:first_order_f_out} and \eqref{eq:exp_approx}, and the simulator, respectively. In (a), we show the collated results for a variety of input states when $\epsilon_{\mathrm{ebit}}$ and $\epsilon_{\mathrm{cnot}}$, respectively are varied with all other errors set to zero. The input states considered have the form given by Eq. \eqref{eq:general_separable_2qubit_input} and are produced from all permutations of $\alpha \in \{0.2, \frac{1}{\sqrt{2}}\}$, $\gamma = \{0.0, 0.2, ..., 1.0\}$, $\phi=0$, $\theta \in \{0, \frac{2\pi}{5},  ... \,, 2\pi\}$. Only Eq. \eqref{eq:exp_approx} and the simulated results are compared---Eq \eqref{eq:first_order_f_out} is not considered. In (b), the only non-zero error parameter is $\epsilon_{\mathrm{ebit}}$ and the input state parameters are: $\alpha=0.2$, $\gamma = 0.6$, $\phi=0$, $\theta=2\pi$. In (c), the only non-zero error parameter is $\epsilon_{\mathrm{cnot}}$ and the input state parameters are: $\alpha=\frac{1}{\sqrt{2}}$, $\gamma = 0.8$, $\phi=0$, $\theta=2\pi$. Whenever two curves appear with the same markers, the top curve uses \eqref{eq:first_order_f_out} for the first-order approximation and the bottom curve uses \eqref{eq:exp_approx}. In all cases, $\Delta_{\mathrm{oe}}$ is calculated using Eq. \eqref{eq:percentage_difference}.}
    \label{fig:percentage_diff_for_different_inputs}
\end{figure}

To verify observation \ref{obs:first_order_upper_bound} from Sec. \ref{subsubsec:first_order_vs_sim} and corroborate comments made about the input state dependence of observation \ref{obs:good_agreement_possible_btwn_approx_and_sim}, we consider the percentage difference, $\Delta_{\mathrm{oe}}$, calculated using Eq. \eqref{eq:percentage_difference}, between the output error computed using the first-order approximation and the simulator, respectively, for various different input states. This is shown in Fig. \ref{fig:percentage_diff_for_different_inputs}. %
Figure \ref{fig:percentage_diff_for_different_inputs}(a) shows the collated results for a variety of input states when $\epsilon_{\mathrm{ebit}}$ and $\epsilon_{\mathrm{cnot}}$, respectively, are varied with all other errors set to zero. The input states considered all have the form given by Eq. \eqref{eq:general_separable_2qubit_input} and parameter values produced using a permutation of the parameters $\alpha \in \{0.2, \frac{1}{\sqrt{2}}\}$, $\gamma = \{0.0, 0.2, ..., 1.0\}$, $\phi=0$, $\theta \in \{0, \frac{2\pi}{5},  ... \,, 2\pi\}$. We average over all of the input states that can be produced in this way. In Fig. \ref{fig:percentage_diff_for_different_inputs}(b), the input state $\ket{\mathrm{input}}_{q_2, \, q_2'} = (0.2 \ket{0}_{q_2} + 0.980 \ket{1}_{q_2}) \otimes
 (0.6 \ket{0}_{q_2'} + 0.8 \ket{1}_{q_2'})$ \footnotemark{} is considered in the presence of a varying non-zero $\epsilon_{\mathrm{ebit}}$ with all other errors set to zero. Similarly, in Fig. \ref{fig:percentage_diff_for_different_inputs} (c), $\ket{\mathrm{input}}_{q_2, \, q_2'} = \frac{1}{\sqrt{2}}( \ket{0}_{q_2} + \ket{1}_{q_2}) \otimes
 (0.8 \ket{0}_{q_2'} + 0.600 \ket{1}_{q_2'})$ \footnotemark[\value{footnote}]  is considered in the presence of a varying non-zero $\epsilon_{\mathrm{cnot}}$ with all other errors set to zero. \footnotetext[\value{footnote}]{All rounded coefficients are displayed here to three significant figures but floating point precision is used for the generation of data.}
 

Figure \ref{fig:percentage_diff_for_different_inputs}(a) corroborates observation \ref{obs:first_order_upper_bound} from Sec. \ref{subsubsec:first_order_vs_sim}, which states that the first order approxmations loosely upper bound the simulated output error. All percentage differences, $\Delta_{\mathrm{oe}}$, in Fig. \ref{fig:percentage_diff_for_different_inputs}(a) are non-negative and so for all of the input states considered, the approximate results are greater than or equal to the exact results. Figures \ref{fig:percentage_diff_for_different_inputs}(b)-(c) differ both qualitatively and quantitatively from Fig. \ref{fig:first_order_diff}(a) and Fig. \ref{fig:first_order_diff}(c), with a different ordering of the $\Delta_{\mathrm{oe}}$ values for the various remote gate schemes and different quantitative values of $\Delta_{\mathrm{oe}}$. This indicates that the agreement between the output error predicted by the first-order approximations and the simulation are highly input state dependent, as suggested in observation \ref{obs:good_agreement_possible_btwn_approx_and_sim} from Sec. \ref{subsubsec:first_order_vs_sim}.

\subsection{Verification of all observations from Sec. \ref{subsubsec:error_type_comparison}}
\label{app:error_type_comparison_robustness}

\begin{figure*}
    \centering
    \begin{overpic}[scale=0.25, trim={0, 0, 3cm, 0}, clip]{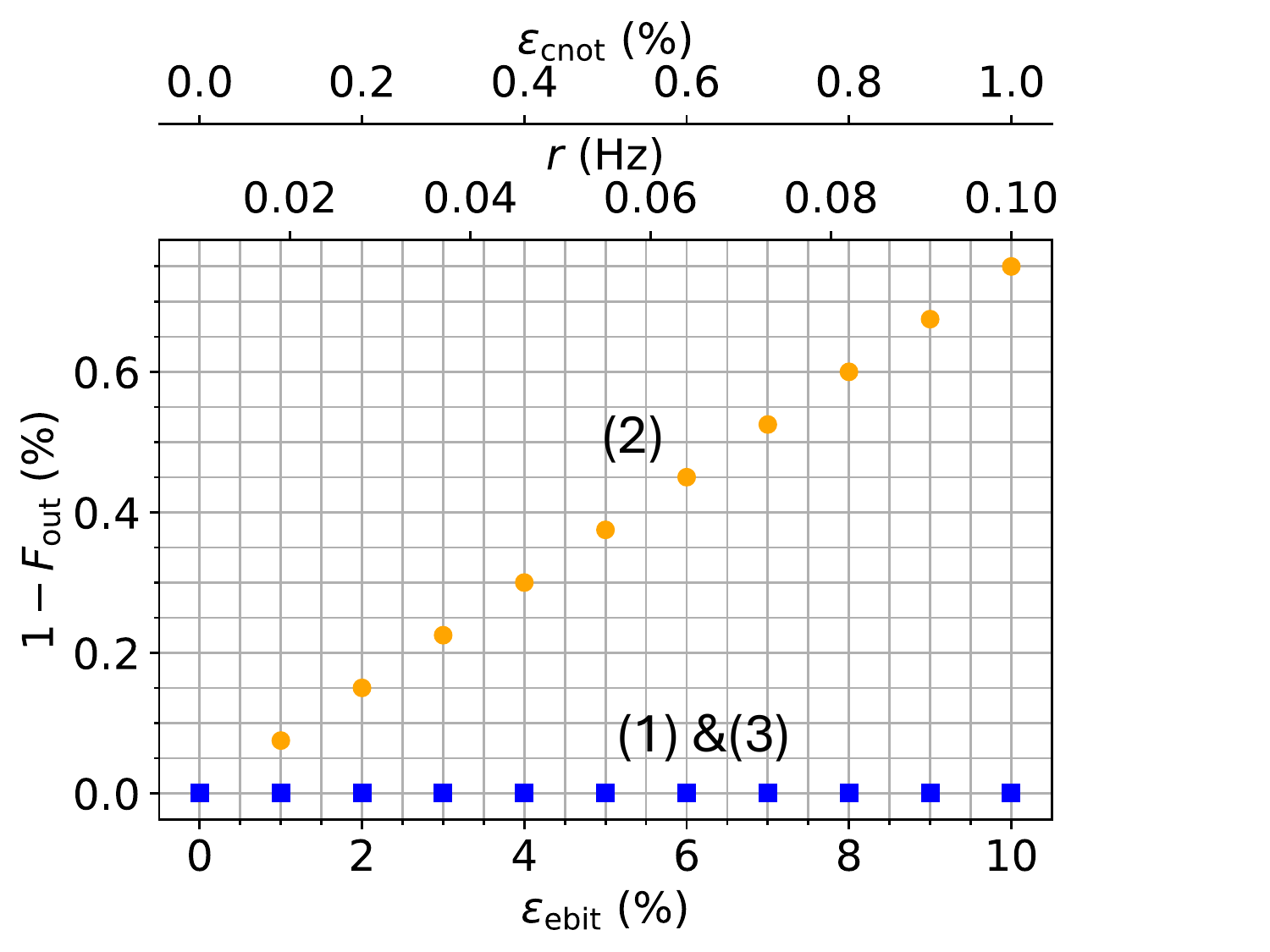}
        \put(20, 55){(a)}    
    \end{overpic}
    \begin{overpic}[scale=0.25, trim={0, 0, 3cm, 0}, clip]{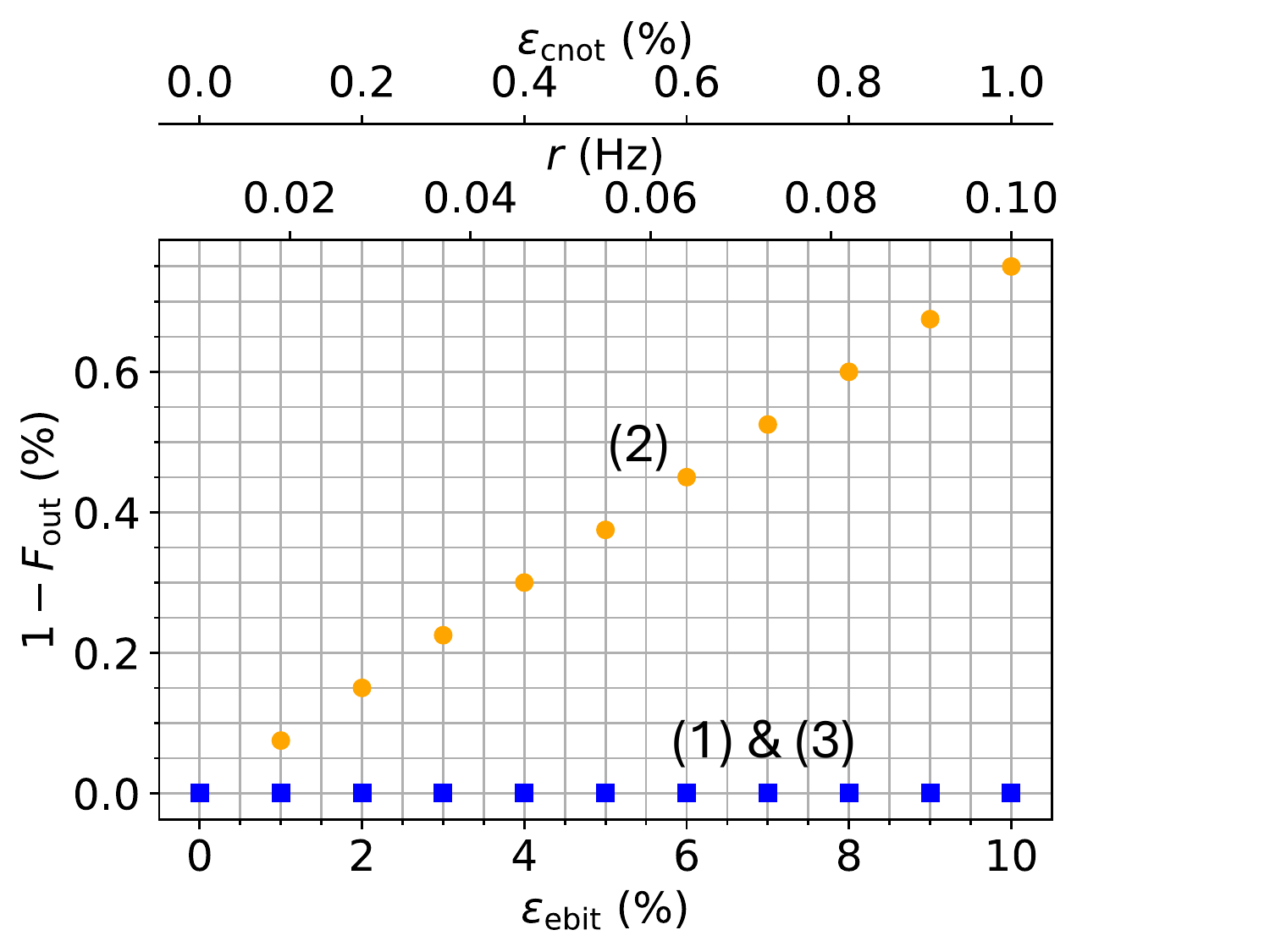}
        \put(20, 55){(b)}    
    \end{overpic}
    \begin{overpic}[scale=0.25, trim={0, 0, 3cm, 0}, clip]{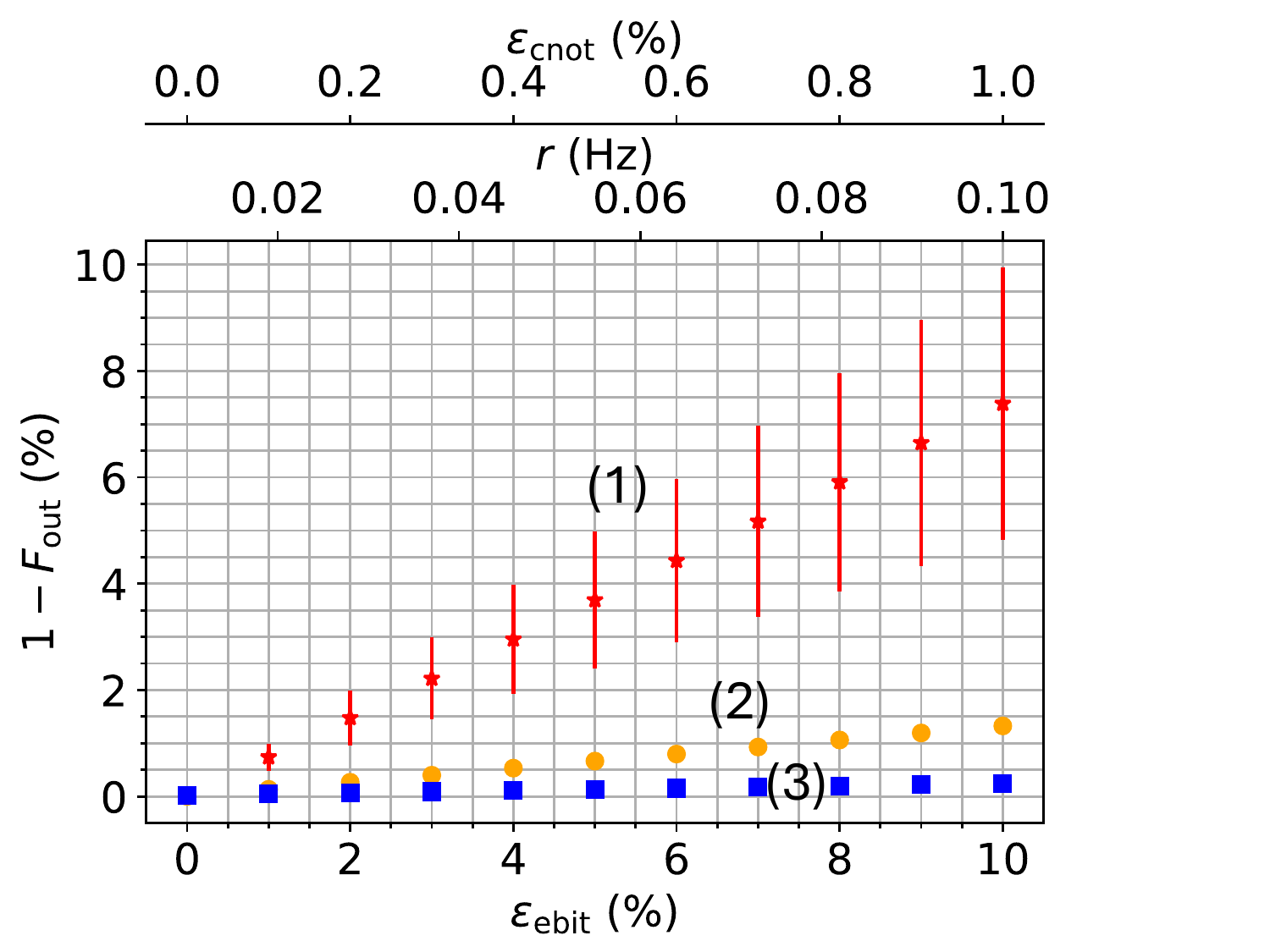}
    \put(20, 55){(c)}    
    \end{overpic}
    \begin{overpic}[scale=0.25, trim={0, 0, 3cm, 0}, clip]{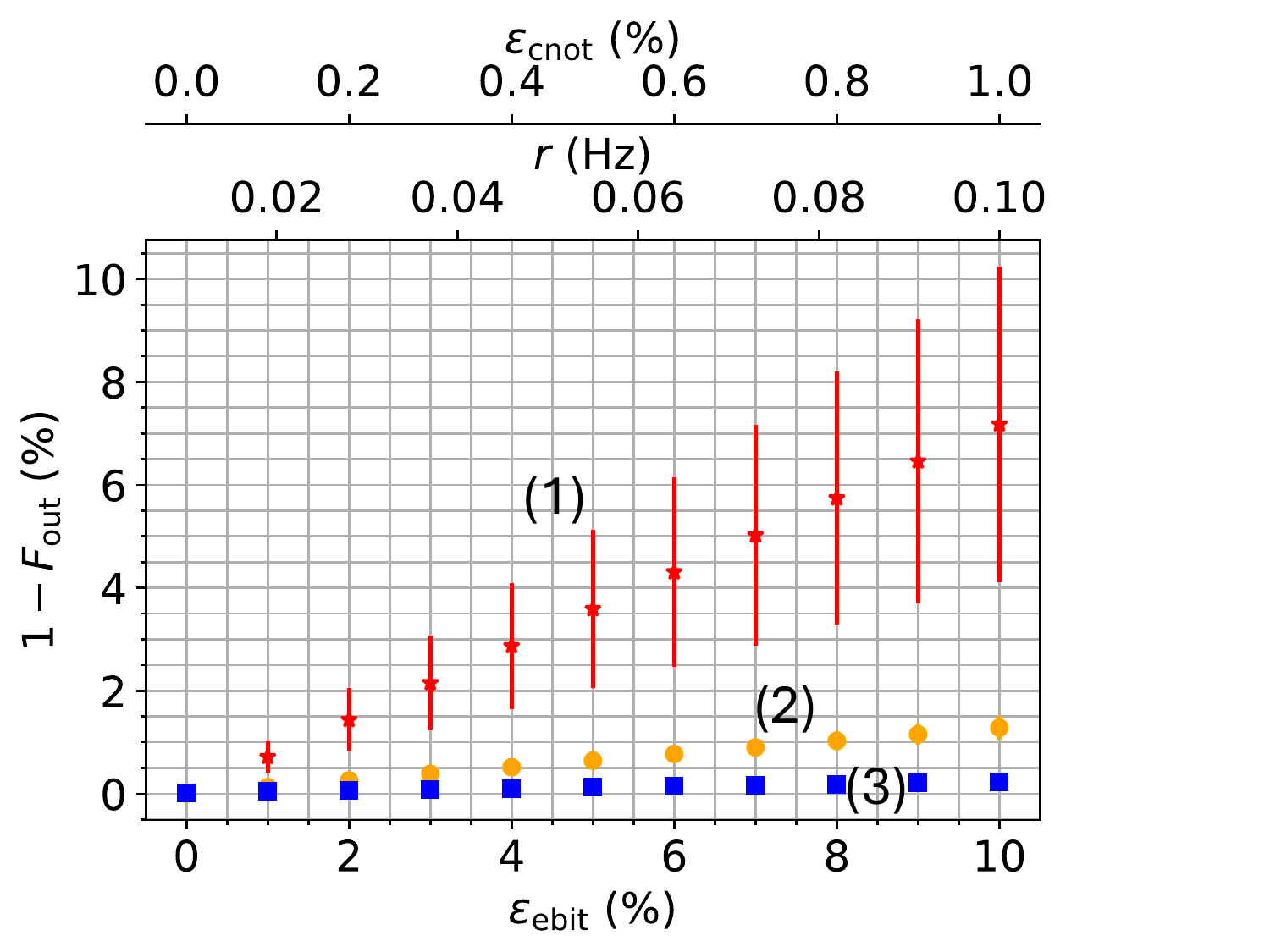}
    \put(20, 55){(d)}    
    \end{overpic}
    \begin{overpic}[scale=0.25, trim={0, 0, 3cm, 0}, clip]{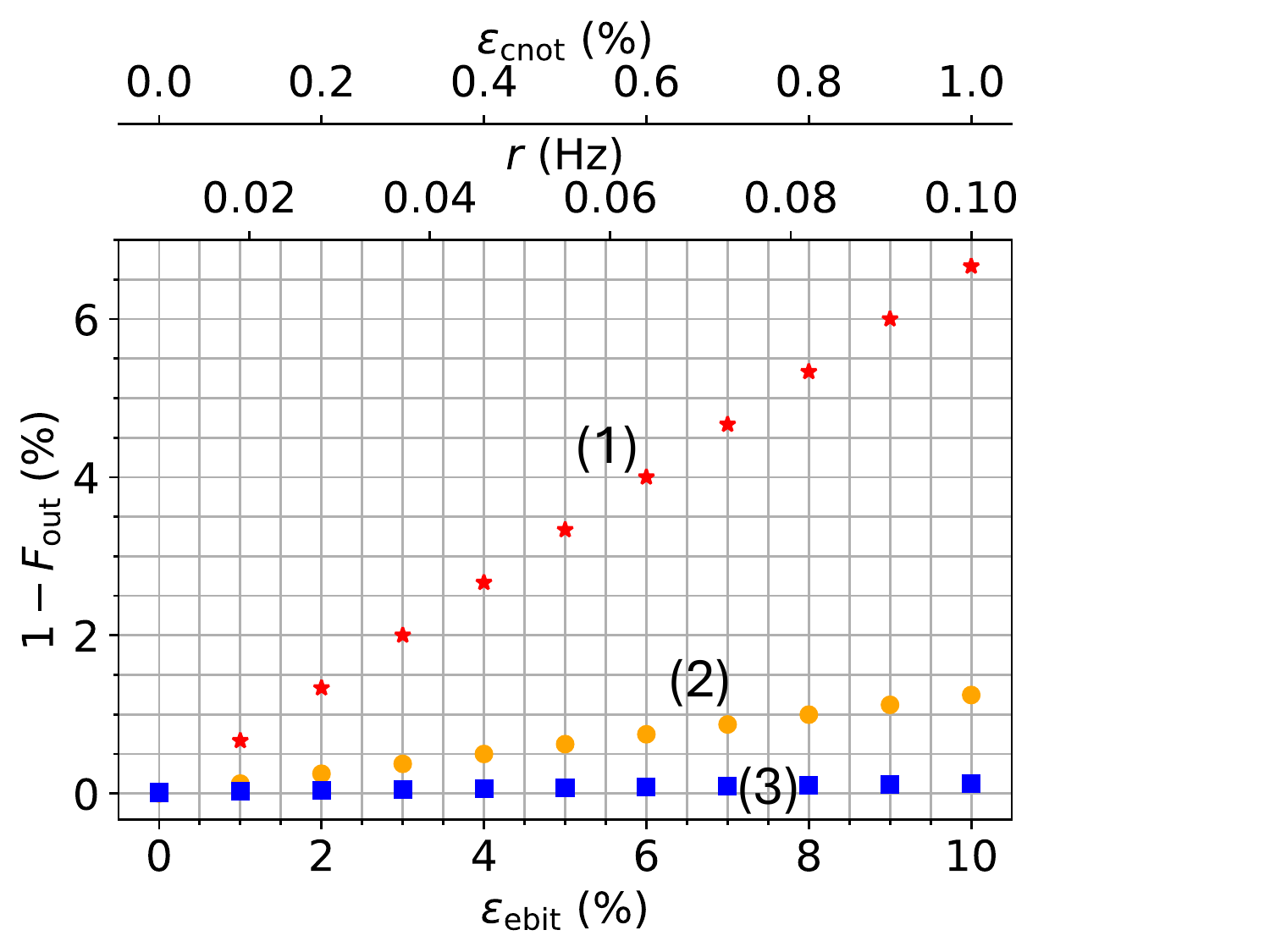}
    \put(20, 55){(e)}    
    \end{overpic}
    \begin{overpic}[scale=0.25, trim={0, 0, 3cm, 0}, clip]{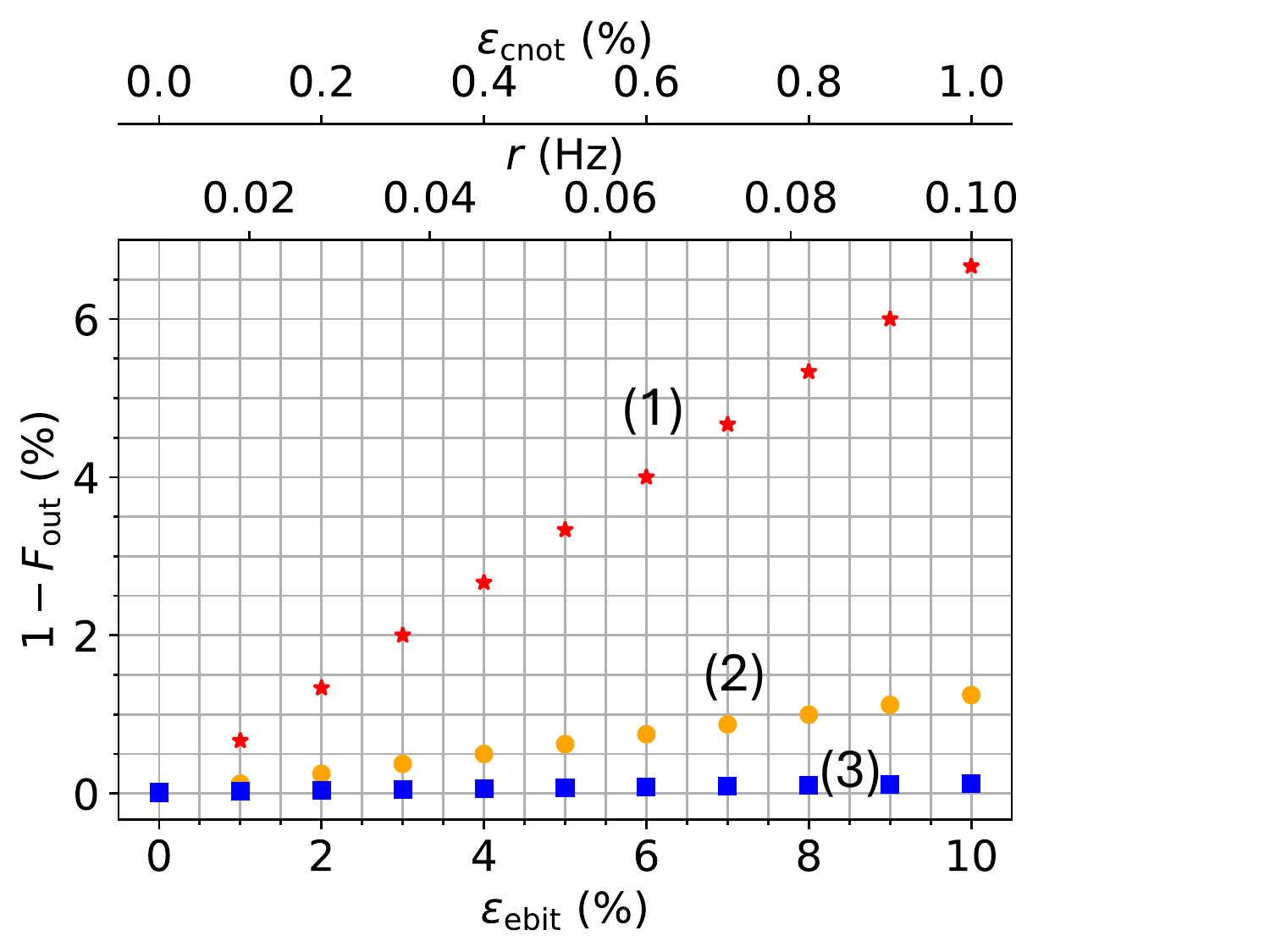}
    \put(20, 55){(f)}    
    \end{overpic}
    \begin{overpic}[scale=0.25, trim={0, 0, 3cm, 0}, clip]{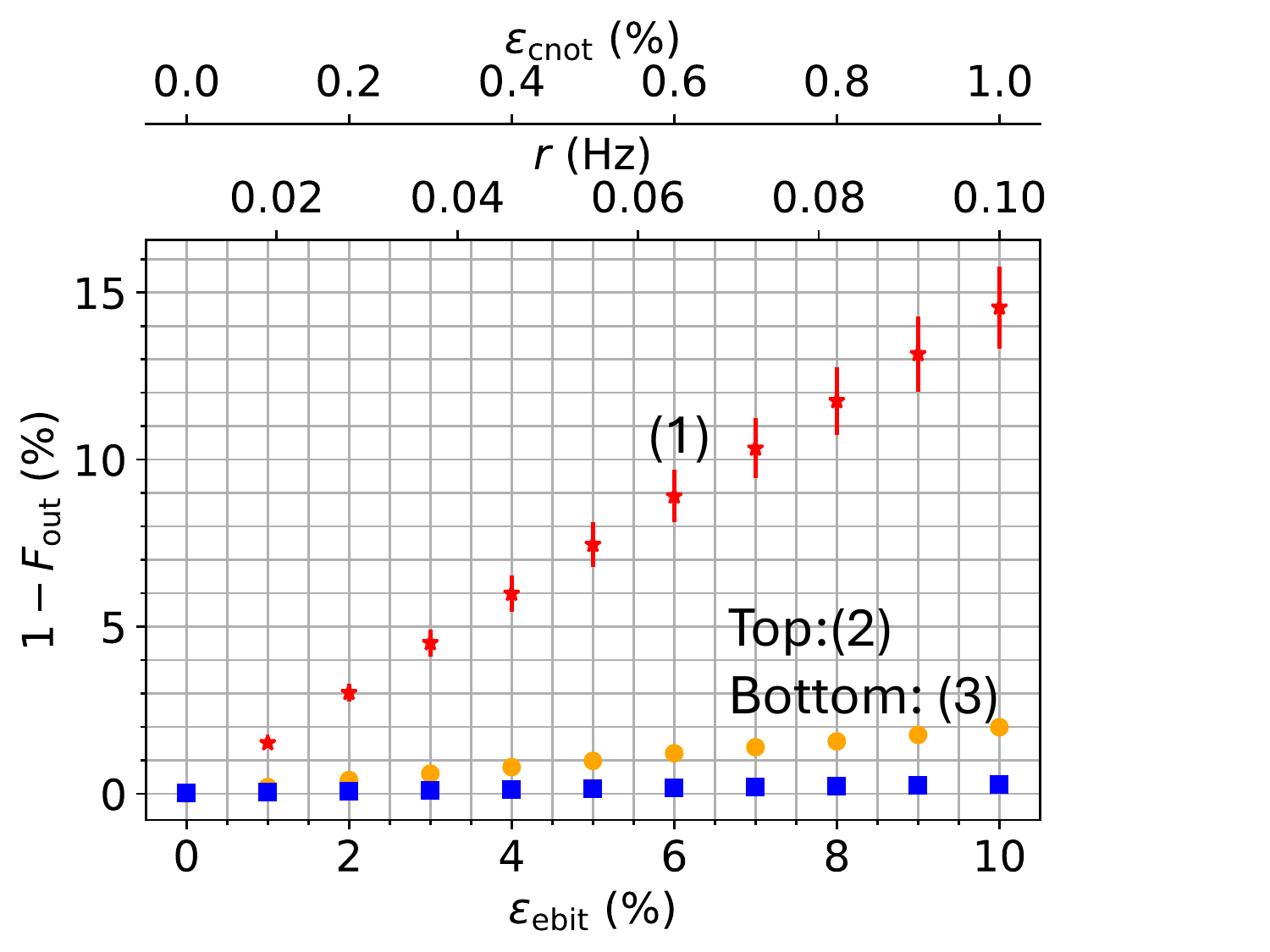}
    \put(20, 55){(g)}    
    \end{overpic}
    \begin{overpic}[scale=0.25, trim={0, 0, 3cm, 0}, clip]{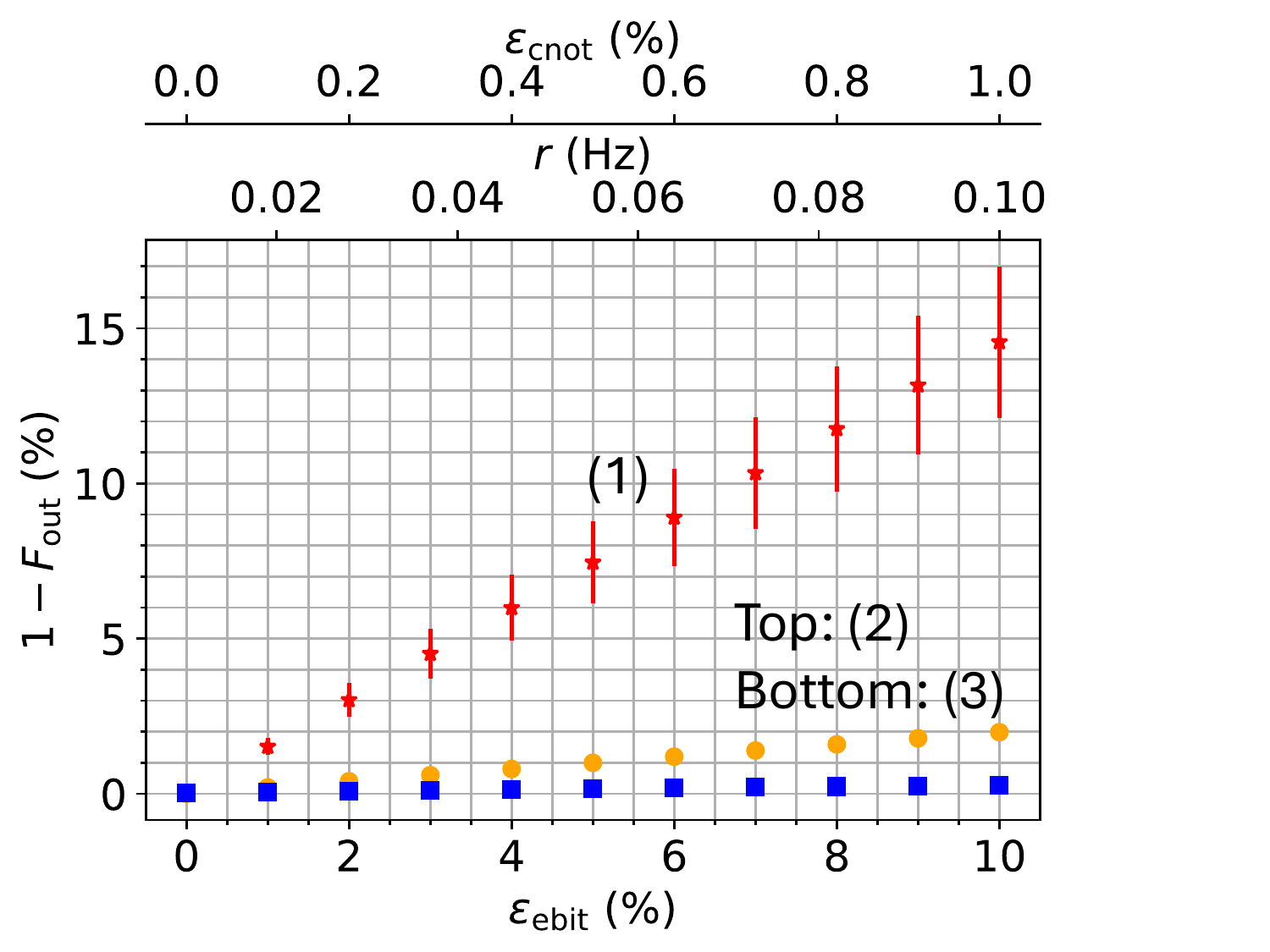}
    \put(20, 55){(h)}    
    \end{overpic}
    \begin{overpic}[scale=0.25, trim={0, 0, 3cm, 0}, clip]{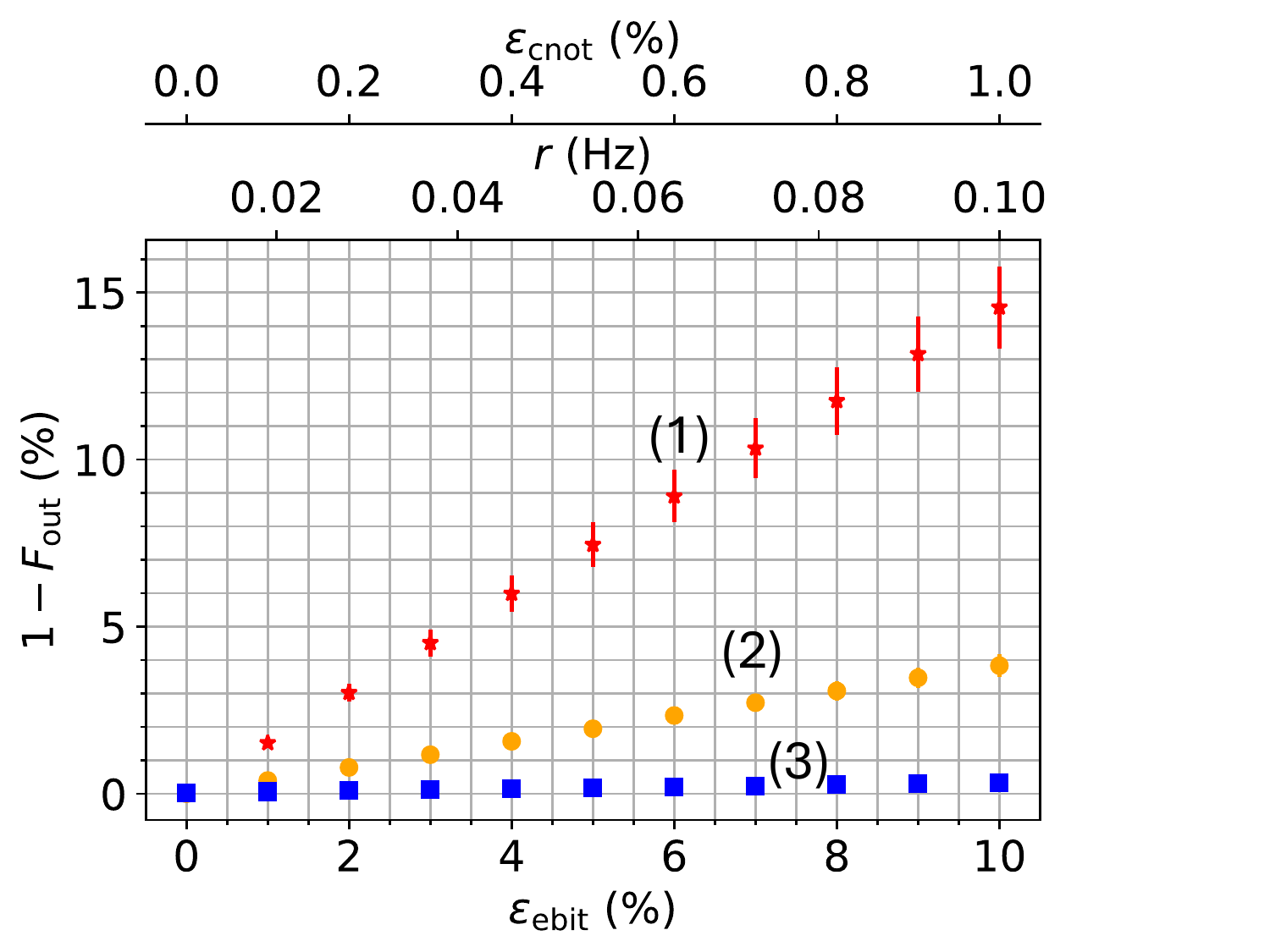}
    \put(20, 55){(i)}    
    \end{overpic}
    \begin{overpic}[scale=0.25, trim={0, 0, 3cm, 0}, clip]{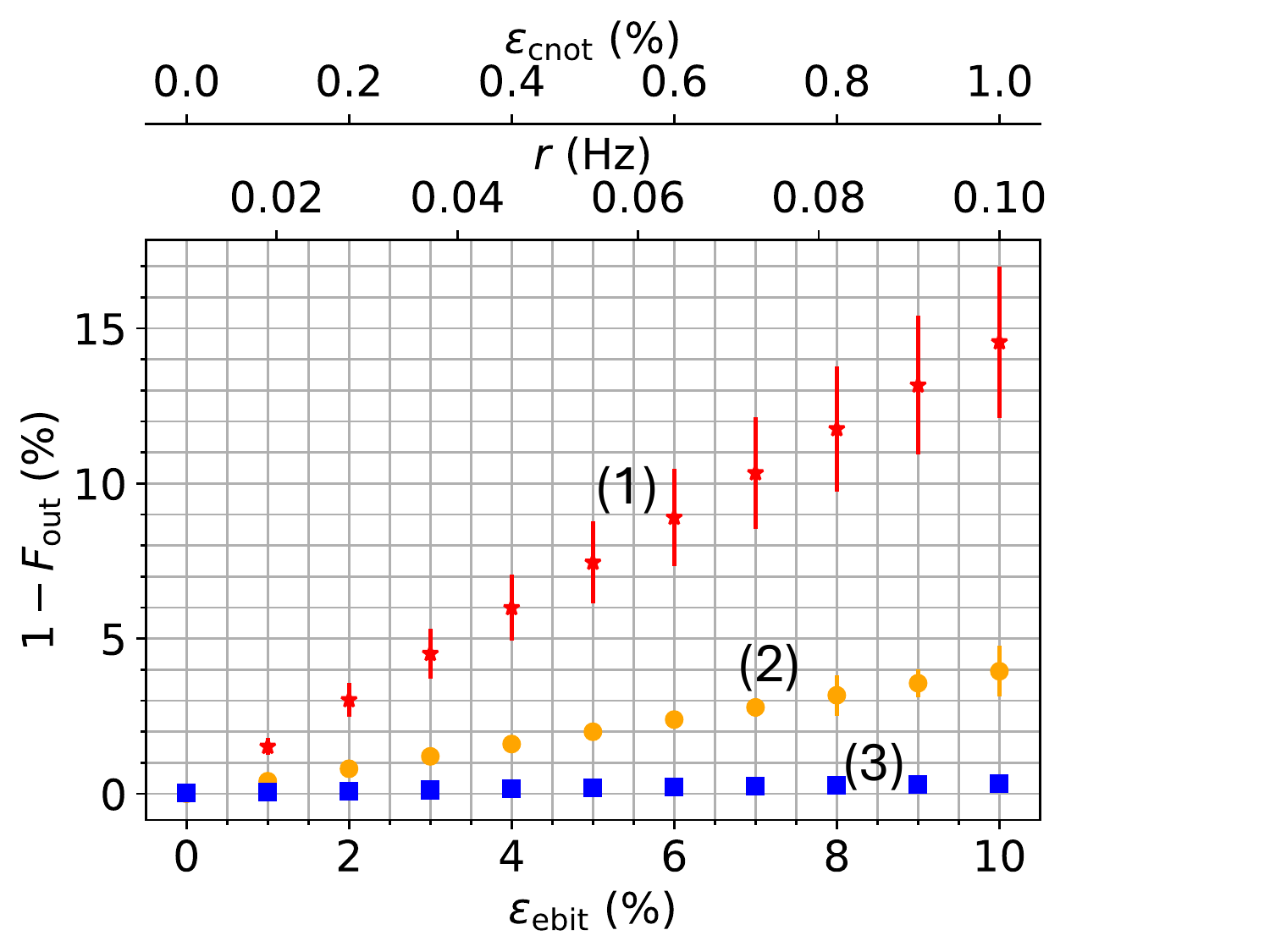}
    \put(20, 55){(j)}    
    \end{overpic}
    \subfloat{%
    \raisebox{0.8\height}{
    \begin{overpic}[scale=0.5, trim={0.5cm, 0.2cm, 0.2cm, 0.2cm}, clip]{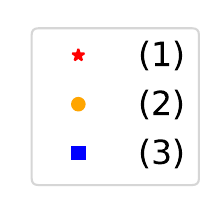}
    \end{overpic}}}
    \caption{The output error as a function of (1) $\epsilon_{\mathrm{ebit}}$ within the state-of-the-art range, (2) $\epsilon_{\mathrm{cnot}}$, (3) $r$, for a single CNOT gate implemented using: (a)-(b) a monolithic processor; (c)-(d) cat-comm; (e)-(f) 1TP; (g)-(h) 2TP; and (i)-(j) TP-safe. The results are averaged over a variety of input states with the form given by Eq. \eqref{eq:general_separable_2qubit_input} and the parameters varied over all permutations of $\alpha \in \{0.2, \frac{1}{\sqrt{2}}\}$, $\gamma = \{0.0, 0.2, ..., 1.0\}$, $\phi=0$, $\theta \in \{0, \frac{2\pi}{5},  ... \,, 2\pi\}$. The average used is the mean for (a), (c), (e), (g), and (i), with error bars indicating the standard deviation, and the median for (b), (d), (f), (h), and (j) remote gates, with error bars indicating the interquartile range. For curves (1), (2), and (3) on each figure, the non-varied error parameters are set to zero. Due to the low standard error previously observed when averaging over simulation runs, only one simulation run is used for each input state.}
    \label{fig:errors_compared_averaged_over_input_state_state_of_art}
\end{figure*}

\begin{figure*}
    \centering
    \begin{overpic}[scale=0.25, trim={0, 0, 3cm, 0}, clip]{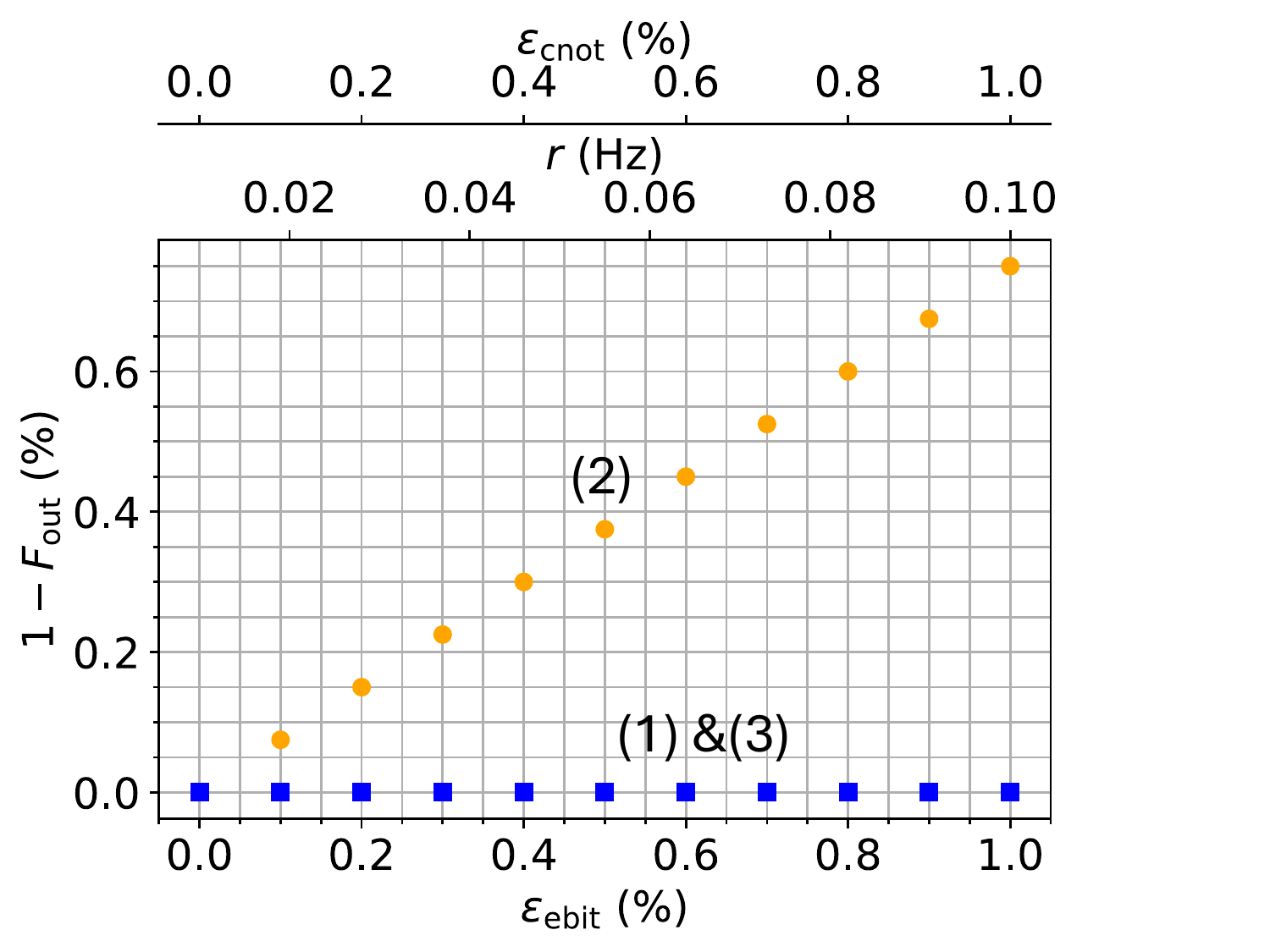}
        \put(20, 55){(a)}    
    \end{overpic}
    \begin{overpic}[scale=0.25, trim={0, 0, 3cm, 0}, clip]{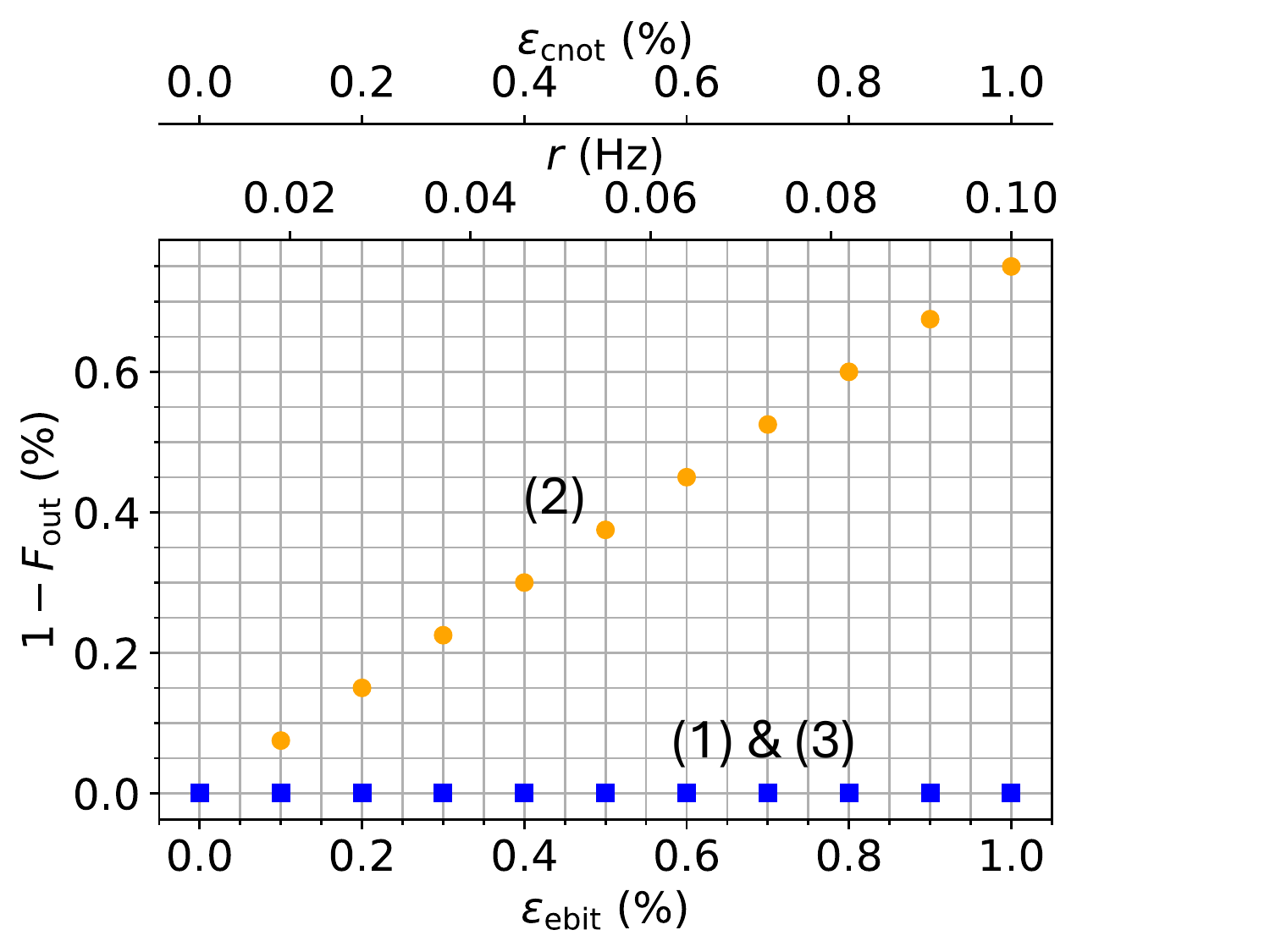}
        \put(20, 55){(b)}    
    \end{overpic}
    \begin{overpic}[scale=0.25, trim={0, 0, 3cm, 0}, clip]{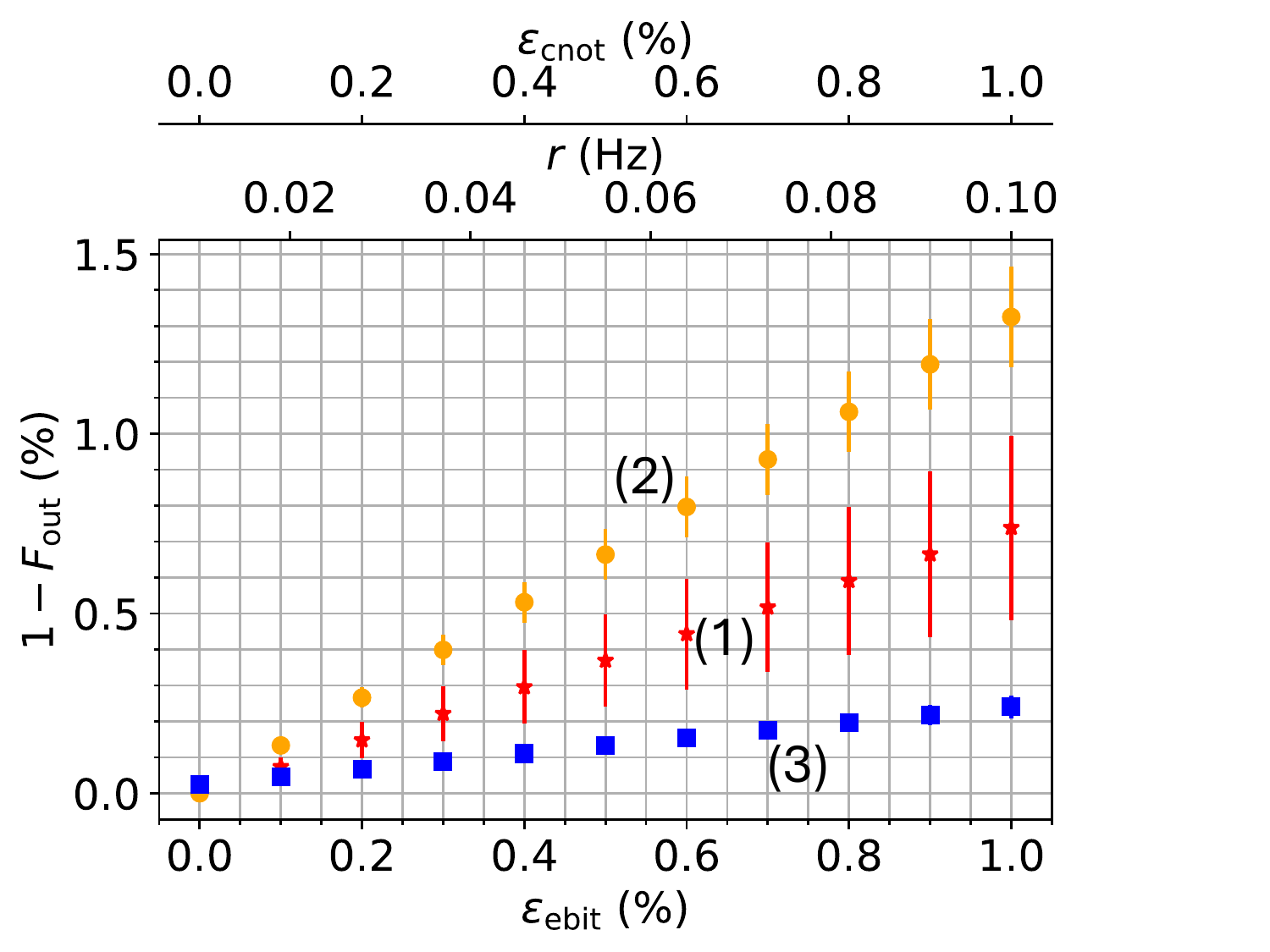}
    \put(20, 55){(c)}    
    \end{overpic}
    \begin{overpic}[scale=0.25, trim={0, 0, 3cm, 0}, clip]{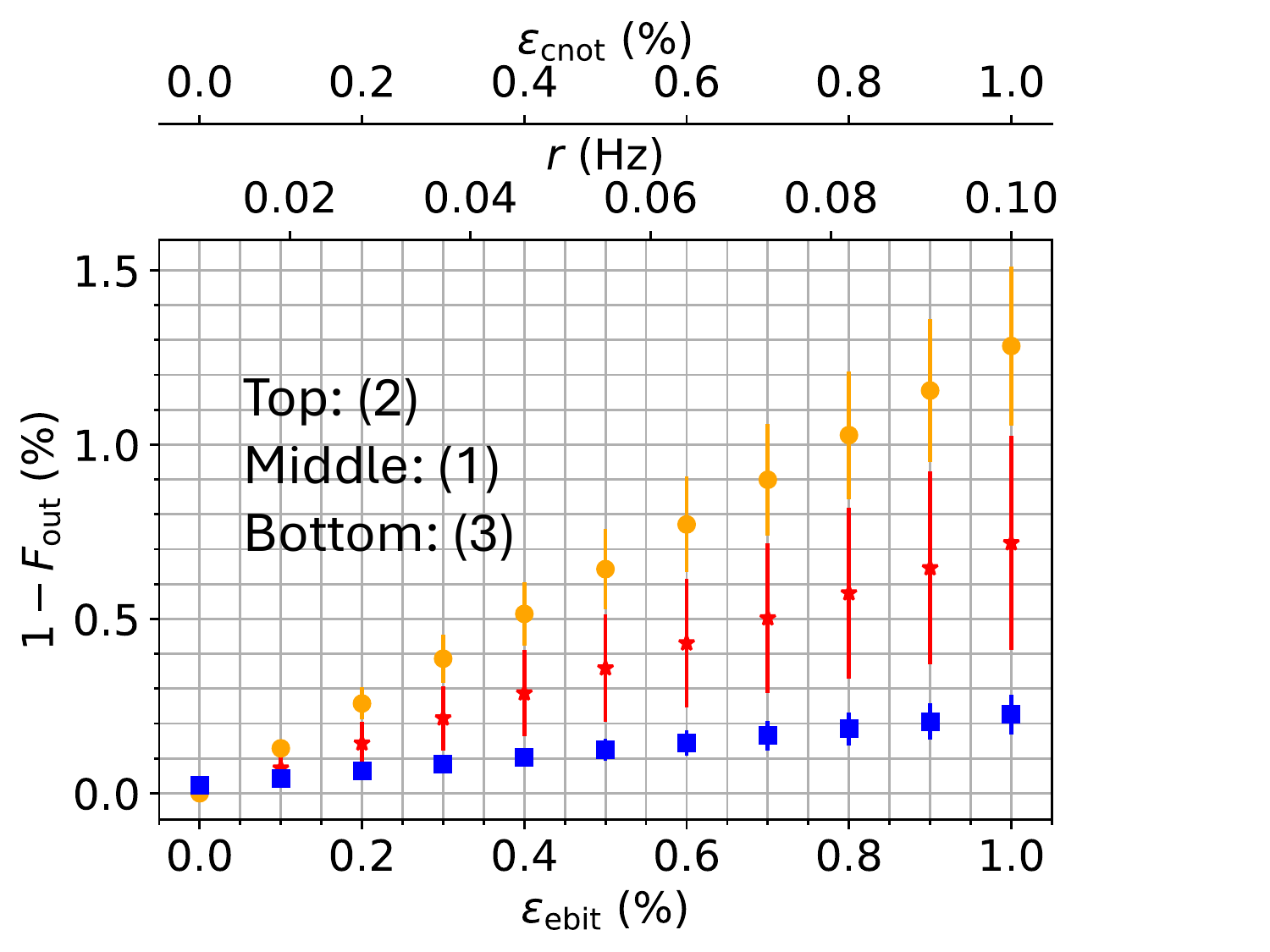}
    \put(20, 55){(d)}    
    \end{overpic}
    \begin{overpic}[scale=0.25, trim={0, 0, 3cm, 0}, clip]{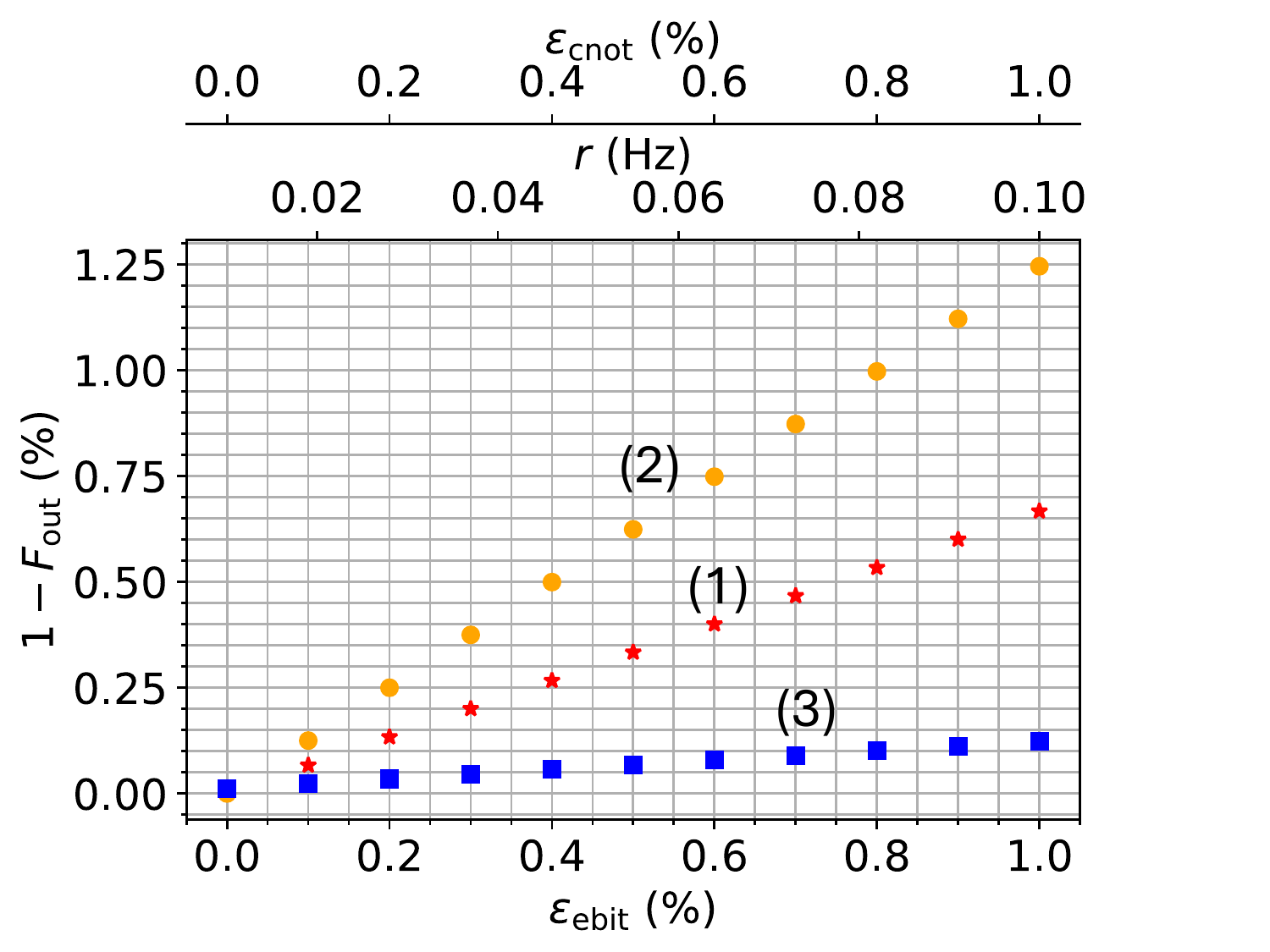}
    \put(20, 55){(e)}    
    \end{overpic}
    \begin{overpic}[scale=0.25, trim={0, 0, 3cm, 0}, clip]{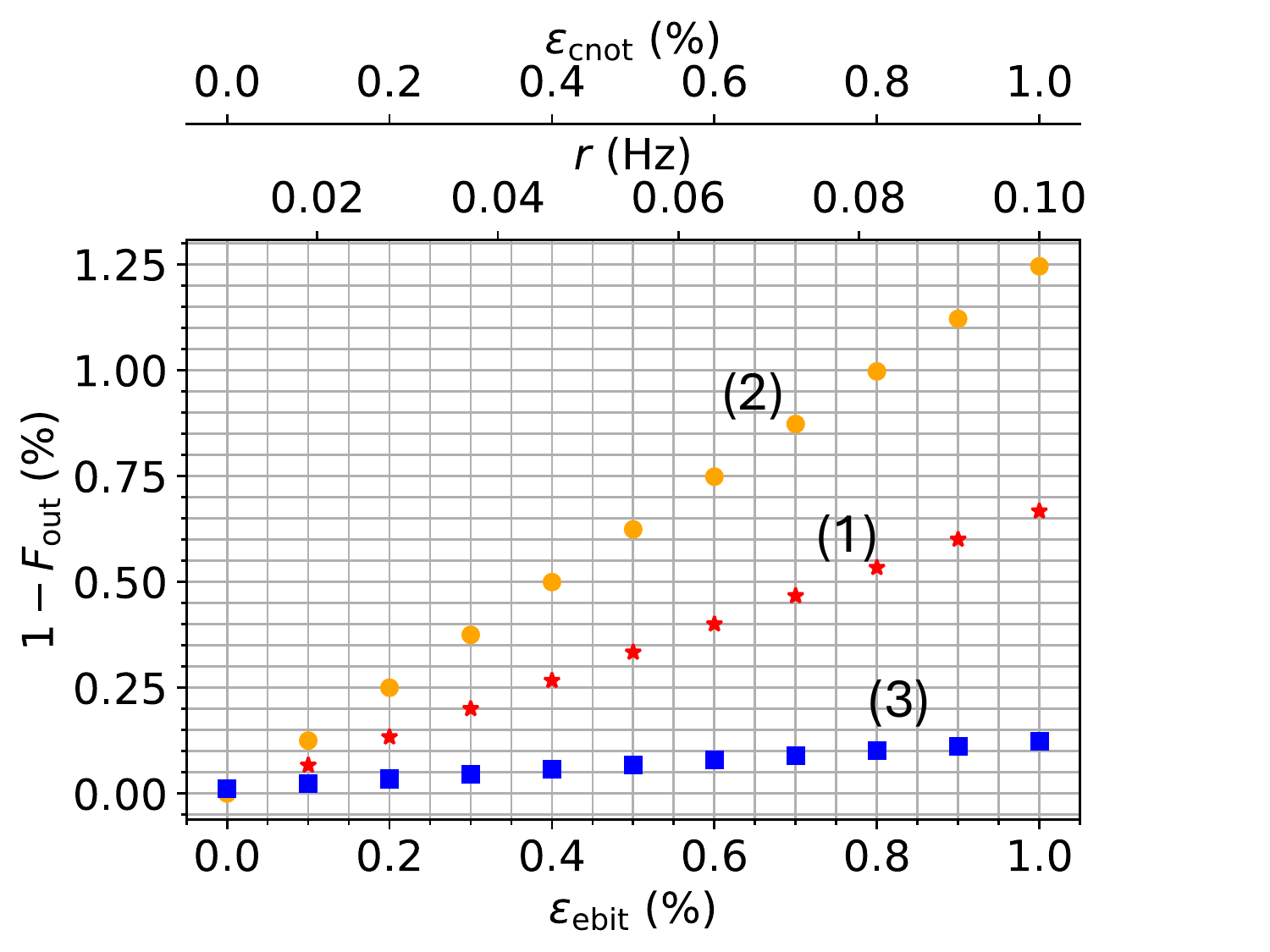}
    \put(20, 55){(f)}    
    \end{overpic}
    \begin{overpic}[scale=0.25, trim={0, 0, 3cm, 0}, clip]{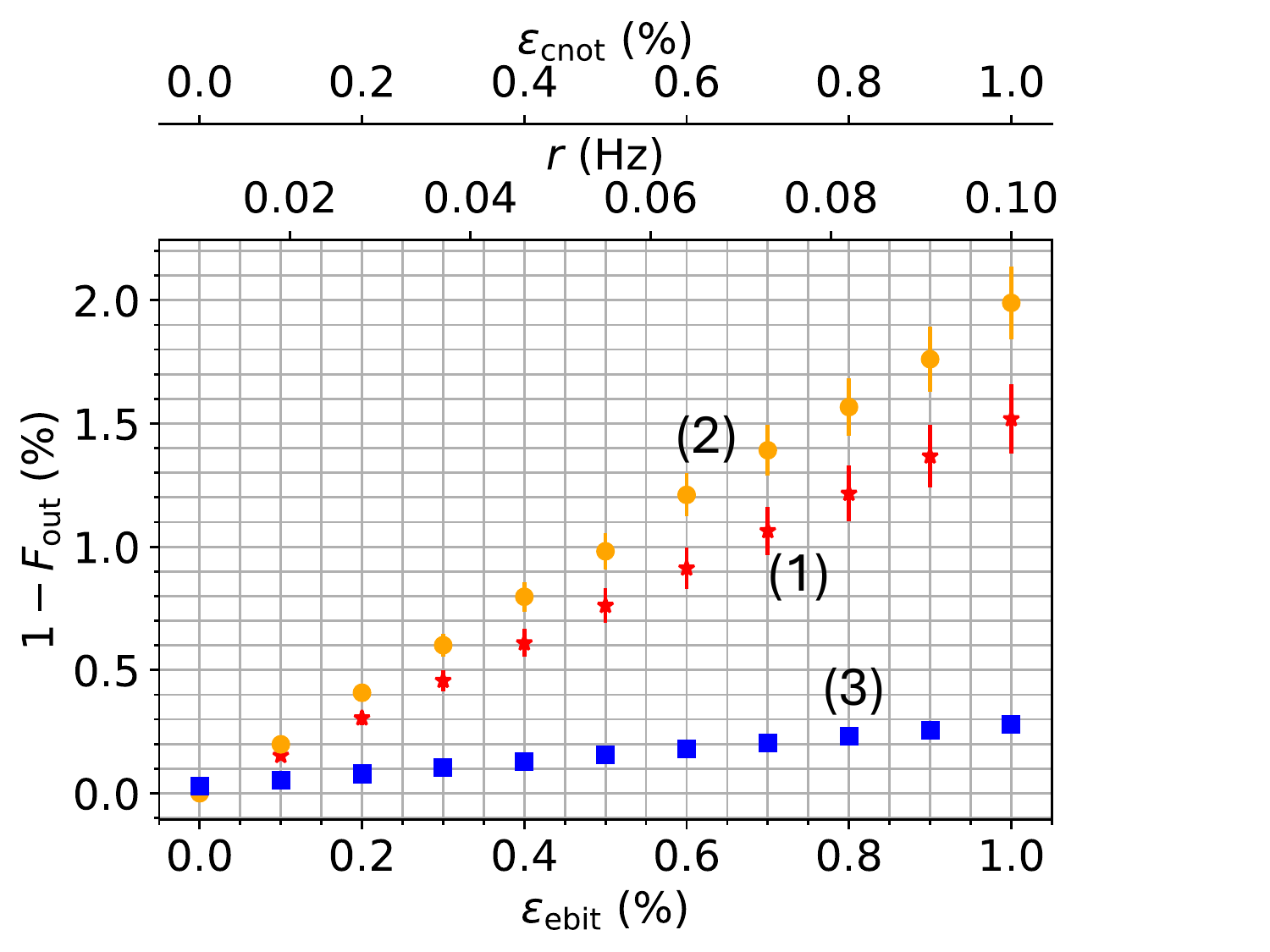}
    \put(20, 55){(g)}    
    \end{overpic}
    \begin{overpic}[scale=0.25, trim={0, 0, 3cm, 0}, clip]{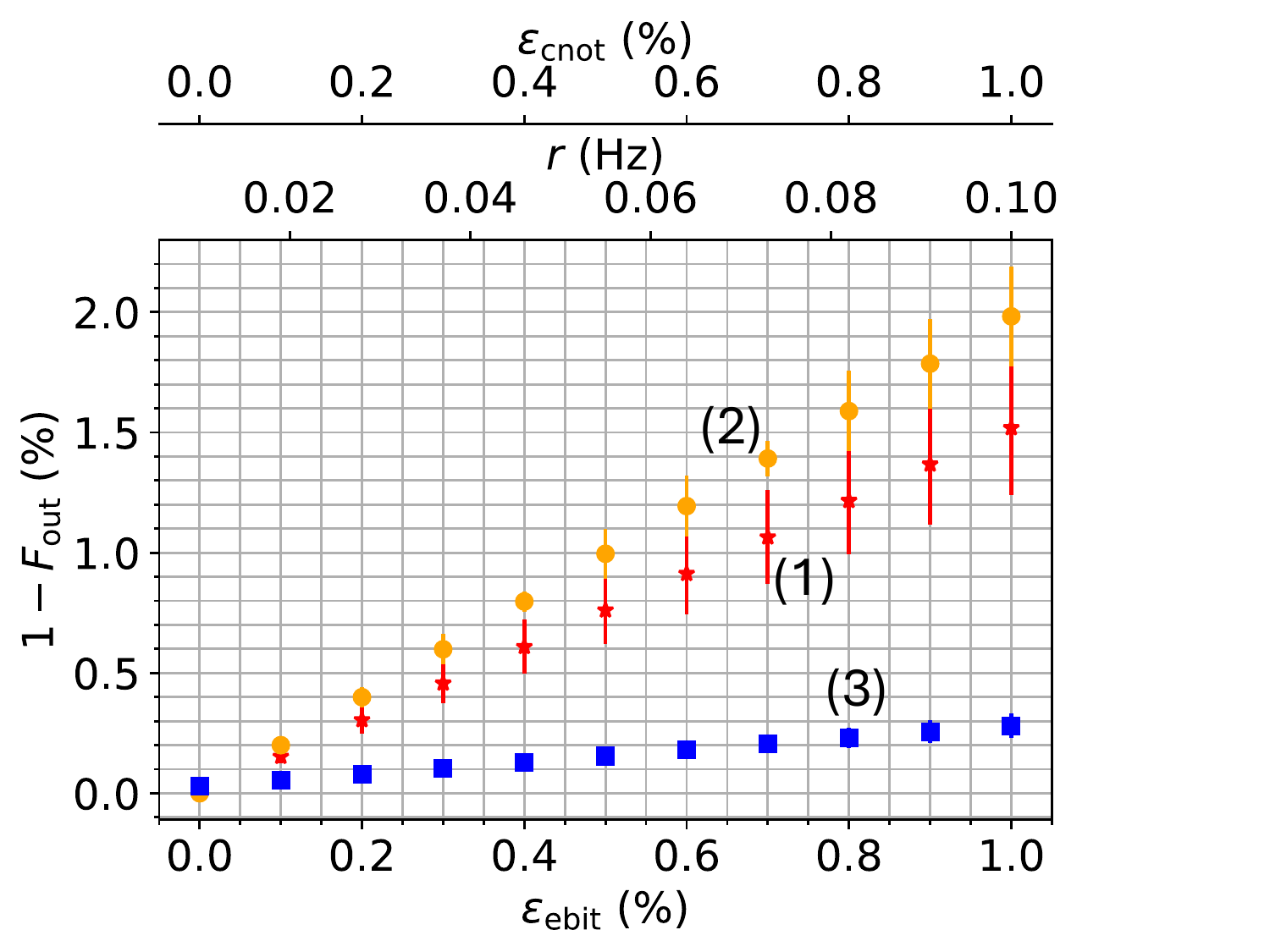}
    \put(20, 55){(h)}    
    \end{overpic}
    \begin{overpic}[scale=0.25, trim={0, 0, 3cm, 0}, clip]{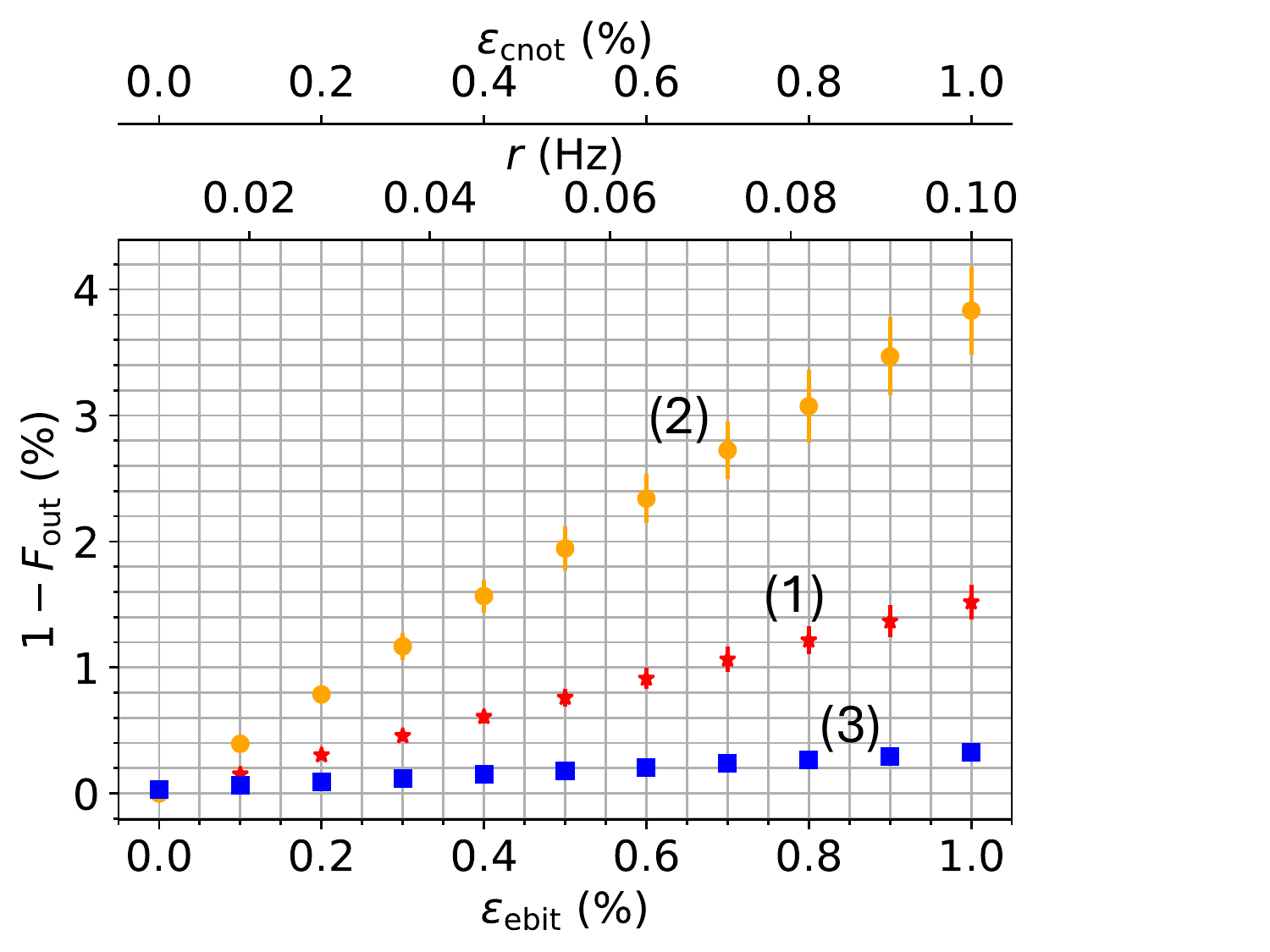}
    \put(20, 55){(i)}    
    \end{overpic}
    \begin{overpic}[scale=0.25, trim={0, 0, 3cm, 0}, clip]{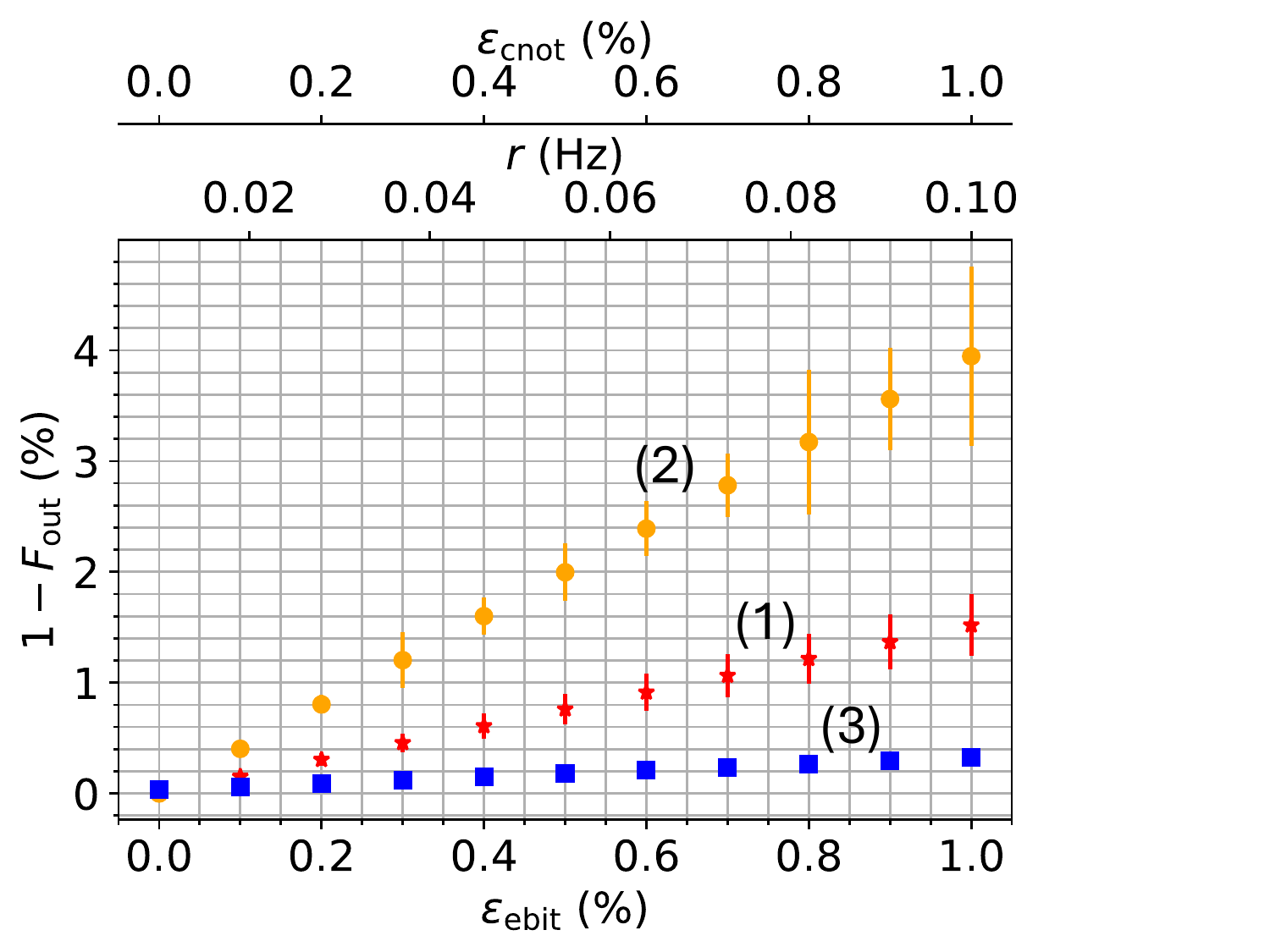}
    \put(20, 55){(j)}    
    \end{overpic}
    \subfloat{%
    \raisebox{0.8\height}{
    \begin{overpic}[scale=0.5, trim={0.5cm, 0.2cm, 0.2cm, 0.2cm}, clip]{comparing_errors_legend_with_numerical_labels.pdf}
    \end{overpic}}}
    \caption{The output error as a function of (1) $\epsilon_{\mathrm{ebit}}$ within the distilled range, (2) $\epsilon_{\mathrm{cnot}}$, (3) $r$, for a single CNOT gate implemented using: (a)-(b) a monolithic processor; (c)-(d) cat-comm; (e)-(f) 1TP; (g)-(h) 2TP; and (i)-(j) TP-safe. The results are averaged over a variety of input states with the form given by Eq. \eqref{eq:general_separable_2qubit_input} and the parameters varied over all permutations of $\alpha \in \{0.2, \frac{1}{\sqrt{2}}\}$, $\gamma = \{0.0, 0.2, ..., 1.0\}$, $\phi=0$, $\theta \in \{0, \frac{2\pi}{5},  ... \,, 2\pi\}$. The average used is the mean for (a), (c), (e), (g), and (i), with error bars indicating the standard deviation, and the median for (b), (d), (f), (h), and (j), with error bars indicating the interquartile range. For curves (1), (2), and (3) on each figure, the non-varied error parameters are set to zero. Due to the low standard error previously observed when averaging over simulation runs, only one simulation run is used for each input state.}
    \label{fig:errors_compared_averaged_over_input_state_distilled}
\end{figure*}

The next observations to verify are observations \ref{obs:state_of_art_ordering} and \ref{obs:distilled_ordering} from Sec. \ref{subsubsec:error_type_comparison}. Both observations are centered around the relative impact of the different error types introduced in Sec. \ref{subsec:error_models} with entanglement error in the state-of-the-art and distilled ranges, respectively. In Fig. \ref{fig:errors_compared_averaged_over_input_state_state_of_art}, %
we show the output error when one non-zero error parameter is varied while all other error parameters are fixed at zero. For entanglement error, the state-of-the-art range of values is considered. Much like Fig. \ref{fig:single_cnot_error_comparison_distilled} from the main text, in each subfigure, we show the output errors generated by three different error types for a given remote gate scheme, but here, we average each data point over the output errors obtained at a given error parameter value for a variety of input states. Again, the input states considered have the form given by Eq. \eqref{eq:general_separable_2qubit_input} and are produced from all permutations of $\alpha \in \{0.2, \frac{1}{\sqrt{2}}\}$, $\gamma = \{0.0, 0.2, ..., 1.0\}$, $\phi=0$, $\theta \in \{0, \frac{2\pi}{5},  ... \,, 2\pi\}$. Averaging is done by taking the mean, in Figs. \ref{fig:errors_compared_averaged_over_input_state_state_of_art}(a), \ref{fig:errors_compared_averaged_over_input_state_state_of_art}(c), \ref{fig:errors_compared_averaged_over_input_state_state_of_art}(e), \ref{fig:errors_compared_averaged_over_input_state_state_of_art}(g), \ref{fig:errors_compared_averaged_over_input_state_state_of_art}(i) and the median, in Figs. \ref{fig:errors_compared_averaged_over_input_state_state_of_art}(b), \ref{fig:errors_compared_averaged_over_input_state_state_of_art}(d), \ref{fig:errors_compared_averaged_over_input_state_state_of_art}(f), \ref{fig:errors_compared_averaged_over_input_state_state_of_art}(h), and \ref{fig:errors_compared_averaged_over_input_state_state_of_art}(i). When the mean is used, the error bars show the standard deviation and when the median is used, the error bars show the interquartile range. Figure \ref{fig:errors_compared_averaged_over_input_state_distilled} %
shows the same thing but with entanglement error varied over the distilled range.

Figures \ref{fig:errors_compared_averaged_over_input_state_state_of_art} and \ref{fig:errors_compared_averaged_over_input_state_distilled} indicate that observations \ref{obs:state_of_art_ordering} and \ref{obs:distilled_ordering} from Sec. \ref{subsubsec:error_type_comparison} do indeed hold for a variety of input states. The output error caused by each error type in Fig. \ref{fig:errors_compared_averaged_over_input_state_state_of_art}, in which the state-of-the-art parameters are used, is consistently smallest for memory depolarisation. Local gate error yields a larger output error but its impact is in turn dominated by entanglement error. Again memory depolarisation is relatively negligible. These findings hold true for all remote gate schemes and are true for both the mean and median over input state with neither the standard deviation nor the interquartile range of any data points overlapping---which would have indicated uncertainty in the relative impact of the different error types. All of this is consistent with observation \ref{obs:state_of_art_ordering}

Similarly, in Fig. \ref{fig:errors_compared_averaged_over_input_state_distilled}, in which the distilled range of entanglement error is used, the output error is smallest when memory depolarisation is non-zero, greater for entanglement error, and greatest for local gate error. This is consistent with observation \ref{obs:distilled_ordering} from Sec. \ref{subsubsec:error_type_comparison}.

The robustness of observations \ref{obs:state_of_art_ordering} and \ref{obs:distilled_ordering} from Sec. \ref{subsubsec:error_type_comparison} can be further verified by an exhaustive search of plots showing the non-averaged output errors for each input state individually. We perform such a search and find that the ordering seen in observations \ref{obs:state_of_art_ordering} and \ref{obs:distilled_ordering} is seen consistently for all of input states with the form of Eq. \eqref{eq:general_separable_2qubit_input} and parameters $\alpha \in \{0.2, \frac{1}{\sqrt{2}}\}$, $\gamma = \{0.0, 0.2, ..., 1.0\}$, $\phi=0$, $\theta \in \{0, \frac{2\pi}{5},  ... \,, 2\pi\}$. The relevant plots can be found in the supplementary information.

\subsection{Comments on Sec. \ref{subsubsec:remote_gate_comparison}}
\label{app:discussion_of_remote_gate_scheme_comparison}

In Sec. \ref{subsubsec:remote_gate_comparison} from the main text, we note that the ordering of remote gate schemes for the input state given by Eq. \eqref{eq:remote_CNOT_input_state}, is, ordered from lowest to highest by their impact on output error: 1TP, cat-comm, 2TP, and then TP-safe. 2TP and TP-safe are equally damaging to the output when only entanglement error is considered. Here, we confirm that the relative positions of 1TP and cat-comm are specific to input state, as alluded to in the main text and Appendix \ref{app:variation_input_state}.

\begin{figure}
    \centering
    \begin{overpic}[scale=0.24, trim={0, 2cm, 0, 0}, clip]{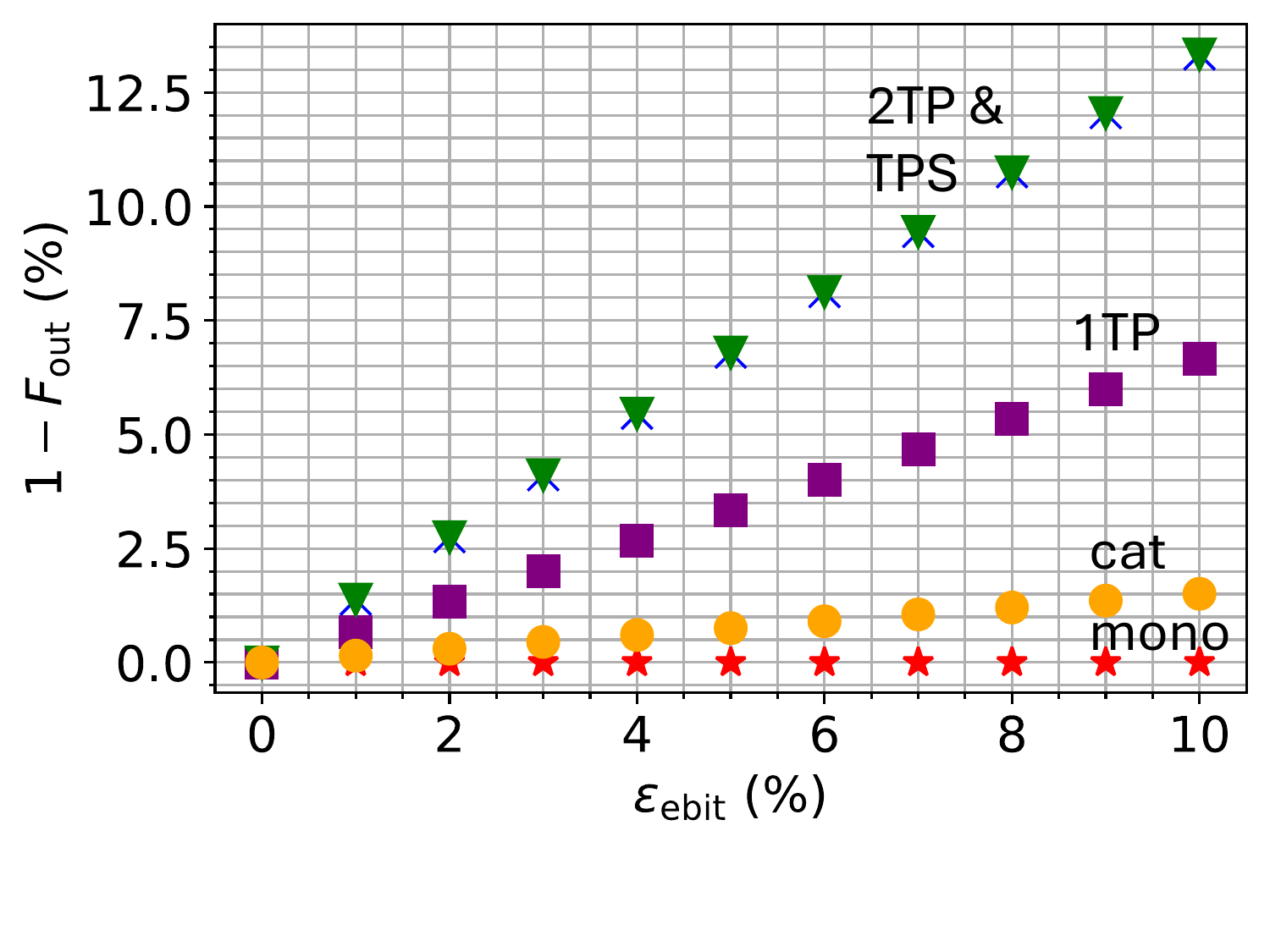}
        \put(50, 60){(a)} 
        \put(20, 34){\includegraphics[scale=0.4, trim={0.5cm, 0.4cm, 0.2cm, 0.4cm}, clip]{comparing_schemes_legend.pdf}}
    \end{overpic}%
    \begin{overpic}[scale=0.24, trim={1.5cm, 2cm, 0, 0}, clip]{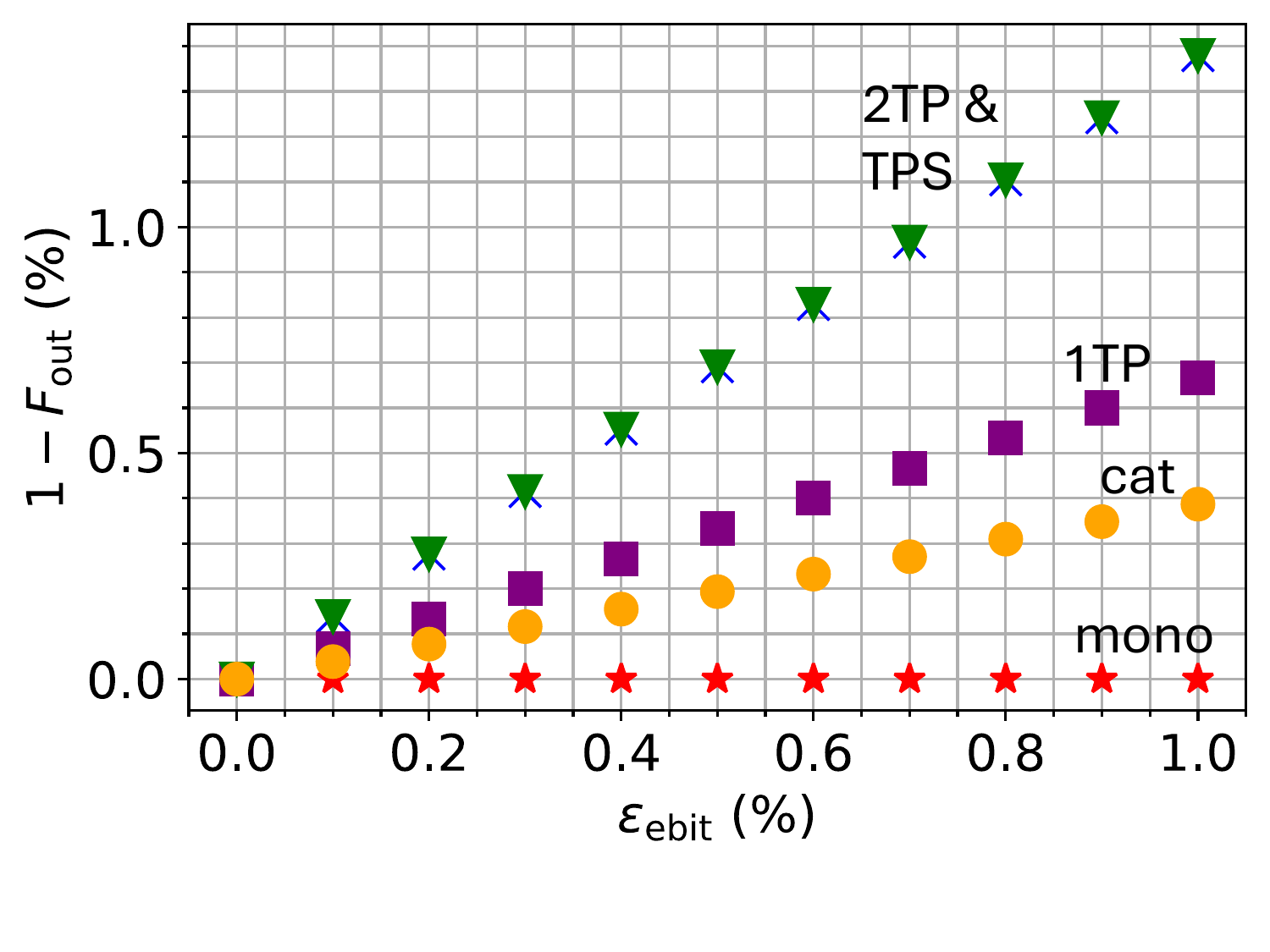}
        \put(45, 64){(b)}    
    \end{overpic}%
    \begin{overpic}[scale=0.24, trim={1.5cm, 1.5cm, 0, 0}, clip]{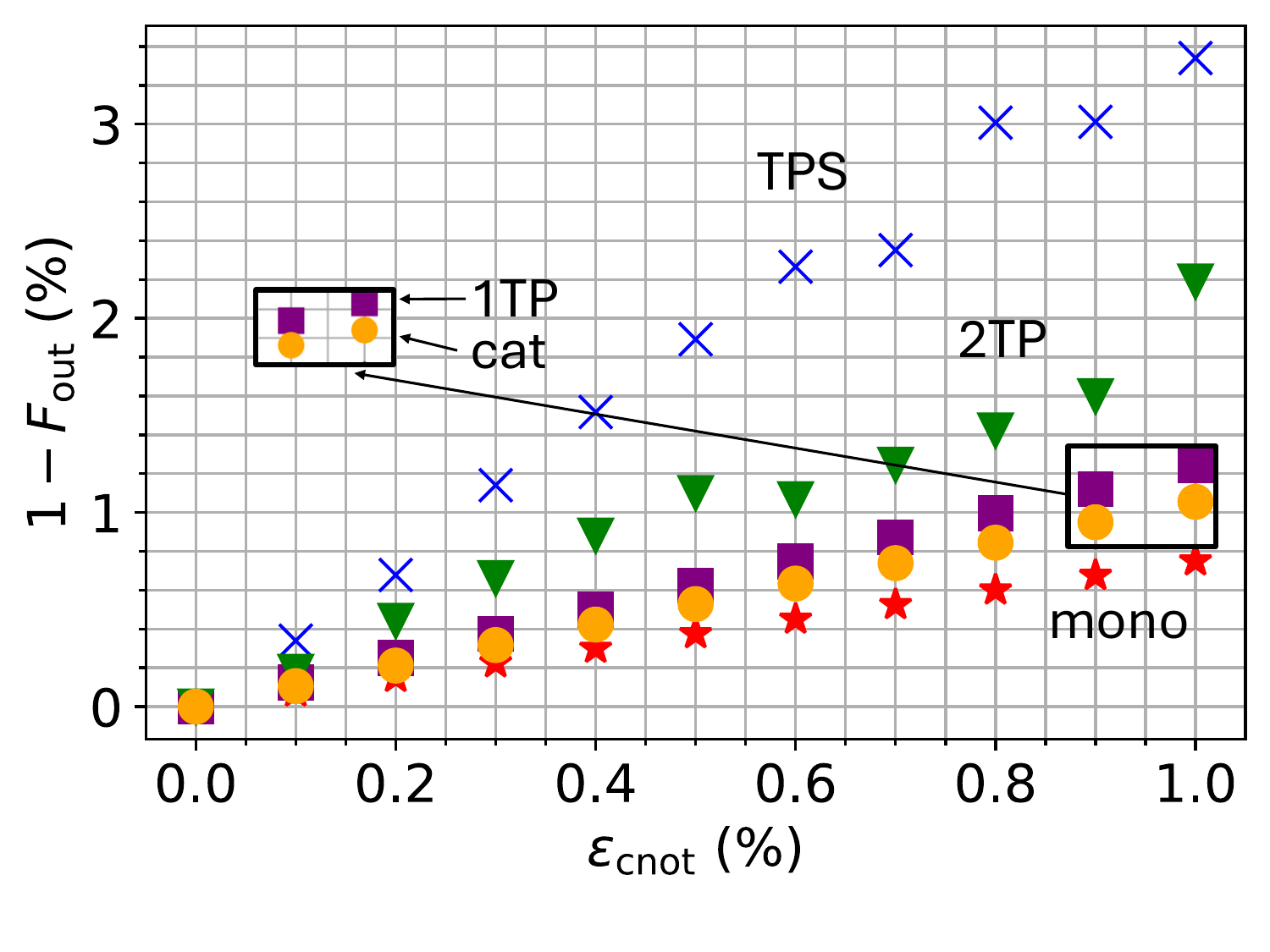}
        \put(45, 66){(c)}  
    \end{overpic}%
    \caption{The output error, $1 - F_{\rm out}$, for an individual remote CNOT gate with non-zero: (a) entanglement error in the state-of-art range and an input state of $\ket{\mathrm{input}}_{q_2, \, q_2'} = (0.2 \ket{0}_{q_2} +  0.980 \ket{1}_{q_2}) \otimes (0.6 \ket{0}_{q_2'} + 0.8 \ket{1}_{q_2'})$; (b) entanglement error in the distilled range and an input state of $\ket{\mathrm{input}}_{q_2, \, q_2'} = (0.2 \ket{0}_{q_2} +  0.980 \ket{1}_{q_2}) \otimes (0.4 \ket{0}_{q_2'} + 0.917 \ket{1}_{q_2'})$; (c) local two-qubit gate error and an input state of $\ket{\mathrm{input}}_{q_2, \, q_2'} = (0.2 \ket{0}_{q_2} +  0.980 \ket{1}_{q_2}) \otimes (0.8 \ket{0}_{q_2'} + 0.6 \ket{1}_{q_2'})$. The output errors when the remote CNOT gate is implemented using cat-comm (cat), 1TP, 2TP, and TP-safe (TPS) are considered. For the monolithic case (mono) a single local CNOT gate is considered. Due to the low standard error previously observed when averaging over simulation runs, only one simulation run is used for each input state. }
    \label{fig:comparing_schemes_different_input_states}
\end{figure}

Figure \ref{fig:comparing_schemes_different_input_states} %
shows a comparison of the output error caused by each of the different remote gate schemes shown in Fig. \ref{fig:remote_cnot_implemented_in_different_ways}. Unlike in the main text, the results for variety of different input states are shown. In Figs. \ref{fig:comparing_schemes_different_input_states}(a)-(b), the entanglement error is varied over the state-of-the-art and  distilled ranges, respectively, and all other errors are set to zero. In Fig. \ref{fig:comparing_schemes_different_input_states}(c), the local two-qubit gate error is varied and all other errors are set to zero. 

In Figs. \ref{fig:comparing_schemes_different_input_states}(a)-(c), 1TP has a larger output error than cat-comm, in contrast to what is seen and discussed in Sec. \ref{subsubsec:remote_gate_comparison} from the main text, in which a different input state, given by Eq. \eqref{eq:remote_CNOT_input_state}, is used. This is indicative of how input state dependent the relative ordering of the schemes can be. Interestingly, when we consider only a non-zero memory depolarisation, we have not yet found an input state for which 1TP does not outperform cat-comm. The most probable reason for this is that cat-comm has a slightly higher latency than 1TP. 1TP communicates the results of both the measurements that occur in the scheme at the same point in the circuit. This means that both measurements can be combined into a single classical message at the same time. By contrast, for cat-comm, the measurements are split between the cat-entanglement and cat-disentanglement protocols, which must occur sequentially. This means that the measurement results must be communicated between the QPUs in separate messages at different times and so the portion of the latency attributable to classical messaging is doubled. The remaining remote gate schemes retain the order quoted in the main text.

\FloatBarrier


\begin{thebibliography}{54}%
\makeatletter
\providecommand \@ifxundefined [1]{%
 \@ifx{#1\undefined}
}%
\providecommand \@ifnum [1]{%
 \ifnum #1\expandafter \@firstoftwo
 \else \expandafter \@secondoftwo
 \fi
}%
\providecommand \@ifx [1]{%
 \ifx #1\expandafter \@firstoftwo
 \else \expandafter \@secondoftwo
 \fi
}%
\providecommand \natexlab [1]{#1}%
\providecommand \enquote  [1]{``#1''}%
\providecommand \bibnamefont  [1]{#1}%
\providecommand \bibfnamefont [1]{#1}%
\providecommand \citenamefont [1]{#1}%
\providecommand \href@noop [0]{\@secondoftwo}%
\providecommand \href [0]{\begingroup \@sanitize@url \@href}%
\providecommand \@href[1]{\@@startlink{#1}\@@href}%
\providecommand \@@href[1]{\endgroup#1\@@endlink}%
\providecommand \@sanitize@url [0]{\catcode `\\12\catcode `\$12\catcode `\&12\catcode `\#12\catcode `\^12\catcode `\_12\catcode `\%12\relax}%
\providecommand \@@startlink[1]{}%
\providecommand \@@endlink[0]{}%
\providecommand \url  [0]{\begingroup\@sanitize@url \@url }%
\providecommand \@url [1]{\endgroup\@href {#1}{\urlprefix }}%
\providecommand \urlprefix  [0]{URL }%
\providecommand \Eprint [0]{\href }%
\providecommand \doibase [0]{https://doi.org/}%
\providecommand \selectlanguage [0]{\@gobble}%
\providecommand \bibinfo  [0]{\@secondoftwo}%
\providecommand \bibfield  [0]{\@secondoftwo}%
\providecommand \translation [1]{[#1]}%
\providecommand \BibitemOpen [0]{}%
\providecommand \bibitemStop [0]{}%
\providecommand \bibitemNoStop [0]{.\EOS\space}%
\providecommand \EOS [0]{\spacefactor3000\relax}%
\providecommand \BibitemShut  [1]{\csname bibitem#1\endcsname}%
\let\auto@bib@innerbib\@empty
\bibitem [{\citenamefont {Bova}\ \emph {et~al.}(2021)\citenamefont {Bova}, \citenamefont {Goldfarb},\ and\ \citenamefont {Melko}}]{commercialApplicationsQC}%
  \BibitemOpen
  \bibfield  {author} {\bibinfo {author} {\bibfnamefont {F.}~\bibnamefont {Bova}}, \bibinfo {author} {\bibfnamefont {A.}~\bibnamefont {Goldfarb}},\ and\ \bibinfo {author} {\bibfnamefont {R.~G.}\ \bibnamefont {Melko}},\ }\bibfield  {title} {\bibinfo {title} {Commercial applications of quantum computing},\ }\href {https://doi.org/10.1140/epjqt/s40507-021-00091-1} {\bibfield  {journal} {\bibinfo  {journal} {EPJ Quantum Technology}\ }\textbf {\bibinfo {volume} {8}},\ \bibinfo {pages} {2} (\bibinfo {year} {2021})}\BibitemShut {NoStop}%
\bibitem [{\citenamefont {Zalka}(1998)}]{quantumSimulation}%
  \BibitemOpen
  \bibfield  {author} {\bibinfo {author} {\bibfnamefont {C.}~\bibnamefont {Zalka}},\ }\bibfield  {title} {\bibinfo {title} {Simulating quantum systems on a quantum computer},\ }in\ \href@noop {} {\emph {\bibinfo {booktitle} {Proceedings of the Royal Society of London. Series A: Mathematical, Physical and Engineering Sciences}}},\ Vol.\ \bibinfo {volume} {454}\ (\bibinfo {year} {1998})\ pp.\ \bibinfo {pages} {313--322}\BibitemShut {NoStop}%
\bibitem [{\citenamefont {Shor}(1997)}]{ShorAlogrithm}%
  \BibitemOpen
  \bibfield  {author} {\bibinfo {author} {\bibfnamefont {P.}~\bibnamefont {Shor}},\ }\bibfield  {title} {\bibinfo {title} {Polynomial-time algorithms for prime factorization and discrete logarithms on a quantum computer},\ }\href@noop {} {\bibfield  {journal} {\bibinfo  {journal} {SIAM journal on Computing}\ }\textbf {\bibinfo {volume} {26}},\ \bibinfo {pages} {1484} (\bibinfo {year} {1997})}\BibitemShut {NoStop}%
\bibitem [{\citenamefont {Grover}(1996)}]{GroverAlgorithm}%
  \BibitemOpen
  \bibfield  {author} {\bibinfo {author} {\bibfnamefont {L.}~\bibnamefont {Grover}},\ }\bibfield  {title} {\bibinfo {title} {A fast quantum-mechanical algorithm for database search},\ }in\ \href@noop {} {\emph {\bibinfo {booktitle} {Proceedings of the Twenty-Eigth Annual ACM Symposium on Theory of Computing}}}\ (\bibinfo {year} {1996})\ pp.\ \bibinfo {pages} {212--219}\BibitemShut {NoStop}%
\bibitem [{\citenamefont {Sarovar}\ \emph {et~al.}(2020)\citenamefont {Sarovar}, \citenamefont {Proctor}, \citenamefont {Rudinger}, \citenamefont {Young}, \citenamefont {Nielsen},\ and\ \citenamefont {Blume-Kohout}}]{Sarovar2020detectingcrosstalk}%
  \BibitemOpen
  \bibfield  {author} {\bibinfo {author} {\bibfnamefont {M.}~\bibnamefont {Sarovar}}, \bibinfo {author} {\bibfnamefont {T.}~\bibnamefont {Proctor}}, \bibinfo {author} {\bibfnamefont {K.}~\bibnamefont {Rudinger}}, \bibinfo {author} {\bibfnamefont {K.}~\bibnamefont {Young}}, \bibinfo {author} {\bibfnamefont {E.}~\bibnamefont {Nielsen}},\ and\ \bibinfo {author} {\bibfnamefont {R.}~\bibnamefont {Blume-Kohout}},\ }\bibfield  {title} {\bibinfo {title} {Detecting crosstalk errors in quantum information processors},\ }\href {https://doi.org/10.22331/q-2020-09-11-321} {\bibfield  {journal} {\bibinfo  {journal} {{Quantum}}\ }\textbf {\bibinfo {volume} {4}},\ \bibinfo {pages} {321} (\bibinfo {year} {2020})}\BibitemShut {NoStop}%
\bibitem [{\citenamefont {Roffe}(2019)}]{QuantumErrorCorrectionIntroductoryGuide}%
  \BibitemOpen
  \bibfield  {author} {\bibinfo {author} {\bibfnamefont {J.}~\bibnamefont {Roffe}},\ }\bibfield  {title} {\bibinfo {title} {Quantum error correction: An introductory guide},\ }\href@noop {} {\bibfield  {journal} {\bibinfo  {journal} {Contemporary Physics}\ }\textbf {\bibinfo {volume} {60}},\ \bibinfo {pages} {226} (\bibinfo {year} {2019})}\BibitemShut {NoStop}%
\bibitem [{\citenamefont {Landauer}(1995)}]{LandauerArgumentAgainstQC}%
  \BibitemOpen
  \bibfield  {author} {\bibinfo {author} {\bibfnamefont {R.}~\bibnamefont {Landauer}},\ }\bibfield  {title} {\bibinfo {title} {Is quantum mechanics useful?},\ }\href {http://www.jstor.org/stable/54534} {\bibfield  {journal} {\bibinfo  {journal} {Philosophical Transactions: Physical Sciences and Engineering}\ }\textbf {\bibinfo {volume} {353}},\ \bibinfo {pages} {367} (\bibinfo {year} {1995})}\BibitemShut {NoStop}%
\bibitem [{\citenamefont {Chuang}\ \emph {et~al.}(1995)\citenamefont {Chuang}, \citenamefont {Laflamme}, \citenamefont {Shor},\ and\ \citenamefont {Zurek}}]{QCFactoringAndDecoherence}%
  \BibitemOpen
  \bibfield  {author} {\bibinfo {author} {\bibfnamefont {I.~L.}\ \bibnamefont {Chuang}}, \bibinfo {author} {\bibfnamefont {R.}~\bibnamefont {Laflamme}}, \bibinfo {author} {\bibfnamefont {P.~W.}\ \bibnamefont {Shor}},\ and\ \bibinfo {author} {\bibfnamefont {W.~H.}\ \bibnamefont {Zurek}},\ }\bibfield  {title} {\bibinfo {title} {Quantum computers, factoring, and decoherence},\ }\href {https://doi.org/10.1126/science.270.5242.1633} {\bibfield  {journal} {\bibinfo  {journal} {Science}\ }\textbf {\bibinfo {volume} {270}},\ \bibinfo {pages} {1633–1635} (\bibinfo {year} {1995})}\BibitemShut {NoStop}%
\bibitem [{\citenamefont {Shor}(1995)}]{9qubitShorErrorCorrection}%
  \BibitemOpen
  \bibfield  {author} {\bibinfo {author} {\bibfnamefont {P.~W.}\ \bibnamefont {Shor}},\ }\bibfield  {title} {\bibinfo {title} {Scheme for reducing decoherence in quantum computer memory},\ }\href {https://doi.org/10.1103/PhysRevA.52.R2493} {\bibfield  {journal} {\bibinfo  {journal} {Physical Review A}\ }\textbf {\bibinfo {volume} {52}},\ \bibinfo {pages} {R2493} (\bibinfo {year} {1995})}\BibitemShut {NoStop}%
\bibitem [{\citenamefont {Liu}\ \emph {et~al.}(2023)\citenamefont {Liu}, \citenamefont {Hann},\ and\ \citenamefont {Jiang}}]{altQDCdef}%
  \BibitemOpen
  \bibfield  {author} {\bibinfo {author} {\bibfnamefont {J.}~\bibnamefont {Liu}}, \bibinfo {author} {\bibfnamefont {C.~T.}\ \bibnamefont {Hann}},\ and\ \bibinfo {author} {\bibfnamefont {L.}~\bibnamefont {Jiang}},\ }\bibfield  {title} {\bibinfo {title} {Data centers with quantum random access memory and quantum networks},\ }\href {https://doi.org/10.1103/PhysRevA.108.032610} {\bibfield  {journal} {\bibinfo  {journal} {Phys. Rev. A}\ }\textbf {\bibinfo {volume} {108}},\ \bibinfo {pages} {032610} (\bibinfo {year} {2023})}\BibitemShut {NoStop}%
\bibitem [{\citenamefont {Baker}\ \emph {et~al.}(2020)\citenamefont {Baker}, \citenamefont {Duckering}, \citenamefont {Hoover},\ and\ \citenamefont {Chong}}]{TimeSlicedPartitioning}%
  \BibitemOpen
  \bibfield  {author} {\bibinfo {author} {\bibfnamefont {J.}~\bibnamefont {Baker}}, \bibinfo {author} {\bibfnamefont {C.}~\bibnamefont {Duckering}}, \bibinfo {author} {\bibfnamefont {A.}~\bibnamefont {Hoover}},\ and\ \bibinfo {author} {\bibfnamefont {F.}~\bibnamefont {Chong}},\ }\bibfield  {title} {\bibinfo {title} {Time-sliced quantum circuit partitioning for modular architectures}\ }in\ {\emph {\bibinfo {booktitle} {Proceedings of the 17th ACM International Conference on Computing Frontiers}}}(\bibinfo {year} {2020})\ pp.\ \bibinfo {pages} {98--107}\BibitemShut {NoStop}%
\bibitem [{\citenamefont {Wu}\ \emph {et~al.}(2022)\citenamefont {Wu}, \citenamefont {Zhang}, \citenamefont {Li}, \citenamefont {Shabani}, \citenamefont {Xie},\ and\ \citenamefont {Ding}}]{AutoComm}%
  \BibitemOpen
  \bibfield  {author} {\bibinfo {author} {\bibfnamefont {A.}~\bibnamefont {Wu}}, \bibinfo {author} {\bibfnamefont {H.}~\bibnamefont {Zhang}}, \bibinfo {author} {\bibfnamefont {G.}~\bibnamefont {Li}}, \bibinfo {author} {\bibfnamefont {A.}~\bibnamefont {Shabani}}, \bibinfo {author} {\bibfnamefont {Y.}~\bibnamefont {Xie}},\ and\ \bibinfo {author} {\bibfnamefont {Y.}~\bibnamefont {Ding}},\ }\bibfield  {title} {\bibinfo {title} {Autocomm: A framework for enabling efficient communication in distributed quantum programs},\ }in\ \href {https://doi.org/10.1109/MICRO56248.2022.00074} {\emph {\bibinfo {booktitle} {2022 55th IEEE/ACM International Symposium on Microarchitecture (MICRO)}}}\ (\bibinfo {year} {2022})\ pp.\ \bibinfo {pages} {1027--1041}\BibitemShut {NoStop}%
\bibitem [{\citenamefont {Ferrari}\ \emph {et~al.}(2021)\citenamefont {Ferrari}, \citenamefont {Cacciapuoti}, \citenamefont {Amoretti},\ and\ \citenamefont {Caleth}}]{Ferrari}%
  \BibitemOpen
  \bibfield  {author} {\bibinfo {author} {\bibfnamefont {D.}~\bibnamefont {Ferrari}}, \bibinfo {author} {\bibfnamefont {A.}~\bibnamefont {Cacciapuoti}}, \bibinfo {author} {\bibfnamefont {M.}~\bibnamefont {Amoretti}},\ and\ \bibinfo {author} {\bibfnamefont {M.}~\bibnamefont {Caleth}},\ }\bibfield  {title} {\bibinfo {title} {Compiler design for distributed quantum computing},\ }\href@noop {} {\bibfield  {journal} {\bibinfo  {journal} {IEEE Transactions on Quantum Engineering}\ }\textbf {\bibinfo {volume} {2}},\ \bibinfo {pages} {1} (\bibinfo {year} {2021})}\BibitemShut {NoStop}%
\bibitem [{\citenamefont {Wu}\ \emph {et~al.}(2023)\citenamefont {Wu}, \citenamefont {Ding},\ and\ \citenamefont {Li}}]{QuComm}%
  \BibitemOpen
  \bibfield  {author} {\bibinfo {author} {\bibfnamefont {A.}~\bibnamefont {Wu}}, \bibinfo {author} {\bibfnamefont {Y.}~\bibnamefont {Ding}},\ and\ \bibinfo {author} {\bibfnamefont {A.}~\bibnamefont {Li}},\ }\bibfield  {title} {\bibinfo {title} {Qucomm: Optimizing collective communication for distributed quantum computing},\ }in\ \href {https://doi.org/10.1145/3613424.3614253} {\emph {\bibinfo {booktitle} {Proceedings of the 56th Annual IEEE/ACM International Symposium on Microarchitecture}}},\ \bibinfo {series and number} {MICRO '23}\ (\bibinfo  {publisher} {Association for Computing Machinery},\ \bibinfo {address} {New York, NY, USA},\ \bibinfo {year} {2023})\ p.\ \bibinfo {pages} {479–493}\BibitemShut {NoStop}%
\bibitem [{\citenamefont {Ferrari}\ \emph {et~al.}(2023)\citenamefont {Ferrari}, \citenamefont {Carretta},\ and\ \citenamefont {Amoretti}}]{ModularDQCcompilingFramework}%
  \BibitemOpen
  \bibfield  {author} {\bibinfo {author} {\bibfnamefont {D.}~\bibnamefont {Ferrari}}, \bibinfo {author} {\bibfnamefont {S.}~\bibnamefont {Carretta}},\ and\ \bibinfo {author} {\bibfnamefont {M.}~\bibnamefont {Amoretti}},\ }\bibfield  {title} {\bibinfo {title} {A modular quantum compilation framework for distributed quantum computing},\ }\href {https://doi.org/10.1109/tqe.2023.3303935} {\bibfield  {journal} {\bibinfo  {journal} {IEEE Transactions on Quantum Engineering}\ }\textbf {\bibinfo {volume} {4}},\ \bibinfo {pages} {1–13} (\bibinfo {year} {2023})}\BibitemShut {NoStop}%
\bibitem [{\citenamefont {JR}\ and\ \citenamefont {Lomonaco}()}]{generalizedGHZandDQC}%
  \BibitemOpen
  \bibfield  {author} {\bibinfo {author} {\bibfnamefont {A.~Y.}\ \bibnamefont {JR}}\ and\ \bibinfo {author} {\bibfnamefont {S.}~\bibnamefont {Lomonaco}},\ }\href@noop {} {\bibinfo {title} {Generalized {GHZ} states and distributed quantum computing}},\ \Eprint {https://arxiv.org/abs/0402148} {arxiv:0402148} \BibitemShut {NoStop}%
\bibitem [{\citenamefont {Yimisiriwattana}\ and\ \citenamefont {Jr}()}]{catCommDistributedShor}%
  \BibitemOpen
  \bibfield  {author} {\bibinfo {author} {\bibfnamefont {A.}~\bibnamefont {Yimisiriwattana}}\ and\ \bibinfo {author} {\bibfnamefont {S.~L.}\ \bibnamefont {Jr}},\ }\href@noop {} {\bibinfo {title} {Distributed quantum computing: A distributed {S}hor alogorithm}},\ \Eprint {https://arxiv.org/abs/0403146v2} {arxiv:0403146v2} \BibitemShut {NoStop}%
\bibitem [{\citenamefont {Bennett}\ \emph {et~al.}(1993)\citenamefont {Bennett}, \citenamefont {Brassard}, \citenamefont {Cr\'epeau}, \citenamefont {Jozsa}, \citenamefont {Peres},\ and\ \citenamefont {Wootters}}]{teleportationProposal}%
  \BibitemOpen
  \bibfield  {author} {\bibinfo {author} {\bibfnamefont {C.~H.}\ \bibnamefont {Bennett}}, \bibinfo {author} {\bibfnamefont {G.}~\bibnamefont {Brassard}}, \bibinfo {author} {\bibfnamefont {C.}~\bibnamefont {Cr\'epeau}}, \bibinfo {author} {\bibfnamefont {R.}~\bibnamefont {Jozsa}}, \bibinfo {author} {\bibfnamefont {A.}~\bibnamefont {Peres}},\ and\ \bibinfo {author} {\bibfnamefont {W.~K.}\ \bibnamefont {Wootters}},\ }\bibfield  {title} {\bibinfo {title} {Teleporting an unknown quantum state via dual classical and {E}instein-{P}odolsky-{R}osen channels},\ }\href {https://doi.org/10.1103/PhysRevLett.70.1895} {\bibfield  {journal} {\bibinfo  {journal} {Physical Review Letters}\ }\textbf {\bibinfo {volume} {70}},\ \bibinfo {pages} {1895} (\bibinfo {year} {1993})}\BibitemShut {NoStop}%
\bibitem [{\citenamefont {Bouwmeester}\ \emph {et~al.}(1997)\citenamefont {Bouwmeester}, \citenamefont {Pan}, \citenamefont {Mattle}, \citenamefont {Eibl}, \citenamefont {Weinfurter},\ and\ \citenamefont {Zeilinger}}]{firstExperimentalTeleportation}%
  \BibitemOpen
  \bibfield  {author} {\bibinfo {author} {\bibfnamefont {D.}~\bibnamefont {Bouwmeester}}, \bibinfo {author} {\bibfnamefont {J.-W.}\ \bibnamefont {Pan}}, \bibinfo {author} {\bibfnamefont {K.}~\bibnamefont {Mattle}}, \bibinfo {author} {\bibfnamefont {M.}~\bibnamefont {Eibl}}, \bibinfo {author} {\bibfnamefont {H.}~\bibnamefont {Weinfurter}},\ and\ \bibinfo {author} {\bibfnamefont {A.}~\bibnamefont {Zeilinger}},\ }\bibfield  {title} {\bibinfo {title} {Experimental quantum teleportation},\ }\href {https://doi.org/10.1038/37539} {\bibfield  {journal} {\bibinfo  {journal} {Nature}\ }\textbf {\bibinfo {volume} {390}},\ \bibinfo {pages} {575} (\bibinfo {year} {1997})}\BibitemShut {NoStop}%
\bibitem [{\citenamefont {Eisert}\ \emph {et~al.}(2000)\citenamefont {Eisert}, \citenamefont {Jacobs}, \citenamefont {Papadopoulos},\ and\ \citenamefont {Plenio}}]{heightIncreaseNonLocalGates}%
  \BibitemOpen
  \bibfield  {author} {\bibinfo {author} {\bibfnamefont {J.}~\bibnamefont {Eisert}}, \bibinfo {author} {\bibfnamefont {K.}~\bibnamefont {Jacobs}}, \bibinfo {author} {\bibfnamefont {P.}~\bibnamefont {Papadopoulos}},\ and\ \bibinfo {author} {\bibfnamefont {M.}~\bibnamefont {Plenio}},\ }\bibfield  {title} {\bibinfo {title} {Optimal local implementaion of nonlocal quantum gates},\ }\href@noop {} {\bibfield  {journal} {\bibinfo  {journal} {Physical Review A}\ }\textbf {\bibinfo {volume} {62}},\ \bibinfo {eid} {052317} (\bibinfo {year} {2000})}\BibitemShut {NoStop}%
\bibitem [{\citenamefont {Dadkhah}\ \emph {et~al.}(2022)\citenamefont {Dadkhah}, \citenamefont {Zomorodi}, \citenamefont {Hosseini}, \citenamefont {Plawiak},\ and\ \citenamefont {Zhou}}]{reordingPartitionDQCcircuitsTPcommOnly}%
  \BibitemOpen
  \bibfield  {author} {\bibinfo {author} {\bibfnamefont {D.}~\bibnamefont {Dadkhah}}, \bibinfo {author} {\bibfnamefont {M.}~\bibnamefont {Zomorodi}}, \bibinfo {author} {\bibfnamefont {S.~E.}\ \bibnamefont {Hosseini}}, \bibinfo {author} {\bibfnamefont {P.}~\bibnamefont {Plawiak}},\ and\ \bibinfo {author} {\bibfnamefont {X.}~\bibnamefont {Zhou}},\ }\bibfield  {title} {\bibinfo {title} {Reordering and partitioning of distributed quantum circuits},\ }\href@noop {} {\bibfield  {journal} {\bibinfo  {journal} {IEEE ACCESS}\ }\textbf {\bibinfo {volume} {10}},\ \bibinfo {pages} {70329} (\bibinfo {year} {2022})}\BibitemShut {NoStop}%
\bibitem [{\citenamefont {Daei}\ \emph {et~al.}(2020)\citenamefont {Daei}, \citenamefont {Navi},\ and\ \citenamefont {Zomorodi-Moghadam}}]{OptimisedQuantumCircuitPartitioning}%
  \BibitemOpen
  \bibfield  {author} {\bibinfo {author} {\bibfnamefont {O.}~\bibnamefont {Daei}}, \bibinfo {author} {\bibfnamefont {K.}~\bibnamefont {Navi}},\ and\ \bibinfo {author} {\bibfnamefont {M.}~\bibnamefont {Zomorodi-Moghadam}},\ }\bibfield  {title} {\bibinfo {title} {Optimized quantum circuit partitioning},\ }\href@noop {} {\bibfield  {journal} {\bibinfo  {journal} {International Journal of Theoretical Physics}\ }\textbf {\bibinfo {volume} {59}},\ \bibinfo {pages} {3804} (\bibinfo {year} {2020})}\BibitemShut {NoStop}%
\bibitem [{\citenamefont {Andres-Martinez}\ and\ \citenamefont {Heunen}(2019)}]{AutomatedDistributionCircuitsViaHypergraph}%
  \BibitemOpen
  \bibfield  {author} {\bibinfo {author} {\bibfnamefont {P.}~\bibnamefont {Andres-Martinez}}\ and\ \bibinfo {author} {\bibfnamefont {C.}~\bibnamefont {Heunen}},\ }\bibfield  {title} {\bibinfo {title} {Automated distribution of quantum circuits via hypergraph partitioning},\ }\href@noop {} {\bibfield  {journal} {\bibinfo  {journal} {Physical Review A}\ }\textbf {\bibinfo {volume} {100}},\ \bibinfo {eid} {032308} (\bibinfo {year} {2019})}\BibitemShut {NoStop}%
\bibitem [{\citenamefont {Sundaram}\ \emph {et~al.}(2021)\citenamefont {Sundaram}, \citenamefont {Gupta},\ and\ \citenamefont {Ramakrishnan}}]{EfficientDistributionQuantumCircuits}%
  \BibitemOpen
  \bibfield  {author} {\bibinfo {author} {\bibfnamefont {R.}~\bibnamefont {Sundaram}}, \bibinfo {author} {\bibfnamefont {H.}~\bibnamefont {Gupta}},\ and\ \bibinfo {author} {\bibfnamefont {R.}~\bibnamefont {Ramakrishnan}},\ }\bibfield  {title} {\bibinfo {title} {Efficient distribution of quantum circuits},\ }in\ \href@noop {} {\emph {\bibinfo {booktitle} {35th International Symposium on Distributed Computing}}}\ (\bibinfo {year} {2021})\ pp.\ \bibinfo {pages} {41:1--41:20}\BibitemShut {NoStop}%
\bibitem [{\citenamefont {Ghodsollahee}\ \emph {et~al.}(2021)\citenamefont {Ghodsollahee}, \citenamefont {Davarzani}, \citenamefont {Zomorodi}, \citenamefont {Plawiak}, \citenamefont {Houshmand},\ and\ \citenamefont {Houshmand}}]{ConnectivityMatrixCompiler}%
  \BibitemOpen
  \bibfield  {author} {\bibinfo {author} {\bibfnamefont {I.}~\bibnamefont {Ghodsollahee}}, \bibinfo {author} {\bibfnamefont {Z.}~\bibnamefont {Davarzani}}, \bibinfo {author} {\bibfnamefont {M.}~\bibnamefont {Zomorodi}}, \bibinfo {author} {\bibfnamefont {P.}~\bibnamefont {Plawiak}}, \bibinfo {author} {\bibfnamefont {M.}~\bibnamefont {Houshmand}},\ and\ \bibinfo {author} {\bibfnamefont {M.}~\bibnamefont {Houshmand}},\ }\bibfield  {title} {\bibinfo {title} {Connectivity matrix model of quantum circuits and its application to distributed quantum circuit optimization},\ }\href@noop {} {\bibfield  {journal} {\bibinfo  {journal} {Quantum Information Processing}\ }\textbf {\bibinfo {volume} {20}} (\bibinfo {year} {2021})}\BibitemShut {NoStop}%
\bibitem [{\citenamefont {Zomorodi-Moghadam}\ \emph {et~al.}(2018)\citenamefont {Zomorodi-Moghadam}, \citenamefont {Houshmand},\ and\ \citenamefont {Houshmand}}]{OptimisingTeleportaionCost}%
  \BibitemOpen
  \bibfield  {author} {\bibinfo {author} {\bibfnamefont {M.}~\bibnamefont {Zomorodi-Moghadam}}, \bibinfo {author} {\bibfnamefont {M.}~\bibnamefont {Houshmand}},\ and\ \bibinfo {author} {\bibfnamefont {M.}~\bibnamefont {Houshmand}},\ }\bibfield  {title} {\bibinfo {title} {Optimizing teleportation cost in distributed quantum circuits},\ }\href@noop {} {\bibfield  {journal} {\bibinfo  {journal} {International Journal of Theoretical Physics}\ }\textbf {\bibinfo {volume} {50}},\ \bibinfo {pages} {848} (\bibinfo {year} {2018})}\BibitemShut {NoStop}%
\bibitem [{\citenamefont {Davis}\ \emph {et~al.}()\citenamefont {Davis}, \citenamefont {Chung}, \citenamefont {Englund},\ and\ \citenamefont {Kettimuthu}}]{davis2023dqcPartitioning}%
  \BibitemOpen
  \bibfield  {author} {\bibinfo {author} {\bibfnamefont {M.~G.}\ \bibnamefont {Davis}}, \bibinfo {author} {\bibfnamefont {J.}~\bibnamefont {Chung}}, \bibinfo {author} {\bibfnamefont {D.}~\bibnamefont {Englund}},\ and\ \bibinfo {author} {\bibfnamefont {R.}~\bibnamefont {Kettimuthu}},\ }\href@noop {} {\bibinfo {title} {Towards distributed quantum computing by qubit and gate graph partitioning techniques}},\ \Eprint {https://arxiv.org/abs/2310.03942} {arXiv:2310.03942 [quant-ph]} \BibitemShut {NoStop}%
\bibitem [{\citenamefont {Jozsa}(1994)}]{fidelityDef}%
  \BibitemOpen
  \bibfield  {author} {\bibinfo {author} {\bibfnamefont {R.}~\bibnamefont {Jozsa}},\ }\bibfield  {title} {\bibinfo {title} {Fidelity for mixed quantum states},\ }\href {https://doi.org/10.1080/09500349414552171} {\bibfield  {journal} {\bibinfo  {journal} {Journal of Modern Optics}\ }\textbf {\bibinfo {volume} {41}},\ \bibinfo {pages} {2315} (\bibinfo {year} {1994})}\BibitemShut {NoStop}%
\bibitem [{\citenamefont {Cirac}\ \emph {et~al.}(1999)\citenamefont {Cirac}, \citenamefont {Ekert}, \citenamefont {Huelga},\ and\ \citenamefont {Macchiavello}}]{DQCoverNoisyChannels}%
  \BibitemOpen
  \bibfield  {author} {\bibinfo {author} {\bibfnamefont {J.~I.}\ \bibnamefont {Cirac}}, \bibinfo {author} {\bibfnamefont {A.~K.}\ \bibnamefont {Ekert}}, \bibinfo {author} {\bibfnamefont {S.~F.}\ \bibnamefont {Huelga}},\ and\ \bibinfo {author} {\bibfnamefont {C.}~\bibnamefont {Macchiavello}},\ }\bibfield  {title} {\bibinfo {title} {Distributed quantum computation over noisy channels},\ }\href {https://doi.org/10.1103/PhysRevA.59.4249} {\bibfield  {journal} {\bibinfo  {journal} {Physical Review A}\ }\textbf {\bibinfo {volume} {59}},\ \bibinfo {pages} {4249} (\bibinfo {year} {1999})}\BibitemShut {NoStop}%
\bibitem [{\citenamefont {Kitaev}()}]{quantumPhaseEstimation}%
  \BibitemOpen
  \bibfield  {author} {\bibinfo {author} {\bibfnamefont {A.~Y.}\ \bibnamefont {Kitaev}},\ }\href@noop {} {\bibinfo {title} {Quantum measurements and the abelian stabilizer problem}},\ \Eprint {https://arxiv.org/abs/quant-ph/9511026} {arXiv:quant-ph/9511026 [quant-ph]} \BibitemShut {NoStop}%
\bibitem [{\citenamefont {Nielsen}\ and\ \citenamefont {Chuang}(2010)}]{NielsenChuang}%
  \BibitemOpen
  \bibfield  {author} {\bibinfo {author} {\bibfnamefont {M.}~\bibnamefont {Nielsen}}\ and\ \bibinfo {author} {\bibfnamefont {I.}~\bibnamefont {Chuang}},\ }\href@noop {} {\emph {\bibinfo {title} {Quantum Computation and Quantum Information}}},\ \bibinfo {edition} {10th}\ ed.\ (\bibinfo  {publisher} {Cambridge University Press},\ \bibinfo {year} {2010})\BibitemShut {NoStop}%
\bibitem [{\citenamefont {Ezzell}\ \emph {et~al.}(2023)\citenamefont {Ezzell}, \citenamefont {Pokharel}, \citenamefont {Tewala}, \citenamefont {Quiroz},\ and\ \citenamefont {Lidar}}]{dynamicalDecouplingSurvey}%
  \BibitemOpen
  \bibfield  {author} {\bibinfo {author} {\bibfnamefont {N.}~\bibnamefont {Ezzell}}, \bibinfo {author} {\bibfnamefont {B.}~\bibnamefont {Pokharel}}, \bibinfo {author} {\bibfnamefont {L.}~\bibnamefont {Tewala}}, \bibinfo {author} {\bibfnamefont {G.}~\bibnamefont {Quiroz}},\ and\ \bibinfo {author} {\bibfnamefont {D.~A.}\ \bibnamefont {Lidar}},\ }\bibfield  {title} {\bibinfo {title} {Dynamical decoupling for superconducting qubits: A performance survey},\ }\href {https://doi.org/10.1103/PhysRevApplied.20.064027} {\bibfield  {journal} {\bibinfo  {journal} {Physical Review Appl.}\ }\textbf {\bibinfo {volume} {20}},\ \bibinfo {pages} {064027} (\bibinfo {year} {2023})}\BibitemShut {NoStop}%
\bibitem [{\citenamefont {Neumann}\ \emph {et~al.}(2020)\citenamefont {Neumann}, \citenamefont {van Houte},\ and\ \citenamefont {Attema}}]{decoherenceSharedEntanglementDQCphaseEstimation}%
  \BibitemOpen
  \bibfield  {author} {\bibinfo {author} {\bibfnamefont {N.~M.~P.}\ \bibnamefont {Neumann}}, \bibinfo {author} {\bibfnamefont {R.}~\bibnamefont {van Houte}},\ and\ \bibinfo {author} {\bibfnamefont {T.}~\bibnamefont {Attema}},\ }\bibfield  {title} {\bibinfo {title} {Imperfect distributed quantum phase estimation}\ }in\ {\emph {\bibinfo {booktitle} {Computational Science – ICCS 2020}}},\ \bibinfo {series and number} {Lecture Notes in Computer Science}\ (\bibinfo  {publisher} {Springer},\  
  \bibinfo {year} {2020}) \BibitemShut {NoStop}%
\bibitem [{\citenamefont {Bennett}\ \emph {et~al.}(1996{\natexlab{a}})\citenamefont {Bennett}, \citenamefont {Brassard}, \citenamefont {Popescu}, \citenamefont {Schumacher}, \citenamefont {Smolin},\ and\ \citenamefont {Wootters}}]{distillation}%
  \BibitemOpen
  \bibfield  {author} {\bibinfo {author} {\bibfnamefont {C.~H.}\ \bibnamefont {Bennett}}, \bibinfo {author} {\bibfnamefont {G.}~\bibnamefont {Brassard}}, \bibinfo {author} {\bibfnamefont {S.}~\bibnamefont {Popescu}}, \bibinfo {author} {\bibfnamefont {B.}~\bibnamefont {Schumacher}}, \bibinfo {author} {\bibfnamefont {J.~A.}\ \bibnamefont {Smolin}},\ and\ \bibinfo {author} {\bibfnamefont {W.~K.}\ \bibnamefont {Wootters}},\ }\bibfield  {title} {\bibinfo {title} {Purification of noisy entanglement and faithful teleportation via noisy channels},\ }\href@noop {} {\bibfield  {journal} {\bibinfo  {journal} {Physical Review Letters}\ }\textbf {\bibinfo {volume} {76}} (\bibinfo {year} {1996}{\natexlab{a}})}\BibitemShut {NoStop}%
\bibitem [{\citenamefont {{Munro}}\ \emph {et~al.}(2010)\citenamefont {{Munro}}, \citenamefont {{Harrison}}, \citenamefont {{Stephens}}, \citenamefont {{Devitt}},\ and\ \citenamefont {{Nemoto}}}]{QMultiplexingToNetwork}%
  \BibitemOpen
  \bibfield  {author} {\bibinfo {author} {\bibfnamefont {W.~J.}\ \bibnamefont {{Munro}}}, \bibinfo {author} {\bibfnamefont {K.~A.}\ \bibnamefont {{Harrison}}}, \bibinfo {author} {\bibfnamefont {A.~M.}\ \bibnamefont {{Stephens}}}, \bibinfo {author} {\bibfnamefont {S.~J.}\ \bibnamefont {{Devitt}}},\ and\ \bibinfo {author} {\bibfnamefont {K.}~\bibnamefont {{Nemoto}}},\ }\bibfield  {title} {\bibinfo {title} {{From quantum multiplexing to high-performance quantum networking}},\ }\href {https://doi.org/10.1038/nphoton.2010.213} {\bibfield  {journal} {\bibinfo  {journal} {Nature Photonics}\ }\textbf {\bibinfo {volume} {4}},\ \bibinfo {pages} {792} (\bibinfo {year} {2010})},\ \Eprint {https://arxiv.org/abs/0910.4038} {arXiv:0910.4038 [quant-ph]} \BibitemShut {NoStop}%
\bibitem [{\citenamefont {Li}\ and\ \citenamefont {Benjamin}(2016)}]{BenjaminHierarchicalSurfaceCode}%
  \BibitemOpen
  \bibfield  {author} {\bibinfo {author} {\bibfnamefont {Y.}~\bibnamefont {Li}}\ and\ \bibinfo {author} {\bibfnamefont {S.~C.}\ \bibnamefont {Benjamin}},\ }\bibfield  {title} {\bibinfo {title} {Hierarchical surface code for network quantum computing with modules of arbitrary size},\ }\href {https://doi.org/10.1103/PhysRevA.94.042303} {\bibfield  {journal} {\bibinfo  {journal} {Physical Review A}\ }\textbf {\bibinfo {volume} {94}},\ \bibinfo {pages} {042303} (\bibinfo {year} {2016})},\ \BibitemShut {NoStop}%
\bibitem [{\citenamefont {Briegel}\ \emph {et~al.}(1998)\citenamefont {Briegel}, \citenamefont {D\"{u}r}, \citenamefont {Cirac},\ and\ \citenamefont {Zoller}}]{imperfectRepeaters}%
  \BibitemOpen
  \bibfield  {author} {\bibinfo {author} {\bibfnamefont {H.}~\bibnamefont {Briegel}}, \bibinfo {author} {\bibfnamefont {W.}~\bibnamefont {D\"{u}r}}, \bibinfo {author} {\bibfnamefont {J.}~\bibnamefont {Cirac}},\ and\ \bibinfo {author} {\bibfnamefont {P.}~\bibnamefont {Zoller}},\ }\bibfield  {title} {\bibinfo {title} {Quantum repeaters: The role of imperfect local operations in quantum communication},\ }\href@noop {} {\bibfield  {journal} {\bibinfo  {journal} {Physical Review Letters}\ }\textbf {\bibinfo {volume} {81}},\ \bibinfo {eid} {5932} (\bibinfo {year} {1998})}\BibitemShut {NoStop}%
\bibitem [{\citenamefont {Razavi}(2023)}]{MohsenBookChapter}%
  \BibitemOpen
  \bibfield  {author} {\bibinfo {author} {\bibfnamefont {M.}~\bibnamefont {Razavi}},\ }\bibinfo {title} {Fiber-based quantum repeaters},\ in\ \href@noop {} {\emph {\bibinfo {booktitle} {Photonic Quantum Technologies}}}\ (\bibinfo  {publisher} {John Wiley \& Sons, Ltd},\ \bibinfo {year} {2023})\ Chap.~\bibinfo {chapter} {24}, pp.\ \bibinfo {pages} {675--691}\BibitemShut {NoStop}%
\bibitem [{\citenamefont {Bennett}\ \emph {et~al.}(1996{\natexlab{b}})\citenamefont {Bennett}, \citenamefont {Brassard}, \citenamefont {Popescu}, \citenamefont {Schumacher}, \citenamefont {Smolin},\ and\ \citenamefont {Wootters}}]{originalDistillationProposal}%
  \BibitemOpen
  \bibfield  {author} {\bibinfo {author} {\bibfnamefont {C.~H.}\ \bibnamefont {Bennett}}, \bibinfo {author} {\bibfnamefont {G.}~\bibnamefont {Brassard}}, \bibinfo {author} {\bibfnamefont {S.}~\bibnamefont {Popescu}}, \bibinfo {author} {\bibfnamefont {B.}~\bibnamefont {Schumacher}}, \bibinfo {author} {\bibfnamefont {J.~A.}\ \bibnamefont {Smolin}},\ and\ \bibinfo {author} {\bibfnamefont {W.~K.}\ \bibnamefont {Wootters}},\ }\bibfield  {title} {\bibinfo {title} {Purification of noisy entanglement and faithful teleportation via noisy channels},\ }\href {https://doi.org/10.1103/PhysRevLett.76.722} {\bibfield  {journal} {\bibinfo  {journal} {Physical Review Letters}\ }\textbf {\bibinfo {volume} {76}},\ \bibinfo {pages} {722} (\bibinfo {year} {1996}{\natexlab{b}})}\BibitemShut {NoStop}%
\bibitem [{\citenamefont {Deutsch}\ \emph {et~al.}(1996)\citenamefont {Deutsch}, \citenamefont {Ekert}, \citenamefont {Jozsa}, \citenamefont {Macchiavello}, \citenamefont {Popescu},\ and\ \citenamefont {Sanpera}}]{PrivacyAmpAndSecurityOverNoisyChannels}%
  \BibitemOpen
  \bibfield  {author} {\bibinfo {author} {\bibfnamefont {D.}~\bibnamefont {Deutsch}}, \bibinfo {author} {\bibfnamefont {A.}~\bibnamefont {Ekert}}, \bibinfo {author} {\bibfnamefont {R.}~\bibnamefont {Jozsa}}, \bibinfo {author} {\bibfnamefont {C.}~\bibnamefont {Macchiavello}}, \bibinfo {author} {\bibfnamefont {S.}~\bibnamefont {Popescu}},\ and\ \bibinfo {author} {\bibfnamefont {A.}~\bibnamefont {Sanpera}},\ }\bibfield  {title} {\bibinfo {title} {Quantum privacy amplification and the security of quantum cryptography over noisy channels},\ }\href {https://doi.org/10.1103/PhysRevLett.77.2818} {\bibfield  {journal} {\bibinfo  {journal} {Physical Review Letters}\ }\textbf {\bibinfo {volume} {77}},\ \bibinfo {pages} {2818} (\bibinfo {year} {1996})}\BibitemShut {NoStop}%
\bibitem [{\citenamefont {Pan}\ \emph {et~al.}(2001)\citenamefont {Pan}, \citenamefont {Simon}, \citenamefont {Brukner},\ and\ \citenamefont {Zeilinger}}]{EntanglementPurificationForQuantumComms}%
  \BibitemOpen
  \bibfield  {author} {\bibinfo {author} {\bibfnamefont {J.-W.}\ \bibnamefont {Pan}}, \bibinfo {author} {\bibfnamefont {C.}~\bibnamefont {Simon}}, \bibinfo {author} {\bibfnamefont {{\v C}.}~\bibnamefont {Brukner}},\ and\ \bibinfo {author} {\bibfnamefont {A.}~\bibnamefont {Zeilinger}},\ }\bibfield  {title} {\bibinfo {title} {Entanglement purification for quantum communication},\ }\href@noop {} {\bibfield  {journal} {\bibinfo  {journal} {Nature}\ }\textbf {\bibinfo {volume} {410}},\ \bibinfo {pages} {1067} (\bibinfo {year} {2001})}\BibitemShut {NoStop}%
\bibitem [{\citenamefont {{Cross}}\ \emph {et~al.}()\citenamefont {{Cross}}, \citenamefont {{Bishop}}, \citenamefont {{Smolin}},\ and\ \citenamefont {{Gambetta}}}]{OpenQASM2.0_paper}%
  \BibitemOpen
  \bibfield  {author} {\bibinfo {author} {\bibfnamefont {A.~W.}\ \bibnamefont {{Cross}}}, \bibinfo {author} {\bibfnamefont {L.~S.}\ \bibnamefont {{Bishop}}}, \bibinfo {author} {\bibfnamefont {J.~A.}\ \bibnamefont {{Smolin}}},\ and\ \bibinfo {author} {\bibfnamefont {J.~M.}\ \bibnamefont {{Gambetta}}},\ }\href@noop {} {\bibinfo {title} {{Open Quantum Assembly Language}}},\ \Eprint {https://arxiv.org/abs/1707.03429} {arXiv:1707.03429 [quant-ph]} \BibitemShut {NoStop}%
\bibitem [{\citenamefont {Flannigan}\ \emph {et~al.}(2022)\citenamefont {Flannigan}, \citenamefont {Pearson}, \citenamefont {Low}, \citenamefont {Buyskikh}, \citenamefont {Bloch}, \citenamefont {Zoller}, \citenamefont {Troyer},\ and\ \citenamefont {Daley}}]{Flannigan_2022}%
  \BibitemOpen
  \bibfield  {author} {\bibinfo {author} {\bibfnamefont {S.}~\bibnamefont {Flannigan}}, \bibinfo {author} {\bibfnamefont {N.}~\bibnamefont {Pearson}}, \bibinfo {author} {\bibfnamefont {G.~H.}\ \bibnamefont {Low}}, \bibinfo {author} {\bibfnamefont {A.}~\bibnamefont {Buyskikh}}, \bibinfo {author} {\bibfnamefont {I.}~\bibnamefont {Bloch}}, \bibinfo {author} {\bibfnamefont {P.}~\bibnamefont {Zoller}}, \bibinfo {author} {\bibfnamefont {M.}~\bibnamefont {Troyer}},\ and\ \bibinfo {author} {\bibfnamefont {A.~J.}\ \bibnamefont {Daley}},\ }\bibfield  {title} {\bibinfo {title} {Propagation of errors and quantitative quantum simulation with quantum advantage},\ }\href {https://doi.org/10.1088/2058-9565/ac88f5} {\bibfield  {journal} {\bibinfo  {journal} {Quantum Science and Technology}\ }\textbf {\bibinfo {volume} {7}},\ \bibinfo {pages} {045025} (\bibinfo {year} {2022})}\BibitemShut {NoStop}%
\bibitem [{\citenamefont {Coopmans}\ \emph {et~al.}(2021)\citenamefont {Coopmans} \emph {et~al.}}]{netsquid}%
  \BibitemOpen
  \bibfield  {author} {\bibinfo {author} {\bibfnamefont {T.}~\bibnamefont {Coopmans}} \emph {et~al.},\ }\bibfield  {title} {\bibinfo {title} {Netsquid, a discrete-event simulation platform for quantum networks},\ }\href@noop {} {\bibfield  {journal} {\bibinfo  {journal} {Communications Physics 4, 164}\ }\textbf {\bibinfo {volume} {4}} (\bibinfo {year} {2021})}\BibitemShut {NoStop}%
\bibitem [{nuq(2022)}]{nuqasm2}%
  \BibitemOpen
    \bibfield  {author} {\bibinfo {author} {\bibfnamefont {J.}~\bibnamefont {Woehr}},}
  \href@noop {} {\bibinfo {title} {nuqasm2}},\ \bibinfo {howpublished} {\url{https://github.com/jwoehr/nuqasm2}} (\bibinfo {year} {2022})\BibitemShut {NoStop}%
\bibitem [{\citenamefont {Stephenson}\ \emph {et~al.}(2020)\citenamefont {Stephenson}, \citenamefont {Nadlinger}, \citenamefont {Nichol}, \citenamefont {An}, \citenamefont {Drmota}, \citenamefont {Ballance}, \citenamefont {Thirumalai}, \citenamefont {Goodwin}, \citenamefont {Lucas},\ and\ \citenamefont {Ballance}}]{ionTrapEntDist94percent2m}%
  \BibitemOpen
  \bibfield  {author} {\bibinfo {author} {\bibfnamefont {L.~J.}\ \bibnamefont {Stephenson}}, \bibinfo {author} {\bibfnamefont {D.~P.}\ \bibnamefont {Nadlinger}}, \bibinfo {author} {\bibfnamefont {B.~C.}\ \bibnamefont {Nichol}}, \bibinfo {author} {\bibfnamefont {S.}~\bibnamefont {An}}, \bibinfo {author} {\bibfnamefont {P.}~\bibnamefont {Drmota}}, \bibinfo {author} {\bibfnamefont {T.~G.}\ \bibnamefont {Ballance}}, \bibinfo {author} {\bibfnamefont {K.}~\bibnamefont {Thirumalai}}, \bibinfo {author} {\bibfnamefont {J.~F.}\ \bibnamefont {Goodwin}}, \bibinfo {author} {\bibfnamefont {D.~M.}\ \bibnamefont {Lucas}},\ and\ \bibinfo {author} {\bibfnamefont {C.~J.}\ \bibnamefont {Ballance}},\ }\bibfield  {title} {\bibinfo {title} {High-rate, high-fidelity entanglement of qubits across an elementary quantum network},\ }\href {https://doi.org/10.1103/PhysRevLett.124.110501} {\bibfield  {journal} {\bibinfo  {journal} {Physical Review Letters}\ }\textbf {\bibinfo {volume} {124}},\ \bibinfo {pages} {110501} (\bibinfo {year}
  {2020})}\BibitemShut {NoStop}%
\bibitem [{ion(2023)}]{ionQAriaSpecs}%
  \BibitemOpen
  \href@noop {}   \bibfield  {author} {\bibinfo {author} {\bibfnamefont {IonQ}~\bibnamefont {staff}} ,\ } {\bibinfo {title} {Ionq aria: Practical performance}},\ \bibinfo {howpublished} {\url{https://ionq.com/resources/ionq-aria-practical-performance}} (\bibinfo {year} {2023})\BibitemShut {NoStop}%
\bibitem [{\citenamefont {Metodi}\ \emph {et~al.}()\citenamefont {Metodi}, \citenamefont {Thaker}, \citenamefont {Cross}, \citenamefont {Chong},\ and\ \citenamefont {Chuang}}]{metodi2005quantum}%
  \BibitemOpen
  \bibfield  {author} {\bibinfo {author} {\bibfnamefont {T.~S.}\ \bibnamefont {Metodi}}, \bibinfo {author} {\bibfnamefont {D.~D.}\ \bibnamefont {Thaker}}, \bibinfo {author} {\bibfnamefont {A.~W.}\ \bibnamefont {Cross}}, \bibinfo {author} {\bibfnamefont {F.~T.}\ \bibnamefont {Chong}},\ and\ \bibinfo {author} {\bibfnamefont {I.~L.}\ \bibnamefont {Chuang}},\ }\href@noop {} {\bibinfo {title} {A quantum logic array microarchitecture: Scalable quantum data movement and computation}},\ \Eprint {https://arxiv.org/abs/0509051} {arXiv:0509051 [quant-ph]} \BibitemShut {NoStop}%
\bibitem [{\citenamefont {Yu}\ and\ \citenamefont {Li}()}]{errorPropagationMonoQC}%
  \BibitemOpen
  \bibfield  {author} {\bibinfo {author} {\bibfnamefont {Z.}~\bibnamefont {Yu}}\ and\ \bibinfo {author} {\bibfnamefont {Y.}~\bibnamefont {Li}},\ }\href@noop {} {\bibinfo {title} {Analysis of error propagation in quantum computers}},\ \Eprint {https://arxiv.org/abs/2209.01699} {arXiv:2209.01699 [quant-ph]} \BibitemShut {NoStop}%
\bibitem [{\citenamefont {Luo}\ and\ \citenamefont {Li}(2016)}]{DQCassistedByRemoteToffoli}%
  \BibitemOpen
  \bibfield  {author} {\bibinfo {author} {\bibfnamefont {M.-X.}\ \bibnamefont {Luo}}\ and\ \bibinfo {author} {\bibfnamefont {H.-R.}\ \bibnamefont {Li}},\ }\bibfield  {title} {\bibinfo {title} {Distributed quantum computation assisted by remote toffoli gate},\ }in\ \href@noop {} {\emph {\bibinfo {booktitle} {Cloud Computing and Security}}},\ \bibinfo {editor} {edited by\ \bibinfo {editor} {\bibfnamefont {X.}~\bibnamefont {Sun}}, \bibinfo {editor} {\bibfnamefont {A.}~\bibnamefont {Liu}}, \bibinfo {editor} {\bibfnamefont {H.-C.}\ \bibnamefont {Chao}},\ and\ \bibinfo {editor} {\bibfnamefont {E.}~\bibnamefont {Bertino}}}\ (\bibinfo  {publisher} {Springer International Publishing},\ \bibinfo {year} {2016})\ pp.\ \bibinfo {pages} {475--485}\BibitemShut {NoStop}%
\bibitem [{\citenamefont {Nickerson}(2015)}]{NickersonDQCfaultTolerantThesis}%
  \BibitemOpen
  \bibfield  {author} {\bibinfo {author} {\bibfnamefont {N.}~\bibnamefont {Nickerson}},\ }\emph {\bibinfo {title} {Practical Fault-Tolerant Quantum Computing}},\ \href@noop {} {Ph.D. thesis},\ \bibinfo  {school} {Imperial College London} (\bibinfo {year} {2015})\BibitemShut {NoStop}%
\bibitem [{\citenamefont {Sarvaghad-Moghaddam}\ and\ \citenamefont {Zomorodi}(2021)}]{GeneralProtocolForDistributedGates}%
  \BibitemOpen
  \bibfield  {author} {\bibinfo {author} {\bibfnamefont {M.}~\bibnamefont {Sarvaghad-Moghaddam}}\ and\ \bibinfo {author} {\bibfnamefont {M.}~\bibnamefont {Zomorodi}},\ }\bibfield  {title} {\bibinfo {title} {A general protocol for distributed quantum gates},\ }\href {https://doi.org/10.1007/s11128-021-03191-0} {\bibfield  {journal} {\bibinfo  {journal} {Quantum Information Processing}\ }\textbf {\bibinfo {volume} {20}},\ \bibinfo {pages} {265} (\bibinfo {year} {2021})}\BibitemShut {NoStop}%
\bibitem [{\citenamefont {{Quetschlich}}\ \emph {et~al.}(2023)\citenamefont {{Quetschlich}}, \citenamefont {{Burgholzer}},\ and\ \citenamefont {{Wille}}}]{MQTBench}%
  \BibitemOpen
  \bibfield  {author} {\bibinfo {author} {\bibfnamefont {N.}~\bibnamefont {{Quetschlich}}}, \bibinfo {author} {\bibfnamefont {L.}~\bibnamefont {{Burgholzer}}},\ and\ \bibinfo {author} {\bibfnamefont {R.}~\bibnamefont {{Wille}}},\ }\bibfield  {title} {\bibinfo {title} {{MQT Bench: Benchmarking Software and Design Automation Tools for Quantum Computing}},\ }\href@noop {} {\bibfield  {journal} {\bibinfo  {journal} {Quantum}\ }\textbf {\bibinfo {volume} {7}},\ \bibinfo {pages} {1062} (\bibinfo {year} {2023})}\BibitemShut {NoStop}%
\bibitem [{\citenamefont {Main}\ \emph {et~al.}(2024)\citenamefont {Main}, \citenamefont {Drmota}, \citenamefont {Nadlinger}, \citenamefont {Ainley}, \citenamefont {Agrawal}, \citenamefont {Nichol}, \citenamefont {Srinivas}, \citenamefont {Araneda},\ and\ \citenamefont {Lucas}}]{firstDeterministicQDC}%
  \BibitemOpen
  \bibfield  {author} {\bibinfo {author} {\bibfnamefont {D.}~\bibnamefont {Main}}, \bibinfo {author} {\bibfnamefont {P.}~\bibnamefont {Drmota}}, \bibinfo {author} {\bibfnamefont {D.~P.}\ \bibnamefont {Nadlinger}}, \bibinfo {author} {\bibfnamefont {E.~M.}\ \bibnamefont {Ainley}}, \bibinfo {author} {\bibfnamefont {A.}~\bibnamefont {Agrawal}}, \bibinfo {author} {\bibfnamefont {B.~C.}\ \bibnamefont {Nichol}}, \bibinfo {author} {\bibfnamefont {R.}~\bibnamefont {Srinivas}}, \bibinfo {author} {\bibfnamefont {G.}~\bibnamefont {Araneda}},\ and\ \bibinfo {author} {\bibfnamefont {D.~M.}\ \bibnamefont {Lucas}},\ }\href {https://arxiv.org/abs/2407.00835} {\bibinfo {title} {Distributed quantum computing across an optical network link}} (\bibinfo {year} {2024}),\ \Eprint {https://arxiv.org/abs/2407.00835} {arXiv:2407.00835 [quant-ph]} \BibitemShut {NoStop}%
\end{thebibliography}
\end{document}